\documentclass[10pt]{iopart}

\usepackage[square,sort&compress, numbers]{natbib}
\newcommand{\ovb}[1]{ \bm{#1} \mkern2mu\vphantom{#1}}
\usepackage{verbatim}
\usepackage{iopams}
\usepackage{slashed}
\usepackage{bm}
\usepackage{graphicx}

\begin{document}

\title{Optical theorem for 2d magnetoelectric quadrupolar arrays }

\author{Sylvia D. Swiecicki$^*$ and J. E. Sipe }

\address{Department of Physics, University of Toronto, Toronto, Ontario, Canada M5S1A7}
\ead{$^*$sswiecic@physics.utoronto.ca}
\vspace{10pt}
\begin{indented}
\item[]\today
\end{indented}

\begin{abstract}
We present all the periodic Green function dyadics that enter a description of a 2d array of emitters at the level that includes the electric dipole, magnetic dipole and electric quadrupole moment of each emitter. We find a concise analytic form for the radiative contributions to the periodic Green function dyadics that give rise to radiation reaction fields, and we give the non-radiative contributions that do not affect energy balance in the form of rapidly converging series. Finally, we present an approximation scheme for evaluating periodic Green function dyadics at long wavelengths that rigorously respects energy conservation. The scheme extends the range of validity of the usual static approximation by the inclusion of a simple dynamic correction. 
\end{abstract}

\vspace{2pc}
\maketitle
\ioptwocol

\section{Introduction}
Resonances in the optical response of metallic and dielectric nano-particles are limited by absorption and radiative losses.  In a collection of nano-particles the radiative losses can exhibit constructive or destructive interference \cite{Lukyanchuk2010}, and for nano-particles arranged in a lattice the radiation can be controlled by adjusting the lattice spacing and structure \cite{DeAbajo2007, Humphrey2014, Lunnemann2013}.  Both sub-radiant \cite{Fedotov2010, Jenkins2012} and super-radiant \cite{Iida2012, Decker2011} excitations have been realized, including surface-lattice resonances with very high quality factors \cite{Auguie2008, Kravets2008, Giannini2010}.  The most widely used description of these effects involves approximating each nano-particle by multipole moments, usually only the electric dipole moment for lattices of metallic nano-particles \cite{Zou2004, Markel2007, Fructos2011, Rasskazov2014}, and the electric and magnetic dipole moments for lattices of dielectric nano-particles \cite{Evlyukhin2010,  Savelev2014,Campione2016, Enevet2017}. However, an electric quadrupole resonance also becomes significant in radiation from large nano-particles \cite{Han2009, Evlyukhin2012,   Alaee2015}, and due to its narrow linewidth it is of interest for sensing applications \cite{Singh2014, Yong2014}, field enhancement \cite{Zhang2011}, and energy waveguiding \cite{Alu2009}.

Calculating the induced multipole moments of nano-particles in a lattice necessitates the evaluation of periodic Green function dyadics \cite{Zou2004, Markel2007,Belov2005,Campione2012,  Savelev2014, Lunnemann2013, Lunnemann2014, Steshenko2011}. They arise because each multipole moment responds not only to incident fields and its own radiation reaction fields, but to radiation scattered from all the other multipoles in the lattice. Periodic Green functions take the form of slowly convergent summations over lattice sites, and so acceleration techniques are required to evaluate them \cite{  Moroz2006, Valerio2007}. The most commonly used include the Ewald summation method \cite{Stevanoviae, Campione2012, Capolino2007, Steshenko2011}, Kummer's, Poisson's, and Shank's transformation \cite{Singh1990, Lampe1985}, the lattice sum methods \cite{Linton2010}, and other hybrid techniques \cite{Guerin2017, Bruno2014, Silveirinha2005}. Explicit implementations of acceleration methods have been given for scalar periodic Green functions \cite{Capolino2007, Linton2010, Silveirinha2005, Guerin2017}, as well as for dyadic functions or the mode dispersion relations that follow from them in a description of 1d, 2d, and 3d electric and magnetoelectric lattices \cite{Belov2005,Campione2012, Shore2012, Savelev2014, Lunnemann2013, Steshenko2011}, and in 1d chains of purely quadrupolar emitters \cite{Alu2009}. 

In general, the electrodynamics underlying the optical response of lattices can be complex, and further approximations are often needed.  A natural requirement is that any approximations respect energy conservation.  For an isolated nano-particle or a finite cluster, the total radiation reaction field acting on each multipole can be written down explicitly \cite{Sipe1974, Alu2009, Sersic2011, Belov2003}; models for the optical response can then be constructed by approximating only the non-radiative part of the response of the nano-particles \cite{DeAbajo2007, Sersic2011}.  Such schemes rigorously satisfy the optical theorem due to an exact treatment of radiative damping \cite{Sipe1974, Sersic2011, Belov2003}.   

The situation for lattices of nano-particles is more complicated. Even after acceleration techniques are applied \cite{Moroz2006, Valerio2007}, radiative contributions to periodic Green function dyadics that give rise to radiation reaction fields involve a summation over an infinite number of lattice sites. Yet analytic expressions for the radiative contributions to mode dispersion relations describing 1d dipolar \cite{Alu2006} and quadrupolar \cite{Alu2009} chains have been found by direct summation in real space. The radiative contributions to dyadics in a 3d lattice from all the planes except the one that involves the field point of interest can be summed exactly using the spectral representations for the periodic dyadics, as has been shown at the electric dipole level \cite{Belov2005} and can easily be generalized to higher multipoles. Determining the radiative contributions at a field point of interest in a 2d lattice is more difficult. Nevertheless, an acceleration method that gives exact radiative contribution to the dipole dyadic describing rectangular 2d lattices has been reported \cite{Belov2005}, and a related approach has been used to evaluate radiative contributions arising in mode dispersion relations for magneto-electric rectangular lattices \cite{Shore2006}. But such a treatment of more generic 2d lattices -- of arbitrary lattice structure, or involving the quadrupole moment or the response of mulipole moments to gradients of electromagnetic fields -- is still lacking.  

Analytic description of radiation damping in a lattice is useful for many reasons. First, in many situations radiative loss surpasses absorption, and expressions for the radiation reaction fields in an analytic form give a prediction for the scaling of a resonance line-width with lattice parameters \cite{Lunnemann2013}. Second, an identification of energy loss channels facilitates the analysis of the normal modes of the lattice, which are associated with poles in the optical response functions \cite{Savelev2014,Alu2009,Fructos2011,Shore2006,Shore2012, Lunnemann2014}.  Normal modes experience loss due to absorption and radiative damping \cite{Fructos2011,Shore2012,Alu2009}, and an analytic description of the energy balance in the system allows the identification of ``ideal" modes that would be supported by the lattice in the absence of loss. Their dispersion relations give insight into the optical response of a lattice, in the same way that the dispersion relation of an ``ideal" surface plasmon at an air/metal interface, calculated with the neglect of absorption loss, can give insight into the response of the system even when loss is included \cite{Barnes2003,Foley2015}. Finally, an identification of radiation reaction terms opens up the possibility of constructing simplified analytic models of the full optical response.

In this paper we derive expressions for all the periodic Green function dyadics arising in a treatment of an arbitrary 2d lattice that includes the electric dipole, magnetic dipole, and electric quadrupole moments of each nano-particle. The method we use is a generalization of an earlier approach used to find the dipole periodic Green function for a rectangular lattice \cite{Belov2005}, and involves a dimensionality reduction with the Poisson transformation and a singularity removal.  We find the radiative contributions to all the periodic Green function dyadics exactly, and we give them in a concise analytic form that identifies radiation reaction terms associated with each diffracted beam. The non-radiative contributions are found in the form of rapidly converging series, and they all involve summations over linear combinations of a few functions. All the singularities associated with the onset of new diffracted orders are easily identified. This is the first time that explicit implementations of all the periodic Green function dyadics describing magnetoelectric quadrupole lattices have been given within one formalism, and the first time a detailed analysis of the energy balance for these lattices has been carried out. 

With the energy balance in the lattice identified, approximate descriptions of the optical response that rigorously respect energy conservation can be introduced in various limiting cases. Here we propose a simple approximation scheme to evaluate all the periodic Green function dyadics for lattices of high symmetry at long wavelengths. The radiative contributions are given by their exact analytic expressions. The non-radiative contributions are given approximately by their static limit and a leading dynamic correction. The correction takes a simple analytic form, but it nevertheless significantly extends the range of validity of the static approximation. 

The manuscript is organized as follows. In section~\ref{model} we introduce multipolar model of a 2d lattice and identify the periodic Green function dyadics. In section~\ref{exact exp} we outline the acceleration method and give the periodic Green dyadics in a form suitable for computations. In section~\ref{approximate} we give approximate periodic Green function dyadics in the long-wavelength limit, and discuss the dynamic corrections to Green functions of different polarity. We conclude in section~\ref{conclusions}.

\section{Optical response of a 2d array}
\label{model}

\subsection{Isolated emitter}
First we consider the optical response of a single emitter. We take the emitter to be embedded in a homogeneous background with an index of refraction $n = \sqrt{\epsilon \mu}$, where $\epsilon$ and $\mu$ are a relative permittivity and permeability of the background medium respectively, which we assume to be real. The emitter is illuminated with incident electromagnetic fields at frequency $\omega$,
\begin{eqnarray}
\quad\bm{E}^{\rm inc}(\bm{r},t) = \bm{E}^{\rm inc}(\bm{r})e^{-\rmi\omega t}+c.c \label{E},\\
\quad\bm{B}^{\rm inc}(\bm{r},t) = \bm{B}^{\rm inc}(\bm{r})e^{-\rmi\omega t}+c.c.\label{B}
\end{eqnarray}
We describe the optical response to the incident fields (\ref{E},\ref{B}) in terms of the emitter's electric dipole moment $\bm{p}$, magnetic dipole moment $\bm{m}$, and electric quadrupole moment $\ovb{q}$, neglecting the higher order multipole moments. The multipole moments are linked to the incident fields and their gradients via the emitter polarizabilities. Rather than working with the usual polarizabilities, it is more convenient to work with the proper polarizabilities, which describe the response of the moments to a sum of the incident fields and the radiation reaction fields,
\begin{eqnarray}
\quad\bm{p} = \ovb{\alpha}^{\rm pE }\cdot \bm{E}'+\ovb{\alpha}^{\rm pB}\cdot \bm{B}'+\ovb{\alpha}^{\rm pF}:\ovb{F}',\label{iso1}\\
\quad\bm{m} = \ovb{\alpha}^{\rm mE}\cdot \bm{E}'+ \ovb{\alpha}^{\rm mB}\cdot \bm{B}' + \ovb{\alpha}^{\rm mF}:\ovb{F}',\\
\quad\ovb{q} = \ovb{\alpha}^{\rm qE}\cdot \bm{E}'+\ovb{\alpha}^{\rm qB}\cdot \bm{B}'+\ovb{\alpha}^{\rm qF}:\ovb{F}',\label{iso2}
\end{eqnarray}
where we identified 
\begin{eqnarray}
\quad\bm{E'} = \bm{E}^{\rm inc} + \rmi\frac{\mu(\tilde{\omega}n)^3}{6\pi \epsilon_0 n^2} \bm{p},\label{rr}\\ 
\quad\bm{B'} = \bm{B}^{\rm inc} + \rmi\frac{\mu(\tilde{\omega}n)^3}{6\pi \epsilon_0 c^2} \bm{m},\\
\quad\ovb{F}' = \ovb{F}^{\rm inc} + \rmi\frac{\mu(\tilde{\omega}n)^5}{20\pi \epsilon_0 n^2} \ovb{q}.\label{rrF}
\end{eqnarray}
Here $\bm{E}^{\rm inc} = \bm{E}^{\rm inc}(0)$ and $\bm{B}^{\rm inc} = \bm{B}^{\rm inc}(0)$ respectively are the electric and magnetic incident fields at the position of the emitter, chosen to be at the origin of coordinates, and $\ovb{F}^{\rm inc} = \ovb{F}^{\rm inc}(0)$ is a symmetrized gradient of the incident electric field at the origin, where 
\begin{equation}
\quad F^{\rm inc}_{ij}(\bm{r}) =\frac{1}{2}\partial_i E^{\rm inc}_j(\bm{r})+\frac{1}{2}\partial_j E^{\rm inc}_i(\bm{r}).
\end{equation}
We neglect all higher order derivatives. The second terms on the right-hand side of   (\ref{rr}-\ref{rrF}) are the radiation reaction fields. Radiation reaction fields describe the radiative damping of the multipole moments, and their inclusion in  (\ref{iso1}-\ref{iso2}) guarantees that the optical theorem is satisfied, irrespective of the approximations made in calculating the proper polarizabilities \cite{Sipe1974, Sersic2011, Belov2003}. Some constraints on the form of the proper polarizabilities are, however, imposed by the reciprocity relations \cite{Lunnemann2013}.

Once the multipole moments are found using \mbox{(\ref{iso1}-\ref{iso2})}, the fields scattered by an emitter are given by the products of the multipoles with the usual free-space Green function dyadics $\ovb{g}(\bm{r})$, the form of which we give for completeness in Appendix A. 

\subsection{2d array of emitters}
We now consider emitters arranged on an arbitrary 2d lattice with one emitter per unit cell. The positions of the emitters are described by the lattice vectors $\bm{R}_{\bm{n}} =  n_1\bm{a}_1+n_2\bm{a}_2$, where $\bm{n} = (n_1,n_2)$ is a vector of integers, $\bm{a}_1$ and $\bm{a}_2$ are the two basis lattice vectors, and we choose the array to be in the $z=0$ plane. The fields incident on the array are of the form (\ref{E},\ref{B}), with the spatial components $\bm{E}^{\rm inc}(\bm{r})$ and $\bm{B}^{\rm inc}(\bm{r})$ satisfying the free-space Maxwell equations in the neighbourhood of the array. We employ the Fourier transform of the spatial components of the fields with respect to  $x$ and $y$ directions in the plane of the array. The Fourier components are characterized by an in-plane wave vector $\bm{\kappa}$, and we consider a response of an array to a single Fourier component at $\bm{\kappa} = \bm{\kappa}_0$. We thus take the spatial components of the fields to be of the form,
\begin{eqnarray}
\quad \bm{E}^{\rm inc}(\bm{r}) = \bm{E}_+^{\rm inc} \rme^{\rmi\bm{v}_+\cdot \bm{r}} + \bm{E}_-^{\rm inc} \rme^{\rmi\bm{v}_-\cdot \bm{r}},\label{Einc} \\
\quad \bm{B}^{\rm inc}(\bm{r}) = \bm{B}_+^{\rm inc} \rme^{\rmi\bm{v}_+\cdot \bm{r}} + \bm{B}_-^{\rm inc} \rme^{\rmi\bm{v}_-\cdot \bm{r}},\label{Binc}
\end{eqnarray}
where $\bm{v}_+$ and $\bm{v}_-$ are respectively the wave vectors of the upward- and downward-propagating fields with the in-plane wave vector $\bm{\kappa}_0$,
\begin{eqnarray}
\quad \bm{v}_{\pm} = \bm{\kappa}_0 \pm w_0\bm{\hat{z}},
\end{eqnarray} 
where $\bm{v}_{\pm}\cdot \bm{v}_{\pm} = (\tilde{\omega}n)^2$, $\tilde{\omega}= \omega/c$, and $w_0 = \sqrt{(\tilde{\omega}n)^2 - \bm{\kappa}_0^2}$, such that $\mathrm{Im} w_0\geq 0$, and $\mathrm{Re}w_0\geq 0$ if $\mathrm{Im}w_0 =0$.

We seek a response of the array to the incident fields identified by (\ref{Einc}-\ref{Binc}). Due to the periodicity of the array, the multipole moments at the lattice site $\bm{R}_{\bm{n}}$ differ from the multipoles at the origin by a phase factor, 
\begin{eqnarray}
\quad \bm{p}_{\bm{n}} = \rme^{\rmi\bm{\kappa}_0\cdot \bm{R}_{\bm{n}}} \bm{p},\label{red_i}\\
\quad \bm{m}_{\bm{n}} = \rme^{\rmi\bm{\kappa}_0\cdot \bm{R}_{\bm{n}}} \bm{m},\\
\quad \ovb{q}_{\bm{n}} = \rme^{\rmi\bm{\kappa}_0\cdot \bm{R}_{\bm{n}}} \ovb{q},\label{red_f}
\end{eqnarray}
where  $\bm{p}_{\bm{n}} $ is the dipole moment at $\bm{R_n}$ and $\bm{p}$ is the dipole moment at the origin, and similarly for the other multipoles. Using the phase relations (\ref{red_i}-\ref{red_f}), the response of the array can be formulated in terms of the multipoles at the origin only, 
\begin{eqnarray}
\quad \bm{p} = \ovb{\alpha}^{\rm pE}\cdot \bm{E}^{\rm tot}+\ovb{\alpha}^{\rm pB}\cdot \bm{B}^{\rm tot}+\ovb{\alpha}^{\rm pF}:\ovb{F}^{\rm tot},\label{iso1b}\\
\quad \bm{m} = \ovb{\alpha}^{\rm mE}\cdot \bm{E}^{\rm tot}+ \ovb{\alpha}^{\rm mB}\cdot \bm{B}^{\rm tot} + \ovb{\alpha}^{\rm mF}:\ovb{F}^{\rm tot},\label{iso2b}\\
\quad \ovb{q} = \ovb{\alpha}^{\rm qE}\cdot \bm{E}^{\rm tot}+\ovb{\alpha}^{\rm qB}\cdot \bm{B}^{\rm tot}+\ovb{\alpha}^{\rm qF}:\ovb{F}^{\rm tot},\label{iso3b}
\end{eqnarray}
where the total fields at the origin -- $\bm{E}^{\rm tot}$, $\bm{B}^{\rm tot}$ and $\ovb{F}^{\rm tot}$ -- are given by a sum of incident fields, radiation reaction fields, and fields scattered by all the multipoles at sites $\bm{R}_{\bm{n}}\neq 0$, and they can be written in the form
\begin{eqnarray}
\quad \bm{E}^{\rm tot} = \bm{E}^{\rm inc} + \ovb{\mathcal{G}}^{\rm Ep} \cdot \bm{p} +\ovb{\mathcal{G}}^{\rm Em}\cdot \bm{m} + \ovb{\mathcal{G}}^{\rm Eq}:\ovb{q},\label{Etot}\\
\quad \bm{B}^{\rm tot} = \bm{B}^{\rm inc} +\ovb{\mathcal{G}}^{\rm Bp}\cdot \bm{p} +\ovb{\mathcal{G}}^{\rm Bm}\cdot\bm{m} + \ovb{\mathcal{G}}^{\rm Bq}:\ovb{q},\\
\quad \ovb{F}^{\rm tot} = \ovb{F}^{\rm inc} + \ovb{\mathcal{G}}^{\rm Fp}\cdot \bm{p} + \ovb{\mathcal{G}}^{\rm Fm}\cdot \bm{m} +\ovb{\mathcal{G}}^{\rm Fq}:\ovb{q}.\label{Ftot}
\end{eqnarray}
We refer to the Green functions $\ovb{\mathcal{G}}$ in (\ref{Etot}-\ref{Ftot}) as the periodic Green functions. Here \mbox{$\ovb{\mathcal{G}}^{\rm Ep}\cdot\bm{p}$} gives the  electric field that is a sum of the field scattered by all the electric dipoles at sites $\bm{R}_{\bm{n}}\neq 0$ and the radiation reaction field of the electric dipole at $\bm{R}_{\bm{n}}= 0$. Similarly $\ovb{\mathcal{G}}^{\rm Bm}\cdot\bm{m}$ describes the magnetic field scattered by the magnetic dipoles together with the radiation reaction field of the magnetic dipole, and $\ovb{\mathcal{G}}^{\rm Fq}:\ovb{q}$ describes the gradient of the electric field scattered by the quadrupoles together with the radiation reaction field of the electric quadrupole at the origin. The remaining terms describe only the scattered fields, and no radiation reaction fields; for example, $\ovb{\mathcal{G}}^{\rm Em}\cdot\bm{m}$ describes the electric field scattered by all the magnetic dipoles at sites $\bm{R}_{\bm{n}}\neq 0$ but does not include any radiation reaction field due to the magnetic dipole, etc. To identify the different contributions to each periodic Green function we write each dyadic as a sum of two terms,
\begin{equation}
\quad \ovb{\mathcal{G}} = \ovb{G} + \ovb{\mathcal{R}}.\label{Gdec}
\end{equation}
The first term in (\ref{Gdec}) gives the scattered field when multiplied by the appropriate moment $\bm{p}$, $\bm{m}$, or $\ovb{q}$, and it is defined in terms of the corresponding free-space Green function dyadic $\ovb{g}(\bm{r})$ as 
\begin{eqnarray}
\quad \ovb{G} =  \lim_{z\rightarrow 0}\sum_{\bm{n}\neq (0,0)} \rme^{\rmi\bm{\kappa}_0\cdot \bm{R_n}} \ovb{g} (-\bm{R_n}+z\bm{\hat{z}}).\label{Gdef}
\end{eqnarray}
The limiting procedure in (\ref{Gdef}) ensures that the Green function is uniquely defined; physically the procedure corresponds to calculating the scattered fields at a small distance away from the plane of the emitters, and at the end of calculation invoking the continuity of the fields at $z=0$. The second term, $\ovb{\mathcal{R}}$, gives rise to the radiation reaction fields in the decomposition of the Green functions $\ovb{\mathcal{G}}^{\rm Ep}$, $\ovb{\mathcal{G}}^{\rm Bm}$ and $\ovb{\mathcal{G}}^{\rm Fq}$, and vanishes in the decomposition of the remaining dyadics.  The  non-vanishing contributions can be identified immediately using (\ref{rr}-\ref{rrF}),
\begin{eqnarray}
\quad \ovb{\mathcal{R}}^{\rm Ep} = \frac{\rmi\mu(\tilde{\omega}n)^3}{6\pi \epsilon_0 n^2}\ovb{U},\label{Rep} \\
\quad \ovb{\mathcal{R}}^{\rm Bm} =  \frac{\rmi\mu(\tilde{\omega}n)^3}{6\pi \epsilon_0 c^2}\ovb{U}, \\
\quad \ovb{\mathcal{R}}^{\rm Fq} = \frac{\rmi\mu(\tilde{\omega}n)^5}{20\pi \epsilon_0 n^2}\bm{\mathcal{U}},\label{Rfq} \end{eqnarray}
where $\ovb{U}$ is a unit dyadic, $U_{ij} = \delta_{ij}$,
and $\bm{\mathcal{U}}$ is a fourth-rank tensor, $\mathcal{U}_{ijkl} = \left(\delta_{ik}\delta_{jl} + \delta_{jk}\delta_{il}\right)/2 -\delta_{ij}\delta_{kl}/3$. All the other dyadics $\ovb{\mathcal{R}}$ vanish as discussed earlier. 

We now briefly discuss the optical theorem for a lattice of emitters. As for a single emitter, conservation of energy in the system is guaranteed by the inclusion of radiation reaction fields within the multipolar model (\ref{iso1b}-\ref{iso3b}). Now, however, each of the multipole moments responds not only to its own radiation reaction field, but also to the fields of all the other multipoles on the lattice. Decomposing each of the periodic Green functions into its total radiative and non-radiative contribution, 
\begin{equation}
\quad \ovb{\mathcal{G}} = {}^{\rm R}\ovb{\mathcal{G}} + {}^{\rm N}\ovb{\mathcal{G}},
\end{equation}
we note that the radiative contribution involves terms of two kinds,
\begin{equation}
\quad {}^{\rm R}\ovb{\mathcal{G}} = {}^{\rm R}\ovb{G} + \ovb{\mathcal{R}}. \label{Grdec}
\end{equation}
The first term in (\ref{Grdec}) includes the fields of all the multipoles at the lattice sites $\bm{R}_{\bm{n}}\neq 0$ associated with radiation reaction, and the second contribution gives the usual radiation reaction of the multipole at the origin. The remaining non-radiative contributions to the peridic Green functions do not affect energy conservation condition, as formally shown in Appendix~B, and they involve only a contribution that is associated with the scattered fields, 
\begin{equation}
\quad {}^{\rm N}\ovb{\mathcal{G}} = {}^{\rm N}\ovb{G}.\label{Gnrdec}
\end{equation}
In what follows we give exact analytic expressions for the radiative Green functions ${}^{\rm R}\ovb{\mathcal{G}}$. We give the non-radiative  contributions ${}^{\rm N}\ovb{\mathcal{G}}$ in the form of rapidly converging series that can be calculated numerically. 

Once the multipole moments are calculated using (\ref{iso1b}-\ref{iso3b}), the fields scattered by an array can be found using the Green function formalism for planar structures \cite{Sipe1987}, which we generalize to quadrupole systems; see (\ref{gepFa}-\ref{gfqFb}) in Appendix~A for the expressions identifying all the free-space Green functions in Fourier space.

\section{Exact expressions for the periodic Green functions}
\label{exact exp}
In this section we transform the expressions for the periodic dyadic Green functions to a form that is suitable for computations. We follow the approach that was introduced earlier to evaluate the electric dipole periodic Green function 
for rectangular lattices \cite{Belov2005}; we generalize the method to arbitrary lattices, and carry out the calculation for all the periodic Green functions that describe magnetoelectric quadrupolar lattices. We start by giving an overview of the method, and then list the results of the calculations, referring the reader to the supplementary material for the detailed derivations. We list the result for the four independent dyadics $\ovb{\mathcal{G}}^{\rm Ep}$, $\ovb{\mathcal{G}}^{\rm Em}$, $\ovb{\mathcal{G}}^{\rm Eq}$, and $\ovb{\mathcal{G}}^{\rm Fq}$.  The remaining dyadics can be found from the relations
\begin{eqnarray}
\quad \ovb{\mathcal{G}}^{\rm Bm} = \frac{n^2}{c^2} \ovb{\mathcal{G}}^{\rm Ep}, \label{i1}\\
\quad \ovb{\mathcal{G}}^{\rm Bp} = -\ovb{\mathcal{G}}^{\rm Em}, \\
\quad \ovb{\mathcal{G}}^{\rm Fp} = -\left(\ovb{\mathcal{G}}^{\rm Eq}\right)^T, \\
\quad \ovb{\mathcal{G}}^{\rm Bq} = -\left(\ovb{\mathcal{G}}^{\rm Fm}\right)^T,
\end{eqnarray}
and
\begin{eqnarray}
\quad \mathcal{G}^{\rm Bq}_{ijk} =-\frac{\rmi\tilde{\omega}n^2}{2c} \left( \epsilon_{ijm} \mathcal{G}^{\rm Ep}_{mk} + \epsilon_{ikm} \mathcal{G}^{\rm Ep}_{mj}\right),\label{i2}
\end{eqnarray}
which directly follow from analogous relations for the free space Green function dyadics. Here $T$ denotes a transposition of the dyadics, $\left(\mathcal{G}^{\rm Eq}\right)^T_{ijk} = \mathcal{G}^{\rm Eq}_{jki}$, and $\left(\mathcal{G}^{\rm Fm}\right)^T_{ijk} = \mathcal{G}^{\rm Fm}_{kij}$; $\epsilon_{ijm}$ is the Levi-Civita symbol.  
 
\subsection{Method overview}
\label{method}
Each periodic Green function $\ovb{\mathcal{G}}$ consists of a contribution that gives rise to the scattered fields, $\ovb{G}$, and in some cases of the non-vanishing contribution $\ovb{\mathcal{R}}$ that gives rise to the radiation reaction field of a single emitter; see (\ref{Rep}-\ref{Rfq}). We evaluate the former contribution using a dimensionality reduction technique that involves transforming a 2d summation into a summation over 1d lines of lattice sites, where either the number of 1d lines is finite or the contributions from different lines converge rapidly; having effectively reduced the 2d summation to a 1d problem, we evaluate the 1d series using the dominant part extraction technique.  The calculation is done for each Cartesian component of each Green function dyadic, $G_{i_1\ldots i_N}$, where the number of Cartesian indices $i_j$ varies from $N=2$ for the magnetoelectric dyadic $\ovb{G}^{\rm Em}$ to $N=4$ for the quadrupolarization dyadic $\ovb{G}^{\rm Fq}$. Even though the components of the dyadics are given in Cartesian coordinates in which the polarizability tensors are usually known, for each Cartesian component $G_{i_1\ldots i_N}$ the summations are performed in a coordinate system associated with the basis vectors of the lattice. Introducing the unit vectors $\bm{\hat{a}}_i = \bm{a}_i/|\bm{a}_i|$, we identify coordinates along the two directions of the lattice as $u_i$, $\bm{r} = \sum_i u_i\bm{\hat{a}}_i +z\bm{\hat{z}}$, and we write $f(u_1,u_2,z)$ for a general function specified in the lattice coordinate system. We can then write the Cartesian component of the periodic Green function as a summation of the form
\begin{eqnarray}
\quad G_{i_1\ldots i_N} = \lim_{z\rightarrow 0}\sum_{(n,n')\neq (0,0)} f_{i_1\ldots i_N}(na_1,n'a_2,z), \label{dec1}
\end{eqnarray}
where $a_i = |\bm{a}_i|$ is the length of the basis vector $\bm{a}_i$, the summand $f_{i_1\ldots i_N}(u_1,u_2,z)$ in general depends on the choice of the dyadic and the dyadic component, and we take the limit $z\rightarrow 0$ at the end of calculations.

We  divide the summation (\ref{dec1}) over a 2d lattice into a summation over a 1d line along the direction of the first lattice vector $\bm{a}_1$ with the site at the origin excluded, and a double summation with one index unrestricted,
\begin{eqnarray}
\quad \sum_{(n,n')\neq (0,0)} f(na_1,&n'a_2,z) = \sum_{n\neq 0} f(na_1,0, z)\nonumber\\
 &+ \sum_{n} \sum_{n'\neq 0} f(na_1, n' a_2,z), \label{dec}
\end{eqnarray}
or an analogous decomposition with the lattice vectors $\bm{a}_1$ and $\bm{a}_2$ interchanged; the choice of the decomposition depends on the form of the dyadic that we evaluate. The first term on the right-hand side of (\ref{dec}) has the form of a trigonometric series, with the series either known in an analytic form \cite{GR} or easily evaluated using the dominant part extraction technique. The second summation in (\ref{dec}) converges slowly in both indices $n$ and $n'$. We accelerate this series using a Poisson summation over the unrestricted index, 
\begin{eqnarray}
\quad \sum_{n}  f(na_1,n'a_2, z) = \frac{1}{a_1}\sum_{n}  f^{\rm F}\left(K^{(1)}_{n} , n'a_2 ,z\right),\label{poisson}
\end{eqnarray}
where the 1d Fourier transform is defined in a usual way,
\begin{eqnarray}
\quad f^{\rm F}\left(K^{(1)}_{n} , n'a_2 ,z\right) =  \int\mathrm{d}t \rme^{\rmi t K^{(1)}_{n} } f(t,n'a_2,z).\label{F}
\end{eqnarray}
and the sum on the right-hand side of (\ref{poisson}) runs over the values of reciprocal vectors of the 1d array, $\bm{K}^{(1)}_{n} = \bm{\hat{a}}_1 K^{(1)}_n$, where $K^{(1)}_n  =\left( 2\pi n /a_1 \right)$. At each value of $n$ the Fourier transform (\ref{F}) takes the form of a cylindrical wave that propagates along the 1d array with a wave vector $Q^{(1)}_{n} = (\bm{\kappa}_0 - \bm{K}^{(1)}_{n})\cdot \bm{\hat{a}}_1 $. The wave is either propagating or evanescent in the directions perpendicular to the 1d array, depending on whether the value of $w^{(1)}_{n} = \sqrt{(\tilde{\omega}n)^2 -\left(Q^{(1)}_{n}\right)^2}$ is real or imaginary; in a typical situation, $w^{(1)}_{n}$ takes real values that correspond to propagating waves only for a small subset of indices $n$. Using (\ref{poisson}) in the summation in the second term on the right-hand side of (\ref{dec}), we have
\begin{eqnarray}
\quad \sum_{n} \sum_{n'\neq 0} f(&na_1, n' a_2,z) \nonumber\\
&= \frac{1}{a_1}\sum_{n}\sum_{n'\neq 0} f^{\rm F}\left(K_n^{(1)}, n'a_2 ,z\right).\label{spat}
\end{eqnarray}
At a fixed value of the index $n$, the spatial summation over index $n'$ in (\ref{spat}) corresponds to evaluating the cylindrical waves at an increasing distance from the 1d array. If the wave is evanescent, $w^{(1)}_{n}\in \mathbb{I}$, the spatial summation in (\ref{spat}) converges exponentially and no further transformations are needed. If the wave is propagating, $w^{(1)}_{n}\in \mathbb{R}$, the spatial summation converges slowly. Thus  we isolate the subset of indices $n$ in the summation (\ref{spat}) that correspond to propagating waves, and for these values of $n$ we carry out a second Poisson transformation,
\begin{eqnarray}
\quad \sum_{n'\neq 0} f^{\rm F}\left( K^{(1)}_n, n'a_2 ,z  \right) = - f^{\rm F}\left(K^{(1)}_n, 0 ,z\right) \nonumber\\
 \qquad\qquad\qquad + \frac{1}{a_2}\sum_{n'} f^{\rm FF}\left(K^{(1)}_n, K^{(2)}_{n'} ,z\right),\label{poisson2}
\end{eqnarray}
and then  accelerate the summation in (\ref{poisson2}) by extracting the dominant part. As a last step, in each term on the right-hand side of (\ref{poisson2}) we identify a contribution that is singular in the $z\rightarrow 0$ limit, and analytically combine the two diverging terms into an expression that is well-behaved at $z=0$. We note that the first expression on the right-hand side of (\ref{poisson2}) is a 2d modal decomposition of the Green function, and for each set of integers $(nn')$ it describes a contribution that propagates in the plane of the array with an in-plane wave vector $\bm{\kappa}^{(12)}_{n n'} = \bm{\kappa}_0 - \bm{K}^{(12)}_{nn'}$, where  $\bm{K}^{(12)}_{nn'} = n\bm{k}_1+n'\bm{k}_2$ is a vector of the 2d reciprocal lattice identified by the basis vectors $\bm{k}_i$, $\bm{k}_i\cdot \bm{a}_j = 2\pi\delta_{ij}$. The modal contributions involve divergent terms in reciprocal space,
\begin{equation}
\quad f^{\rm FF}\left(K^{(1)}_n, K^{(2)}_{n'} ,z\right) \propto \frac{1}{w^{(12)}_{nn'}}, \label{pole}
\end{equation}
where $w^{(12)}_{nn'} = \sqrt{(\tilde{\omega}n)^2 - \left[\bm{\kappa}^{(12)}_{nn'}\right]^2}$ vanishes at the onset of the $(nn')$-th diffraction order. All the remaining contributions to the periodic Green functions are non-divergent for all values of frequencies and the in-plane wave vectors $\bm{\kappa}_0$. Having found the contribution due to the scattered fields, $\ovb{G}$, we add the contribution due to the radiation reaction fields, $\ovb{\mathcal{R}}$, to obtain the full periodic Green function $\ovb{\mathcal{G}}$. 

In what follows we use the procedure outlined above to find the four independent periodic Green functions, $\ovb{\mathcal{G}}^{\rm Ep}$, $\ovb{\mathcal{G}}^{\rm Em}$, $\ovb{\mathcal{G}}^{\rm Eq}$, and $\ovb{\mathcal{G}}^{\rm Fq}$. We give the final results, explicitly identifying the total radiative and non-radiative contribution to each of the periodic Green function.

\subsection{Preliminary expressions}
\label{prel}

In this section we introduce functions that describe the non-radiative components of the periodic Green function dyadics. We start by introducing functions that describe contributions from a 1d summation over a line, given by the first term in the decomposition (\ref{dec}). Depending on the dyadic chosen the summation is done along the first or the second basis lattice vector, and to distinguish between the two directions we introduce a superscript $i$ that indicates a summation along $\bm{a}_i$. The 1d summation over a line takes the form of a trigonometric series and consists of two kind of terms: a fast converging series with the dominant part removed, and an analytic expression that gives the dominant part which can be summed exactly. The fast converging series are identified by the following functions:
\begin{eqnarray}
\quad A_{\pm}^{(i)}  = \sum_{n=1}^{\infty} \frac{\sin \left(n s_{\pm}^{(i)}\right)}{n^2(n+1)(n+2)},\label{ga}\\
\quad B_{\pm}^{(i)}   = \sum_{n=1}^{\infty} \frac{(3n+2) \cos \left(n s_{\pm}^{(i)}\right) }{n^3(n+1)(n+2)},\label{gb}\\
\quad C_{\pm}^{(i)}  = \sum_{n=1}^{\infty} \frac{\sin \left(n s_{\pm}^{(i)}\right)}{n^4},\\
\quad D_{\pm}^{(i)}  = \sum_{n=1}^{\infty} \frac{\cos \left(n s_{\pm}^{(i)}\right)}{n^5},\label{gd} 
\end{eqnarray}
where $s^{(i)}_{\pm}$ is a reduction of an argument $S^{(i)}_{\pm} = \tilde{\omega}na_i\pm \bm{\kappa}_0\cdot\bm{a}_i$ to an interval $\left<-\pi,\pi\right)$; that is,
\begin{equation}
\quad S_{\pm}^{(i)} = s_{\pm}^{(i)} \pm 2\pi n_{\pm}^{(i)},\label{s}
\end{equation}
where $s_{\pm}^{(i)} \in \left<-\pi,\pi\right)$ and the $n_{\pm}^{(i)}$ are integers. The dominant part contributions are given by the following functions,
\begin{eqnarray}
\quad \tilde{\mathcal{A}}_{\pm}^{(i)}&  = \cos s_{\pm}^{(i)}-\frac{3}{2}   \nonumber\\
&+ \left( \cos 2s_{\pm}^{(i)} - 2\cos s_{\pm}^{(i)} +1 \right) \mathfrak{l} \left( s_{\pm}^{(i)} \right) \nonumber\\
&+\frac{\pi}{2}  \left( 2\sin s_{\pm}^{(i)}-\sin 2s_{\pm}^{(i)} \right) \mathfrak{s} \left( s_{\pm}^{(i)} \right), \label{f1}\\
\quad \tilde{\mathcal{B}}_{\pm}^{(i)} & = \sin s_{\pm}^{(i)}  \nonumber\\ 
& + \left( \sin 2s_{\pm}^{(i)}-4\sin  s_{\pm}^{(i)} \right) \mathfrak{l} \left( s_{\pm}^{(i)} \right) \nonumber\\
&+\frac{\pi}{2}  \left( \cos 2s_{\pm}^{(i)}-4\cos s_{\pm}^{(i)}+3\right) \mathfrak{s} \left( s_{\pm}^{(i)} \right),
\end{eqnarray}
where we introduced a notation
\begin{eqnarray}
\quad \mathfrak{l} \left( s_{\pm}^{(i)} \right) = \ln  \left( 2\left| \sin \frac{s_{\pm}^{(i)}}{2} \right| \right), \\
\quad \mathfrak{s} \left( s_{\pm}^{(i)} \right) =  \mathrm{ sgn} \left( s_{\pm}^{(i)}\right) -\frac{s_{\pm}^{(i)}}{\pi}, 
\end{eqnarray}
and  $\mathrm{sgn}(x)$ is the usual sign function. 

Next we introduce functions that describe contributions to the periodic Green function dyadics originating from a spatial summation over index $n'$ in (\ref{spat}) at a fixed value of the reciprocal index $n$, where the value of $n$ is such that it corresponds to an evanescent wave. We choose the reciprocal 1d array to be along the lattice vector $\bm{a}_i$, with the reciprocal lattice identified by the reciprocal vectors $\bm{K}^{(i)}_{n} = \bm{\hat{a}}_i \left( 2\pi n /a_i \right)$. We identify  $\bm{\kappa}^{\left(i\right)}_{n} = \bm{\kappa}_0 - \bm{K}^{(i)}_{n}$ as the in-plane incident wave vector translated over a 1d reciprocal lattice,  $Q^{(i)}_{n} = \bm{\kappa}^{(i)}_{n}\cdot \bm{\hat{a}}_i$ as the propagation wave number in the direction of the 1d reciprocal array, and $w^{(i)}_{n} = \sqrt{(\tilde{\omega}n)^2 -\left(Q^{(i)}_{n}\right)^2}$ as the value of the wave vector in the  direction perpendicular to the array. The real space 1d array we choose  to be along $\bm{a}_{\bar{i}}$, where $\bar{i}$ denotes the index that is not $i$, with the lattice vectors of the form $\bm{R}^{(\bar{i})}_{n'} = n' \bm{a}_{\bar{i}}$. We identify $d^{(\bar{i})}_{n'}$ as the distance from the lattice site $\bm{R}^{(\bar{i})}_{n'}$ to the line $\bm{\hat{a}}_i$, $d^{(\bar{i})}_{n'} = |\bm{a}_{\bar{i}} n'\sin \phi|$, where $\phi$ is the angle between the basis vectors of the lattice. Finally we introduce a wave number $ \tilde{\kappa}^{(i\bar{i})}_{n} = \left[ \bm{\kappa}_0 - \bm{\hat{a}}_i Q^{(i)}_{n} \right]\cdot \bm{\hat{a}}_{\bar{i}}$ describing a plane wave propagation along the $\bm{R}^{(\bar{i})}_{n'}$ array. In this notation, the functions describing the contributions from a summation of the cylindrical wave identified by the index $n$ over the positions in space identified by the lattice sites $\bm{R}_{n'}^{(\bar{i})}$ are as follows, 
\begin{eqnarray}
\quad M^{(i)}_{n}&= \sum_{n'=1}^{\infty}M^{(i)}_{nn'} ,\label{M}\\
\quad N^{(i)}_{n}&= \sum_{n'=1}^{\infty}N^{(i)}_{nn'} ,\\
\quad L^{(i)}_{n} &=\sum_{n'=1}^{\infty}L^{(i)}_{nn'}, \label{L}
\end{eqnarray}
where 
\begin{eqnarray}
\quad M^{(i)}_{nn'}&= 4\tau^{(i)}K_0\left( \chi^{(i)}_{nn'} \right) \cos \psi^{(i)}_{nn'},\\
\quad N^{(i)}_{nn'} &=  4\tau^{(i)} |\sin \phi| \rmi^{-1} w^{(i)}_{n} K_1\left(\chi^{(i)}_{nn'} \right)  \sin \psi^{(i)}_{nn'}  ,\\
\quad L^{(i)}_{nn'} &=2\tau^{(i)} \left(w^{(i)}_{n} \sin \phi \right)^2 K_2\left( \chi^{(i)}_{nn'} \right) \cos \psi^{(i)}_{nn'}. 
\end{eqnarray}
Here $K_{\nu} (z)$ is a modified Bessel function of the second kind \cite{AS}, we have defined a plane wave propagation factor $\psi^{(i)}_{nn'} = \tilde{\kappa}^{(i\bar{i})}_{n} n'a_{\bar{i}}$, a cylindrical wave propagation factor $\chi^{(i)}_{nn'} = \rmi^{-1}w^{(i)}_{n} d^{(\bar{i})}_{n'}$, and a constant $\tau^{(i)} = A_c/ (2\pi a_i)$. We use expressions (\ref{M}-\ref{L}) for the values of $n$ satisfying $n>n^{(i)}_+$ and $n < n^{(i)}_-$, with the integers $n_{\pm}^{(i)}$ as defined in (\ref{s}); for these values we have $a_i^2\left(w^{(i)}_{n}\right)^2 \leq -\pi^2 $ and the equations (\ref{M}-\ref{L}) converge exponentially due to the asymptotic properties of the Bessel functions. 

Indices taking values in the interval $n\in \left(n^{(i)}_-,n^{(i)}_+\right)$ correspond to propagating cylindrical waves, $a_i^2\left(w^{(i)}_{n}\right)^2\geq\pi^2$, and thus (\ref{M}-\ref{L}) cannot be used. For these values of $n$ we use a second Poisson transformation (\ref{poisson2}), and recast the spatial summation into a summation over a reciprocal space. The resulting double summation in reciprocal space is taken over the in-plane wave vectors that are a translation of the incident in-plane wave vector over a 2d reciprocal lattice, $\bm{\kappa}^{(i\bar{i})}_{n n'} = \bm{\kappa}_0 - \bm{K}^{i\bar{i}}_{nn'}$, where $\bm{K}^{(i\bar{i})}_{nn'} = n\bm{k}_i+n'\bm{k}_{\bar{i}}$. In addition to the translated in-plane wave vector $\bm{\kappa}^{(i\bar{i})}_{n n'}$ we also need a notation for its $\bar{i}$-th component in the coordinate system associated with the lattice, which we identify as
\begin{equation}
\quad \kappa^{(i\bar{i})}_{nn';\bar{i}} = \frac{1}{\sin^2 \phi} \left[ \bm{\kappa}^{(i\bar{i})}_{nn'} - \left(\bm{\kappa}^{(i\bar{i})}_{nn'} \cdot \bm{\hat{a}}_i\right) \bm{\hat{a}}_i \right]\cdot\bm{\hat{a}}_{\bar{i}}.\label{k2d}
\end{equation}
Finally we identify the out-of-plane component of the wave vector as $w_{nn'}^{(i\bar{i})} = \sqrt{(\tilde{\omega}n)^2 - \left(\kappa^{(i\bar{i})}_{nn'}\right)^2}$. Now we introduce the functions that describe the contributions to periodic Green functions from a summation over the reciprocal index $n'$ at a fixed reciprocal index $n$. The summation over $n'$ takes a form of a fast converging series after the dominant part is extracted, and the summation is described by the following functions:
\begin{eqnarray}
\quad P^{(i)}_n =& \sum_{n'=1}^{\infty}\left(P^{(i)}_{nn'}+P^{(i)}_{n,-n'}\right), \label{Pfc}\\
\quad T^{(i)}_n =& \sum_{n'=1}^{\infty}\left(T^{(i)}_{nn'}+T^{(i)}_{n,-n'} \right),\\
\quad O^{(i)}_n =& \sum_{n'=1}^{\infty}\left(O^{(i)}_{nn'}+O^{(i)}_{n,-n'}\right), 
\end{eqnarray}
where
\begin{eqnarray}
\quad P^{(i)}_{nn'} &= \mathrm{Re} \frac{\rmi}{w^{(i\bar{i})}_{nn'}} - \frac{\tau^{(i)}}{|n'|} -\frac{1}{2}\left( \frac{\tau^{(i)}}{|n'|} \right)^3 p^{(i\bar{i})}_{n}, \\
\quad T^{(i)}_{nn'} &= \sin^2 \phi \left[  \mathrm{Re}\frac{\rmi\kappa^{(i\bar{i})}_{nn';\bar{i}} }{w^{(i\bar{i})}_{nn'}}+  \left( \frac{\tau^{(i)}}{|n'|} \right)^3 t^{(i\bar{i})}_{n} \right],\\
\quad O^{(i)}_{nn'} &= \mathrm{Re}\left(\frac{ \rmi \sin^2 \phi }{w^{(i\bar{i})}_{nn'}}\right) \left[ \left(w^{(i\bar{i})}_{nn'}\right)^2-\frac{1}{2}\left(w^{(i)}_{n}\right)^2 \right]  \nonumber\\
&+\sin^2 \phi \left[ \frac{ |n'|}{\tau^{(i)}}+ \frac{1}{2}\left( \frac{\tau^{(i)}}{|n'|} \right)^3 o^{(i\bar{i})}_{n} \right], 
\end{eqnarray}
and where the coefficients associated with the extracted dominant parts are given by
\begin{eqnarray}
\quad p^{(i\bar{i})}_{n} =  \left(w^{(i\bar{i})}_{n0}\right)^2- 3\left(Q_{n}^{(i)}\right)^2+3\left(\kappa^{(i\bar{i})}_{n0}\right)^2, \\
\quad t^{(i\bar{i})}_{n} = \kappa^{(i\bar{i})}_{n0;\bar{i}} \left(w^{(i)}_{n}\right)^2, \\
\quad o^{(i\bar{i})}_{n} = \frac{1}{4}\left(w^{(i)}_{n}\right)^4.
\end{eqnarray}
The contributions from the dominant parts together with the contributions from the lattice site at the origin are given by
\begin{eqnarray}
\quad \tilde{\mathcal{P}}^{(i)}_n&= \mathrm{Re}\frac{\rmi}{w^{(i\bar{i})}_{n0}} +  2\tau^{(i)} \ln \left(  \frac{\tau^{(i)}}{2}  |w^{(i)}_{n}|  \right)  \nonumber\\
&+2\gamma\tau^{(i)} +(\tau^{(i)})^3 \zeta(3) p^{(i\bar{i})}_{n}, \label{singular} \\
\quad \tilde{\mathcal{T}}^{(i)}_n &= \kappa^{(i\bar{i})}_{n0;\bar{i}} \sin^2 \phi  \left(\mathrm{Re}\frac{\rmi }{w^{(i\bar{i})}_{n0}} - 2 \tau^{(i)} \right) \nonumber\\
&-2\sin^2 \phi \left(\tau^{(i)}\right)^3 \zeta(3) t^{(i\bar{i})}_{n},  
\end{eqnarray}
\begin{eqnarray}
\quad \tilde{\mathcal{O}}^{(i)}_n&=\mathrm{Re}\left( \frac{\rmi \sin^2 \phi}{ w^{(i\bar{i})}_{{n}0}} \right)  \left[  \left(w^{(i\bar{i})}_{n0}\right)^2 - \frac{1}{2}\left(w^{(i)}_{n}\right)^2 \right] \nonumber\\
&- \tau_i \sin^2 \phi \left[ \left(w^{(i\bar{i})}_{n0}\right)^2 -\frac{1}{2} \left(w^{(i)}_{n}\right)^2 \right]    \nonumber\\
&+\sin^2 \phi\left[ \frac{1}{6\tau^{(i)}} -  \left(\tau^{(i)}\right)^3 \zeta(3)o^{(i\bar{i})}_{n} \right], \label{flast}
\end{eqnarray}
where $\zeta(k) = \sum_{n=1}^{\infty} 1/n^k$ is the Riemann-zeta function and $\gamma$ is the Euler-Mascheroni constant.

A cylindrical wave with the wave vector identified by the index $n = n^{(i)}_+$ or $n = n^{(i)}_-$  can be propagating ($w_n^{(i)}\in \mathbb{R}$), evanescent ($w_n^{(i)}\in \mathbb{I}$), or grazing along the direction of $\bm{a}_i$ ($w_n^{(i)}=0$), depending on the value of the frequency and the incident in-plane wave vector $\bm{\kappa}_0$. While the formulas (\ref{Pfc}-\ref{flast}) hold in all cases, the function (\ref{singular}) has a logarithmic singularity in the limit $w^{(i)}_{n}\rightarrow 0$. The singularity is an artefact of the dimensionality reduction technique, and describes the Rayleigh anomaly of the 1d array $\bm{R}^{(i)}_{\bm{n}}$. As the array is really 2d the singularity is only apparent, and in the dyadics in which it arises we remove it by combining the singular term in (\ref{singular}) with another singularity originating from the real-space line summation over the 1d array $\bm{R}^{(i)}_{\bm{n}}$ (of the form similar to  (\ref{ga}-\ref{gd}) but not listed). We introduce a notation for the contribution (\ref{singular}) with the singularity removed,
\begin{equation}
\quad \tilde{\mathcal{V}}^{(i)}_n = \tilde{\mathcal{P}}^{(i)}_n - 2\tau^{(i)} \ln \left(\frac{\tau^{(i)}}{2}\left|w^{(i)}_{n}\right|\right), 
\end{equation}
and a notation for the additional finite contributions that arise when the two singular terms are combined, 
\begin{eqnarray}
\quad H_{\pm}^{(i)}  &= \ln \left( \frac{\tau^{(i)}}{4 a_i} \left| \frac{s_{\pm}^{(i)}}{\sin \left(s_{\pm}^{(i)}/2\right)} \right|  \right) \nonumber\\
 &+ \mathrm{ sgn}\left( |\Delta^{(i)}|\right) \ln  \left(\frac{\tau^{(i)}\left| \slashed{S}^{(i)}_{\pm} \right|}{2 a_i }\right), \\
\quad J_{\pm}^{(i)} &= - \tilde{\omega}na_i H_{\pm}^{(i)} +s_{\pm}^{(i)}\ln \left( \frac{\tau^{(i)}}{2a_i}\left|s_{\pm}^{(i)} \right| \right) \nonumber\\
&+ \slashed{S}^{(i)}_{\pm}  \mathrm{sgn}\left(|\Delta^{(i)}|\right) \ln \left(  \frac{\tau^{(i)}\left| \slashed{S}^{(i)}_{\pm} \right| }{2a_i}\right),
\end{eqnarray}
where we defined $\Delta^{(i)} = n^{(i)}_+ - n^{(i)}_-$ and $\slashed{S}^{(i)}_{\pm} = s_{\pm}^{(i)}+2\pi\Delta^{(i)}$.

We give simple analytic expressions for the radiative periodic Green functions. They take the form of a 2d Fourier sum over the in-plane wave vectors $\bm{\kappa}_{\bm{n}} =  \bm{\kappa}^{(i\bar{i})}_{nn'}$ that correspond to propagating waves, $w_{\bm{n}} = w^{(i\bar{i})}_{nn'}\in \mathbb{R}$, and where we find it convenient to introduce a short-hand notation with the indices $i$ and $\bar{i}$ suppressed; in the expressions describing the radiative contributions both lattice directions are treated  on an equal footing and thus the order of the indices is irrelevant. We thus write each radiative periodic Green function as
\begin{equation}
\quad {}^{\rm R}\ovb{\mathcal{G}} = \sum_{\bm{n}:w_{\bm{n}}\in \mathbb{R}} {}^{\rm R}\ovb{\mathcal{G}}_{\bm{n}},\label{FDG}
\end{equation}
where ${}^{\rm R}\ovb{\mathcal{G}}_{\bm{n}} = {}^{\rm R}\ovb{\mathcal{G}}\left(\bm{\kappa} _{\bm{n}}\right)$ are Fourier components characterized by the in-plane wave vectors $\bm{\kappa}_{\bm{n}}$. A Fourier component ${}^{\rm R}\ovb{\mathcal{G}}_{\bm{n}}$, when multiplied by the appropriate moment, gives the radiative field associated with a beam diffracted at an angle $\theta_{\bm{n}} = \sin ^{-1}\left[ (\bm{\kappa}_{\bm{n}})/(\tilde{\omega}n)\right]$ with respect to the normal, and it takes an especially simple form in the basis associated with the in-plane wave vector $\bm{\kappa}_{\bm{n}}$. For each index $\bm{n}$ we identify a normalized wave vector 
 $\bm{\hat{\kappa}}_{\bm{n}} =\bm{\kappa}_{\bm{n}}/|\bm{\kappa}_{\bm{n}}|$, an s-polarization vector that is orthogonal to it, $\bm{\hat{s}}_{\bm{n}} = \bm{\hat{\kappa}}_{\bm{n}}\times \bm{\hat{z}}$, and we form the triad $(\bm{\hat{s}}_{\bm{n}}, \bm{\hat{\kappa}}_{\bm{n}},\bm{\hat{z}})$; this constitutes an orthonormal basis and in it we give the results for ${}^{\rm R}\ovb{\mathcal{G}}_{\bm{n}}$. The transformation to the Cartesian basis in which the non-radiative dyadic components are given is straightforward, and we do not give it explicitly.
 
\subsection{Periodic Green functions} 

\subsubsection{Scalar Green function}

In this section we give results for the periodic scalar Green function $\mathcal{G}_0$, defined in terms of the usual free-space scalar Green function $g_0$ (see Appendix A) as
\begin{eqnarray}
\quad \mathcal{G}_0 =  \lim_{z\rightarrow 0}\sum_{\bm{n}\neq (0,0)} \rme^{\rmi\bm{\kappa}_0\cdot \bm{R_n}} g_0 (-\bm{R_n}+z\bm{\hat{z}})+ \rmi\frac{\tilde{\omega}n}{4\pi},\label{G0}
\end{eqnarray}
where the last term in (\ref{G0}) plays a role similar to the radiation reaction fields in the dyadic periodic Green functions. Although the scalar function (\ref{G0}) does not explicitly enter the description of a 2d array, we calculate it for completeness and to minimize the number of components of the dyadic Green functions that we need to evaluate. The scalar Green function relates the in-plane and out-of-plane components of the dyadic Green functions through the Helmholtz equation. The radiative contribution ${}^{\rm R}\mathcal{G}_0$ is purely imaginary and it takes a form of a 2d Fourier sum over propagating waves (see (\ref{FDG})) with the Fourier components 
\begin{equation}
\quad {}^{\rm R}\mathcal{G}_{0\bm{n}} =  \frac{\rmi}{2A_cw_{\bm{n}}},
\end{equation}
where $A_c$ is the area of a unit cell. The non-radiative contribution is purely real and is given by the rapidly converging sum,
\begin{equation}
\quad {}^{\rm N}\mathcal{G}_{0} = \frac{1}{2}\left( \mathcal{S}^{(1)}+\mathcal{S}^{(2)}\right),
\end{equation}
where
\newpage
\begin{eqnarray}
\quad \mathcal{S}^{(i)} &=   \sum_{n\notin (n^{(i)}_-,n^{(i)}_+)}  \frac{
M^{(i)}_n}{2A_c} + \sum_{n=n^{(i)}_-}^{n^{(i)}_+} \frac{P^{(i)}_n}{2A_c}  \nonumber\\
&+ \sum_{n = n^{(i)}_-+1}^{n^{(i)}_+-1} \frac{\tilde{\mathcal{P}}^{(i)}_n}{2A_c}  + \sum_{j=\pm} \frac{H^{(i)}_j }{4\pi a_i}  \nonumber\\
&+ \sum_{n=\{n^{(i)}_+, n^{(i)}_-\} } \frac{\tilde{\mathcal{V}}^{(i)}_n}{2A_c} ,\label{G0NR}
\end{eqnarray}
and where the functions in (\ref{G0NR}) are defined in section \ref{prel}.

\subsubsection{Evaluation of $\mathcal{G}^{\rm Ep}_{ij}$.}
\label{Gepexact}
The radiative contribution is purely imaginary and can be written as the Fourier sum (\ref{FDG}) with the Fourier components
\begin{eqnarray}
\quad {}^{\rm R} \ovb{\mathcal{G}}^{\rm Ep}_{\bm{n}} =  \frac{\rmi\mu}{\epsilon_0n^2}\frac{ w_{\bm{n}}^2 \bm{\hat{\kappa}}_{\bm{n}} \bm{\hat{\kappa}}_{\bm{n}} + (\tilde{\omega}n)^2\bm{\hat{s}}_{\bm{n}} \bm{\hat{s}}_{\bm{n}} + \kappa_{\bm{n}}^2\bm{\hat{z}} \bm{\hat{z}}}{2w_{\bm{n}}A_c} .\label{GEPR}
\end{eqnarray}
The non-radiative contribution is purely real and we write it in the Cartesian basis. We note that the dyadic $\ovb{\mathcal{G}}^{\rm Ep}$ is symmetric in its indices, and we have $\mathcal{G}^{\rm Ep}_{xz} = \mathcal{G}^{\rm Ep}_{yz} =0$ immediately from its definition. Furthermore, from the Helmholtz equation we have
\begin{eqnarray}
\quad \mathcal{G}^{\rm Ep}_{zz} = -\mathcal{G}^{\rm Ep}_{xx} -\mathcal{G}^{\rm Ep}_{yy}  +2\frac{\mu \tilde{\omega}^2}{\epsilon_0} \mathcal{G}_0,
\end{eqnarray}
with the equality holding for both the radiative and non-radiative contributions separately. Thus we only list the independent in-plane components  ${}^{\rm N}\mathcal{G}^{\rm Ep}_{\alpha\beta}$, where $\alpha,\beta = x,y$. We find  
\begin{equation}
\quad {}^{\rm N}\mathcal{G}^{\rm Ep}_{\alpha\beta} = \frac{\mu}{\epsilon_0 n^2}\sum_{(lm)} \frac{C^{(lm)}_{\alpha\beta} \mathcal{S}_{(lm)}}{\left[\left(\bm{\hat{a}}_1\times\bm{\hat{a}}_2\right) \cdot\bm{\hat{z}}\right]^2},\label{Gepex}
\end{equation}
where the sum is over the three independent sets of indices, $(lm)=\{(11),(22),(12)\}$, with the coefficients $C^{(lm)}_{\alpha\beta}$ listed in Table \ref{TabEP}. There and in all later tables we write $\hat{a}_{ix} \equiv \boldsymbol{\hat{a}}_i\cdot \boldsymbol{\hat{x}}$ and $\hat{a}_{iy} \equiv \boldsymbol{\hat{a}}_i\cdot \boldsymbol{\hat{y}}$.
\begin{table}[htb]
\caption{\label{TabEP}The coefficients $C^{(lm)}_{\alpha\beta}$.}
\begin{indented}
\lineup
\item[]\begin{tabular}{@{}llll}
\br                              
 & $(11)$ &$(22)$ & $(12)$\\ \mr
$xx$ & $\hat{a}_{2y}^2$ & $\hat{a}_{1y}^2$ & $-2\hat{a}_{1y} \hat{a}_{2y}$ \\
$yy$ & $\hat{a}_{2x}^2$ & $\hat{a}_{1x}^2$ & $-2\hat{a}_{1x} \hat{a}_{2x}$ \\ 
$xy$ & $-\hat{a}_{2x}\hat{a}_{2y}$ & $-\hat{a}_{1x}\hat{a}_{1y}$ & $\hat{a}_{1x}\hat{a}_{2y} + \hat{a}_{2x} \hat{a}_{1y}$    \\ 
\br
\end{tabular}
\end{indented}
\end{table}
The  $\mathcal{S}_{(lm)}$ are partial sums; for $l=m$ we find 
\newpage
\begin{eqnarray}
\quad \mathcal{S}_{(ii)} &= \sum_{n\notin (n^{(i)}_-,n^{(i)}_+)}  \frac{\left(w^{(i)}_{n}\right)^2}{2A_c}   M^{(i)}_n \nonumber\\
&+ \sum_{n=n^{(i)}_-}^{n^{(i)}_+} \frac{\left(w^{(i)}_{n}\right)^2}{2A_c} \left(P^{(i)}_n+\tilde{\mathcal{P}}^{(i)}_n \right)\nonumber\\
&+ \sum_{j=\pm}\left( \tilde{\omega}n \frac{ \tilde{\mathcal{B}}^{(i)}_j +4A^{(i)}_j  }{4\pi a_i^2}-\frac{\tilde{\mathcal{A}}^{(i)}_j - 2B^{(i)}_j}{4\pi a_i^3} \right),
\end{eqnarray}
and for $l\neq m$ we find,
\begin{equation}
\quad \mathcal{S}_{(12)} = \frac{1}{2}\left( \mathcal{S}^{(1)}_{(12)}+\mathcal{S}^{(2)}_{(12)} \right),
\end{equation}
where 
\begin{eqnarray}
\quad \mathcal{S}^{(i)}_{(12)} = &-\sum_{n\notin (n^{(i)}_-,n^{(i)}_+)} \frac{Q^{(i)}_{n}}{2A_c} N^{(i)}_n + \mathcal{S}_{ii} \cos \phi \nonumber\\
& - \sum_{n=n^{(i)}_-}^{n^{(i)}_+}\frac{Q^{(i)}_{n}}{2A_c} \left( T^{(i)}_n+\tilde{\mathcal{T}}^{(i)}_n \right).
\end{eqnarray}

\subsubsection{Evaluation of $\mathcal{G}^{\rm Em}_{ij}$.}
The radiative contribution is purely imaginary and is given by the Fourier sum (\ref{FDG}) with the Fourier components
\begin{eqnarray}
\quad {}^{\rm R}\ovb{\mathcal{G}}^{\rm Em}_{\bm{n}}&= \frac{\rmi\mu\tilde{\omega}}{\epsilon_0 c} \frac{\kappa_{\bm{n}}}{2A_c w_{\bm{n}}}\bigg( \bm{\hat{z}}\bm{\hat{s}}_{\bm{n}} -\bm{\hat{s}}_{\bm{n}}\bm{\hat{z}} \bigg).\label{GEMR}
\end{eqnarray}
The non-radiative contribution is purely real and we write it in the Cartesian basis. We note that the dyadic is antisymmetric in its indices, so $\mathcal{G}^{\rm Em}_{ii}=0$, and we find that $\mathcal{G}^{\rm Em}_{xy}=0$ directly from the periodic Green function definition. We thus list only the two independent non-vanishing components, ${}^{\rm N} \mathcal{G}^{\rm Em}_{\alpha z}$ for $\alpha=x,y$. We find
\begin{equation}
\quad {}^{\rm N} \mathcal{G}^{\rm Em}_{\alpha z} =  \frac{\mu\tilde{\omega}}{\epsilon_0 c}  \sum_{(l)} \frac{C^{(l)}_{\alpha} \mathcal{S}_{(l)}}{\left(\bm{\hat{a}}_1\times\bm{\hat{a}}_2\right)\cdot \bm{\hat{z}}} ,
\end{equation}
where the sum is over $l=\{(1),(2)\}$ and the coefficients $C^{(l)}_{\alpha}$ are listed in Table \ref{tabEM}. The $S_{(l)}$ are partial sums found to be given by
\begin{table}[hbt]
\caption{\label{tabEM}The coefficients $C^{(l)}_{\alpha}$.}
\begin{indented}
\lineup
\item[]\begin{tabular}{@{}lll}
\br                              
 & $(1)$ & $(2)$ \\ \mr
$x$ & $\hat{a}_{2x}$ & $-\hat{a}_{1x}$ \\
$y$ & $\hat{a}_{2y}$ & $-\hat{a}_{1y}$ \\  
\br
\end{tabular}
\end{indented}
\end{table}
\newpage
\begin{eqnarray}
\quad \mathcal{S}_{(i)} &= \sum_{n\notin (n^{(i)}_-,n^{(i)}_+)}  \frac{Q^{(i)}_{n}}{2A_c} M^{(i)}_n + \sum_{n=n^{(i)}_-}^{n^{(i)}_+}\frac{Q^{(i)}_{n}}{2A_c} P^{(i)}_n \nonumber\\
 &+ \sum_{n = n^{(i)}_-+1}^{n^{(i)}_+-1}\frac{Q^{(i)}_{n}}{2A_c} \tilde{\mathcal{P}}^{(i)}_n+\sum_{n=n^{(i)}_+, n^{(i)}_-}\frac{Q^{(i)}_{n}}{2A_c} \tilde{\mathcal{V}}^{(i)}_n \nonumber\\
&+ \frac{1}{4\pi a_i^2}\sum_{j=\pm}j\left( \frac{1}{2}\tilde{\mathcal{B}}^{(i)}_j + 2A^{(i)}_j +J_j^{(i)}  \right).
\end{eqnarray}

\subsubsection{Evaluation of $\mathcal{G}^{\rm Eq}_{ijk}$.}
The radiative contribution is purely real and is given by a Fourier sum (\ref{FDG}), with the Fourier components 
\begin{eqnarray}
\quad {}^{\rm R}\ovb{\mathcal{G}}^{\rm Eq}_{\bm{n}} = \frac{\mu}{\epsilon_0n^2}\frac{\kappa_{\bm{n}}}{2A_c w_{\bm{n}}} \bigg[ w^2_{\bm{n}} \bm{\hat{\kappa}}_{\bm{n}}\left( \bm{\hat{\kappa}}_{\bm{n}} \bm{\hat{\kappa}}_{\bm{n}} - \bm{\hat{z}}\bm{\hat{z}}  \right) \nonumber\\
\quad+ (\tilde{\omega}n)^2 \bm{\hat{s}}_{\bm{n}}\left\{ \bm{\hat{s}}_{\bm{n}},\bm{\hat{\kappa}}_{\bm{n}} \right\}  - \left( w^2_{\bm{n}} - \bm{\kappa_n}^2 \right) \bm{\hat{z}} \left\{ \bm{\hat{z}},\bm{\hat{\kappa}}_{\bm{n}} \right\} \bigg],
\end{eqnarray}
where we introduce a notation for a symmetrized dyadic,
\begin{equation}
\quad \left\{ \bm{\hat{a}},\bm{\hat{b}} \right\} = \frac{1}{2} \left( \bm{\hat{a}}\bm{\hat{b}} + \bm{\hat{b}} \bm{\hat{a}} \right).
\end{equation}
The non-radiative contribution is purely imaginary and we write it in the Cartesian basis. We note that the periodic Green function dyadic is symmetric in its last two indices, $\mathcal{G}^{\rm Eq}_{ijk} = \mathcal{G}^{\rm Eq}_{ikj}$, but is not symmetric with respect to other permutation of indices. To list the non-radiative contributions we find it convenient to introduce a fully symmetric dyadic, ${}^{\rm N}\mathcal{G}^{\rm Eq}_{[ijk]}$, in terms of which the usual periodic Green function ${}^{\rm N}\mathcal{G}^{\rm Eq}_{ijk}$ is given by
\begin{eqnarray}
\quad {}^{\rm N}\mathcal{G}^{\rm Eq}_{ijk} &= {}^{\rm N}\mathcal{G}^{\rm Eq}_{[ijk]}  +\frac{\rmi\omega}{6} \delta_{jk}\epsilon_{ilm} {}^{\rm N}\mathcal{G}^{\rm Em}_{lm} \nonumber\\
 &-\frac{\rmi\omega}{12}\left(\delta_{ij}\epsilon_{klm} + \delta_{ik}\epsilon_{jlm}  \right) {}^{\rm N}\mathcal{G}^{\rm Em}_{lm}.\label{GEqex}
\end{eqnarray}
We consider a number of independent components of the symmetrized Green function ${}^{\rm N}\mathcal{G}^{\rm Eq}_{[ijk]}$:  Directly from the definition we find that ${}^{\rm N}\mathcal{G}^{\rm Eq}_{[zzz]} = {}^{\rm N}\mathcal{G}^{\rm Eq}_{[z\alpha\beta]} = 0$, for $\alpha,\beta = x,y$, and from the Helmholtz equation we have
\begin{eqnarray}
\quad {}^{\rm N}\mathcal{G}^{\rm Eq}_{[\alpha zz]} = &-{}^{\rm N}\mathcal{G}^{\rm Eq}_{[\alpha xx]} -{}^{\rm N}\mathcal{G}^{\rm Eq}_{[\alpha yy]} + \frac{\omega}{3\rmi} \epsilon_{\alpha lm} {}^{\rm N}\mathcal{G}^{\rm Em}_{lm}.
\end{eqnarray}
We thus need only list the in-plane components of the dyadic, ${}^{\rm N}\mathcal{G}^{\rm Eq}_{[\alpha\beta\gamma]}$ for $\alpha,\beta,\gamma = x,y$. We have
\begin{equation}
\quad {}^{\rm N}\mathcal{G}^{\rm Eq}_{[\alpha\beta\gamma]} =  \frac{\rmi\mu}{\epsilon_0 n^2} \sum_{(lmn)} \frac{C^{(lmn)}_{\alpha\beta\gamma} \mathcal{S}_{(lmn)}}{\left[ \left(\bm{\hat{a}}_1\times\bm{\hat{a}}_2 \right) \cdot \bm{\hat{z}} \right]^3},
\end{equation}
where the sum runs over the sets of indices $(lmn) = \{(111),(222),(112),(221) \}$, and the coefficients $C^{(lmn)}_{\alpha\beta\gamma}$ are given in Table \ref{tabEQ}. 
\begin{table*}[htb]
\caption{\label{tabEQ} The coefficients $C^{(lmn)}_{\alpha\beta\gamma}$.}
\begin{indented}
\lineup
\item[]\begin{tabular}{@{}lllll}
\br                              
 & $(111)$ & $(222)$ & $(112)$ & $(221)$ \\ \mr
$xxx$ &$-\hat{a}_{2y}^3$ & $\hat{a}_{1y}^3$ & $3\hat{a}_{1y} \hat{a}^2_{2y}$  & $-3\hat{a}_{1y}^2\hat{a}_{2y}$  \\
$yyy$ & $\hat{a}_{2x}^3$ & $-\hat{a}_{1x}^3$ & $-3\hat{a}_{1x} \hat{a}^2_{2x}$ & $3\hat{a}_{1x}^2\hat{a}_{2x}$ \\
$xxy$ & $\hat{a}^2_{2y}\hat{a}_{2x}$ & $-\hat{a}^2_{1y}\hat{a}_{1x}$ & $-\hat{a}_{2y}(\hat{a}_{1x}\hat{a}_{2y} + 2\hat{a}_{1y} \hat{a}_{2x})$ & $\hat{a}_{1y}(\hat{a}_{2x}\hat{a}_{1y} + 2\hat{a}_{2y} \hat{a}_{1x})$ \\
$xyy$ & $-\hat{a}^2_{2x}\hat{a}_{2y}$ & $\hat{a}^2_{1x}\hat{a}_{1y}$ & $\hat{a}_{2x}(\hat{a}_{1y}\hat{a}_{2x} + 2\hat{a}_{1x} \hat{a}_{2y})$ & $-\hat{a}_{1x}(\hat{a}_{2y}\hat{a}_{1x} + 2\hat{a}_{2x} \hat{a}_{1y})$ \\
\br
\end{tabular}
\end{indented}
\end{table*}
The $\mathcal{S}_{(lmn)}$ are partial sums; for $l=m=n$ we have
\begin{eqnarray}
\quad \mathcal{S}_{(iii)}&=\sum_{n \notin (n^{(i)}_-,n^{(i)}_+)} \frac{Q^{(i)}_{n}\left(w^{(i)}_{n}\right)^2 }{2A_c}M^{(i)}_n \nonumber\\
&+\sum_{n=n^{(i)}_-}^{n^{(i)}_+}\frac{Q^{(i)}_{n}\left(w^{(i)}_{n}\right)^2}{2A_c} \left( P^{(i)}_n+\tilde{\mathcal{P}}^{(i)}_n \right) \nonumber\\
&+ \frac{1}{4\pi a_i^4}\sum_{j=\pm}j \bigg[ 3\tilde{\omega}na_i\left(\tilde{\mathcal{A}}_j^{(i)}-2B_j^{(i)}\right)\nonumber\\
&\qquad +6C^{(i)}_j-(\tilde{\omega}na_i)^2\left(\tilde{\mathcal{B}}_j^{(i)}+4A_j^{(i)}\right)  \bigg],
\end{eqnarray}
and for $l =m \neq n$,
\begin{eqnarray}
\quad \mathcal{S}_{(ii\bar{i})} &= \bm{\hat{a}}\cdot \bm{\hat{b}} \mathcal{S}_{(iii)} + \sum_{n\notin (n^{(i)}_-,n^{(i)}_+)}\frac{\mathfrak{c}^{(i)}_n }{6A_c} N^{(i)}_n\nonumber\\
& + \sum_{n=n^{(i)}_-}^{n^{(i)}_+}\frac{\mathfrak{c}^{(i)}_n}{6A_c} \left( T^{(i)}_n+\tilde{\mathcal{T}}^{(i)}_n \right),
\end{eqnarray}
where we introduced 
\begin{equation}
\quad \mathfrak{c}^{(i)}_n = (\tilde{\omega}n)^2 - 3\left(Q^{(i)}_{n}\right)^2.
\end{equation}

\subsubsection{Evaluation of $\mathcal{G}^{\rm Fq}_{ijkl}$. }
\label{Gfqexact}
The radiative contribution is purely imaginary and is given by the Fourier sum (\ref{FDG}) with the Fourier components
\begin{eqnarray}
\quad {}^{\rm R} \ovb{\mathcal{G}}^{\rm Fq}_{\bm{n}} &=\frac{\rmi\mu}{\epsilon_0n^2} \frac{1}{2A_c w_{\bm{n}}} \bigg[ w_{\bm{n}}^2 (\tilde{\omega}n)^2  \{ \bm{\hat{s}}_{\bm{n}},\bm{\hat{z}}\}\{ \bm{\hat{s}}_{\bm{n}},\bm{\hat{z}} \} \nonumber\\
& + (\tilde{\omega}n)^2\kappa_{\bm{n}}^2  \{ \bm{\hat{s}}_{\bm{n}},\bm{\hat{\kappa}}_{\bm{n}}\}
\{\bm{\hat{s}}_{\bm{n}}, \bm{\hat{\kappa}}_{\bm{n}}\}\nonumber\\
&+\kappa_{\bm{n}}^2 w_{\bm{n}}^2 \left( \bm{\hat{z}}\bm{\hat{z}} -\bm{\hat{\kappa}}_{\bm{n}} \bm{\hat{\kappa}}_{\bm{n}}\right) \left( \bm{\hat{z}}\bm{\hat{z}}
-\bm{\hat{\kappa}}_{\bm{n}} \bm{\hat{\kappa}}_{\bm{n}}\right)  \nonumber\\
&+  (\bm{\kappa_n}^2 - w_{\bm{n}}^2)^2 \{ \bm{\hat{\kappa}}_{\bm{n}},\bm{\hat{z}}\} \{\bm{ \hat{\kappa}}_{\bm{n}},{\bm{\hat{z}}} \} \bigg].
\end{eqnarray}
The non-radiative contribution is purely real and we give it in the Cartesian basis. We note that the dyadic $\mathcal{G}^{\rm Fq}_{ijkl}$ is symmetric in its first two indices $\mathcal{G}^{\rm Fq}_{ijkl} = \mathcal{G}^{\rm Fq}_{jikl}$, the last two indices $\mathcal{G}^{\rm Fq}_{ijkl} = \mathcal{G}^{\rm Fq}_{ijlk}$, as well as with respect to interchanging the first two and last two indices, $\mathcal{G}^{\rm Fq}_{ijkl} = \mathcal{G}^{\rm Fq}_{klij}$, but it is not  symmetric with respect to all the permutations of indices. Thus to list the non-radiative contributions we find it convenient to introduce a fully symmetric dyadic ${}^{\rm N}\mathcal{G}^{\rm Fq}_{[ijkl]}$, in terms of which the usual dyadic ${}^{\rm N}\mathcal{G}^{\rm Fq}_{ijkl}$ is given by  
\begin{eqnarray}
\quad {}^{\rm N}\mathcal{G}^{\rm Fq}_{ijkl} &=  {}^{\rm N}\mathcal{G}^{\rm Fq}_{[ijkl]}- \frac{\mu (\tilde{\omega}n)^4}{3 \epsilon_0 n^2} \delta_{ij}\delta_{kl} {}^{\rm N} \mathcal{G}_0 \nonumber\\
&+ \frac{\mu (\tilde{\omega}n)^4}{6 \epsilon_0 n^2} \left(\delta_{il}\delta_{jk} + \delta_{ik}\delta_{jl}  \right) {}^{\rm N} \mathcal{G}_0 \nonumber\\
&+ \frac{(\tilde{\omega}n)^2}{3} \left( \delta_{kl}{}^{\rm N}\mathcal{G}^{\rm Ep}_{ij} +\delta_{ij} {}^{\rm N}\mathcal{G}^{\rm Ep}_{kl} \right) \nonumber\\
&+ \frac{(\tilde{\omega}n)^2}{12} \left(\delta_{jl}{}^{\rm N}\mathcal{G}^{\rm Ep}_{ik} + \delta_{ik} {}^{\rm N}\mathcal{G}^{\rm Ep}_{jl} \right)\nonumber\\
&+ \frac{(\tilde{\omega}n)^2}{12} \left(\delta_{jk} {}^{\rm N}\mathcal{G}^{\rm Ep}_{il} +\delta_{il} {}^{\rm N}\mathcal{G}^{\rm Ep}_{jk} \right).\label{GFqex}
\end{eqnarray} 
Directly from the Green function definition we find that ${}^{\rm N}\mathcal{G}^{\rm Fq}_{[zzz\alpha]} = {}^{\rm N}\mathcal{G}^{\rm Fq}_{[z\alpha\beta\gamma]} =0 $ for $\alpha,\beta,\gamma = x,y$. From the Helmholtz equation we find
\begin{eqnarray}
\quad {}^{\rm N}\mathcal{G}^{\rm Fq}_{[zz\alpha\beta]} = &-{}^{\rm N}\mathcal{G}^{\rm Fq}_{[xx\alpha\beta]} - {}^{\rm N}\mathcal{G}^{\rm Fq}_{[yy\alpha\beta]}\nonumber\\
&- 4\frac{(\tilde{\omega}n)^2}{3} {}^{\rm N}\mathcal{G}^{\rm Ep}_{\alpha\beta}.
\end{eqnarray}
We thus need to only find the in-plane components of the dyadic, ${}^{\rm N}\mathcal{G}^{\rm Fq}_{\alpha\beta\gamma\delta}$ for $\alpha,\beta,\gamma,\delta = x,y$. We find
\begin{equation}
\quad {}^{\rm N} \mathcal{G}^{\rm Fq}_{[\alpha\beta\gamma\delta]} =\frac{\mu}{\epsilon_0 n^2}  \sum_{(lmnp)} \frac{C^{(lmnp)}_{\alpha\beta\gamma\delta} \mathcal{S}_{(lmnp)}}{\left[\left(\bm{\hat{a}}_1\times\bm{\hat{a}}_2\right)\cdot \bm{\hat{z}} \right]^4},
\end{equation}
where the sum runs over the set of indices $(lmnp) = \{(1111), (2222), (1112), (2221), (1122) \}$ with the coefficients $C^{(lmnp)}_{\alpha\beta\gamma\delta}$ given in Table \ref{tabFQ}. 
\begin{table*}[htb]
\caption{\label{tabFQ} The coefficients $C^{(lmnk)}_{\alpha\beta\gamma\delta}$.}
\begin{indented}
\lineup
\item[]\begin{tabular}{@{}llllll}
\br                              
 & $(1111)$ & $(2222)$ & $(1112)$ & $(2221)$ & $(1122)$ \\ \mr
$xxxx$ & $-\hat{a}_{2y}^4$ & $-\hat{a}_{1y}^4$ & $4\hat{a}_{1y} \hat{a}^3_{2y}$  & $4\hat{a}_{1y}^3\hat{a}_{2y}$ & $-6 \hat{a}_{1y}^2\hat{a}_{2y}^2$  \\
$yyyy$ & $-\hat{a}_{2x}^4$ & $-\hat{a}_{1x}^4$ & $4\hat{a}_{1x} \hat{a}^3_{2x}$  & $4\hat{a}_{1x}^3\hat{a}_{2x}$ & $-6 \hat{a}_{1x}^2\hat{a}_{2x}^2$  \\
$xxxy$ & $\hat{a}_{2x}\hat{a}_{2y}^3$ & $\hat{a}_{1x}\hat{a}_{1y}^3$ & $-\hat{a}_{2y}^2\left[3\hat{a}_{1y}\hat{a}_{2x} + \hat{a}_{2y} \hat{a}_{1x}\right]$ & $-\hat{a}_{1y}^2\left[3\hat{a}_{1x}\hat{a}_{2y} + \hat{a}_{2x} \hat{a}_{1y}\right]$ & $3\hat{a}_{1y}\hat{a}_{2y}\left[\hat{a}_{1x}\hat{a}_{2y} + \hat{a}_{2x} \hat{a}_{1y}\right]$ \\
$xyyy$ & $\hat{a}_{2y}\hat{a}_{2x}^3$ & $\hat{a}_{1y}\hat{a}_{1x}^3$ & $-\hat{a}_{2x}^2\left[3\hat{a}_{1x}\hat{a}_{2y} + \hat{a}_{2x} \hat{a}_{1y}\right]$ & $-\hat{a}_{1x}^2\left[3\hat{a}_{1y}\hat{a}_{2x} + \hat{a}_{2y} \hat{a}_{1x}\right]$ & $3\hat{a}_{1x}\hat{a}_{2x}\left[\hat{a}_{1x}\hat{a}_{2y} + \hat{a}_{2x} \hat{a}_{1y}\right]$ \\
$xxyy$ & $-\hat{a}^2_{2x}\hat{a}^2_{2y}$ & $-\hat{a}^2_{1x}\hat{a}_{1y}^2$ & $2\hat{a}_{2x}\hat{a}_{2y}\left[\hat{a}_{1x}\hat{a}_{2y} + \hat{a}_{2x} \hat{a}_{1y}\right]$ & $2\hat{a}_{1x}\hat{a}_{1y}\left[\hat{a}_{1x}\hat{a}_{2y} + \hat{a}_{2x} \hat{a}_{1y}\right]$ & $-\left[\hat{a}_{1x}\hat{a}_{2y} + \hat{a}_{2x} \hat{a}_{1y}\right]^2 -2 \hat{a}_{1x}\hat{a}_{1y} \hat{a}_{2x} \hat{a}_{2y}$ \\
\br
\end{tabular}
\end{indented}
\end{table*}
The $\mathcal{S}_{(lmnp)}$ are partial sums. For $l=m=n=p$ we have
\begin{eqnarray}
\quad \mathcal{S}_{(iiii)} &= \sum_{n=n^{(i)}_-}^{n^{(i)}_+}\frac{\left(w^{(i)}_{n}\right)^4}{2A_c}\left( P^{(i)}_n + \tilde{\mathcal{P}}^{(i)}_n \right) \nonumber\\
&+\sum_{n \notin (n^{(i)}_-,n^{(i)}_+)} \frac{ \left(w^{(i)}_{n}\right)^4}{2A_c} M^{(i)}_n + \sum_{j=\pm} \frac{ 6D_j^{(i)}}{\pi a_i^5} \nonumber\\
& + \sum_{j=\pm} \left( \tilde{\omega}^2n^2 \frac{ \tilde{\mathcal{A}}^{(i)}_j - 2B_j^{(i)} }{\pi a_i^3}  + \tilde{\omega}n \frac{6 C_j^{(i)}}{\pi a_i^4} \right),
\end{eqnarray}
and for $l=m=n\neq p$ we have
\newpage
\begin{eqnarray}
\mathcal{S}_{(iii\bar{i})} =&-\sum_{n\notin (n_-^{(i)},n_+^{(i)})} \frac{Q^{(i)}_{n} \left(w^{(i)}_{n}\right)^2}{2A_c} N^{(i)}_n  \nonumber\\
&-\sum_{n=n_-^{(i)}}^{n_+^{(i)}}\frac{Q^{(i)}_{n}\left(w^{(i)}_{n}\right)^2}{2A_c}\left( T^{(i)}_n + \tilde{\mathcal{T}}^{(i)}_n \right)  \nonumber\\
&+ \mathcal{S}_{(iiii)} \mathrm{cos}\phi.  
\end{eqnarray}
For $l=m$ and $n=p$ we write
\begin{equation}
\quad S_{(1122)} = \frac{1}{2} \left( S^{(1)}_{(1122)} + S^{(2)}_{(1122)} \right),
\end{equation}
where
\begin{eqnarray}
\quad \mathcal{S}^{(i)}_{(1122)} &=  \sum_{n=n^{(i)}_-}^{n^{(i)}_+}\frac{\mathfrak{c}^{(i)}_n}{6 A_c} \left( O^{(i)}_n +  \tilde{\mathcal{O}}^{(i)}_n \right) \nonumber\\
&-\sum_{n\notin (n^{(i)}_-,n^{(i)}_+)} \frac{\mathfrak{c}^{(i)}_n}{6A_c} L^{(i)}_n\nonumber\\
& + 2\mathcal{S}_{(iii\bar{i})}\cos \phi  - \mathcal{S}_{(iiii)}\frac{ \cos^2 \phi +1}{2} \nonumber\\
& + \frac{2}{3} (\tilde{\omega}n)^2  \mathcal{S}_{(ii)}\sin^2 \phi .
\end{eqnarray}

\section{Approximate expressions for the periodic Green function dyadics}
\label{approximate}
The periodic Green function dyadics can all be evaluated using the expressions given in the previous section, together with (\ref{i1}-\ref{i2}). The non-radiative contributions, however, take the form of summations which, while converging rapidly, do not give an immediate physical insight into the nature of the interactions between the multipoles. In the long wavelength limit, a simplified description of an array can be obtained by approximating the non-radiative periodic Green functions by their static ($\omega=0$) limit. The static periodic Green functions describe only the short-range contributions to the non-radiative interactions, and can be evaluated directly in real space without the need to use the summation acceleration techniques. But the neglect of the mid- and long-range interactions described by the dynamic contributions to periodic Green function dyadics is only justified when the wavelength of light is orders of magnitude larger than a lattice spacing, a condition that is rarely satisfied in metamaterial systems. For that reason we give a simple approximate scheme that extends the range of validity of the static approximation of the non-radiative periodic Green function dyadics; the radiative contributions to periodic Green function dyadics can easily be evaluated exactly. For the non-radiative contributions we identify leading dynamic corrections to the static limit of the periodic Green function dyadics, valid for dyadics describing lattices of high symmetry. The dynamic corrections we identify take a simple analytic form. The approximation scheme gives an excellent description of all the periodic Green function dyadics for frequencies up to $\tilde{\omega}\sqrt{A_c}\approx 1$. We compare the importance of dynamic corrections in a calculation of the four independent periodic Green function dyadics.

We illustrate the approximation method for the dipole periodic Green function, and then give the results for the remaining functions. We begin by decomposing the dipole dyadic into its static and dynamic contributions,
\begin{equation}
\quad \ovb{\mathcal{G}}^{\rm Ep} = \ovb{\mathcal{L}}^{\rm Ep} + \ovb{\mathcal{T}}^{\rm Ep},\label{Gdecom}
\end{equation}
where we identified the static contribution as $\ovb{\mathcal{L}}^{\rm Ep} = \lim_{\omega\rightarrow 0} \ovb{\mathcal{G}}^{\rm Ep}$, and the remaining dynamic contribution as $\ovb{\mathcal{T}}^{\rm Ep}$. The static contribution involves summation over terms that drop-off as $1/r^3$, and this summation is done in real space. The dynamic contribution involves a radiation reaction term $\ovb{\mathcal{R}}^{\mathrm{Ep}}$ from the dipole at the origin, and a contribution that involves a summation over terms that drop-off as $1/r$ and $1/r^2$. Due to a slow convergence of the latter summation in real space we do the summation in Fourier space. We arrive at 
\begin{equation}
\quad \ovb{\mathcal{T}}^{\rm Ep} = \sum_{\bm{n}} \ovb{\mathcal{T}}^{\rm Ep}_{\bm{n}} + \ovb{\mathcal{R}}^{\rm Ep},\label{T}
\end{equation}
where 
\begin{eqnarray}
\quad \ovb{\mathcal{T}}^{\rm Ep}_{\bm{n}} &= \frac{1}{A_c} \ovb{g}^{\rm Ep}_{\omega}(\bm{\kappa}_{\bm{n}}) \nonumber\\
&-\int_{BZ}\frac{\mathrm{d}^2\kappa}{(2\pi)^2} \ovb{g}^{\rm Ep}_{\omega}(\bm{\kappa}_{\bm{n}}+\bm{\kappa}). \label{Tn}
\end{eqnarray}
The first contribution to (\ref{Tn}) is the Poisson term that originates from a Fourier transform of the summation taken over all the lattice sites including the origin. The second term in (\ref{Tn}), when multiplied by the electric dipole moment, gives Fourier contributions at wave vectors within one Brillouin zone from $\boldsymbol{\kappa}_{\boldsymbol{n}}$ to the electric field that describes the interaction of the dipole at the origin with itself. The second term is subtracted from the first in order to arrive at a well-defined periodic Green function dyadic, which involves a summation over all the lattice sites except the origin; see (\ref{Gdef}). Here we have identified a frequency-dependent component of the free-space Green function evaluated in the plane of the array,
\begin{equation}
\quad \ovb{g}_{\omega}(\bm{\kappa}) = \lim_{z\rightarrow 0}\left[\ovb{g}(\bm{\kappa},z) -\lim_{\omega\rightarrow 0}\ovb{g}(\bm{\kappa}, z)\right], \label{eqn1}
\end{equation}
where the spatial limit is taken in a symmetric way, $\lim_{z\rightarrow 0} = \frac{1}{2} \lim_{z \rightarrow 0^+} + \frac{1}{2}\lim_{z\rightarrow 0^-}$. The Fourier transform in the x,y plane of an arbitrary function $\mathcal{O}(\bm{r})$ is understood in a usual way,
\begin{equation}
\quad \mathcal{O}(\bm{r}) = \int \frac{d\bm{\kappa}}{(2\pi)^2} \mathcal{O}(\bm{\kappa},z) \rme^{\rmi\bm{\kappa}\cdot \bm{R}}.
\end{equation}

Next we separate the radiative and non-radiative contributions to $\ovb{\mathcal{G}}^{\mathrm{Ep}}$. The static term $\ovb{\mathcal{L}}^{\mathrm{Ep}}$ is purely non-radiative, so the radiative contributions are exclusively from $\ovb{\mathcal{T}}^{\mathrm{Ep}}$, and we consider their different components. Radiative contributions to the total self-interaction term that are identified by a sum of all the contributions given by the second terms on the right-hand side of (\ref{Tn}) exactly cancels the radiative dyadic $\ovb{\mathcal{R}}^{\rm Ep}$ on the right-hand side of (\ref{T}), as can be easily verified by a direct calculation. The only non-vanishing radiative contribution is then identified by the radiative contributions to the Poisson terms, ${}^{\rm R}\ovb{\mathcal{T}}^{\rm Ep}_{\bm{n}} = \rmi\mathrm{Im}\ovb{g}^{\rm Ep}_{\omega}(\bm{\kappa}_{\bm{n}})/A_c$,  which are of course equal to the Fourier components of the radiative periodic Green function identified in section \ref{exact exp},
\begin{equation}
\quad {}^{\rm R} \ovb{\mathcal{G}}^{\rm Ep}_{\bm{n}} = {}^{\rm R}\ovb{\mathcal{T}}^{\rm Ep}_{\bm{n}}.
\end{equation}
At frequencies such that diffraction does not occur the only non-vanishing radiative contribution is from Poisson term at $\bm{\kappa}_{\bm{n}} = \bm{\kappa}_0$, and at these frequencies we have 
\begin{equation}
\quad {}^{\rm R}\ovb{\mathcal{G}}^{\rm Ep} = {}^{\rm R}\ovb{\mathcal{G}}^{\rm Ep}_{0}.
\end{equation}

The non-radiative contribution to the dynamic correction, on the other hand, does not admit a simple analytic form, even in the absence of diffraction. But at frequencies well-below the diffraction limit we expect the contributions at small values of in-plane wave vector to be dominant. We thus  approximate the total non-radiative dynamic correction by a term associated with contributions at $\boldsymbol{\kappa}_0$ and at wave vectors within one Brillouin zone from it,
\begin{equation}
\quad {}^{\rm N} \ovb{\mathcal{T}}^{\rm Ep} \approx {}^{\rm N} \ovb{\mathcal{T}}^{\rm Ep}_0. \label{effmed}
\end{equation}
However, even within the approximation (\ref{effmed}) the dynamic correction is non-analytic, as the evaluation of the self-interaction term (see (\ref{Tn})) in general requires a numerical integration. The correction (\ref{effmed}) can be simplified further by noting that at frequencies well below the diffraction limit the self-interaction term is dominated by contributions at wave vectors close to the center of the Brillouin zone, with the contributions at wave vectors close to the boundaries of the Brillouin zone less important. When evaluating self-interaction term we thus approximate the integral over the Brillouin zone with an integral over a circle with the area equal to the area of the Brillouin zone,  
\begin{eqnarray}
\quad \int_{\rm BZ}\frac{\mathrm{d}^2\kappa}{(2\pi)^2} \ovb{g}^{\rm Ep}_{\omega}(\bm{\kappa}_{\bm{0}}&+\bm{\kappa}) \nonumber\\
 &\approx \int_{C(0,R)}\frac{\mathrm{d}^2\kappa}{(2\pi)^2} \ovb{g}^{\rm Ep}_{\omega}(\bm{\kappa}),\label{gepI}
\end{eqnarray}
where $R = (2\pi)/\sqrt{\pi A_c}$. Using approximations (\ref{effmed},\ref{gepI}) we arrive at an analytic approximation to the non-radiative dynamic correction,
\begin{eqnarray}
\quad{}^{\rm N} \ovb{\mathcal{T}}^{\rm Ep}_{\rm app} &= \frac{\mu \kappa_0}{2A_c n^2 \epsilon_0} \left[ \bm{\hat{\kappa}}_0\bm{\hat{\kappa}}_0  - \bm{\hat{z}}\bm{\hat{z}} \right] \nonumber\\
&-\frac{\mu}{4\pi} \frac{ R^3 - w_R (R^2 - 4\tilde{\omega}^2n^2) }{6\epsilon_0 n^2}  \ovb{U}_{\parallel} \nonumber\\
&+ \frac{\mu}{4\pi} \frac{ R^3 - w_R (R^2 + 2\tilde{\omega}^2n^2) }{3\epsilon_0 n^2} \bm{\hat{z}}\bm{\hat{z}}, \label{Gepapp}
\end{eqnarray}
where we have put $w_R = \sqrt{R^2 - \left(\tilde{\omega}n\right)^2}$, and  identified a unit dyadic in the plane of the array, $\ovb{U}_{\parallel} =  \bm{\hat{x}}\bm{\hat{x}}+\bm{\hat{y}}\bm{\hat{y}}$. The first line in (\ref{Gepapp}) is a Poisson term at wave vector $\bm{\kappa}_{\bm{n}} = \bm{\kappa}_0$, and the remaining lines approximate the self-interaction term of the dipole at the origin. Having identified an approximate dynamic contribution we arrive at an approximation to the non-radiative dipole periodic Green function, 
\begin{equation}
\quad \ovb{\mathcal{G}}^{\rm Ep} \approx  {}^{\rm R}\ovb{\mathcal{G}}^{\rm Ep}_0 + \ovb{\mathcal{L}}^{\rm Ep} + {}^{\rm N} \ovb{\mathcal{T}}^{\rm Ep}_{\rm app}. \label{GEPAPP}
\end{equation} 

We find approximate expressions for the remaining periodic Green functions in the same manner. For each of the dyadics we identify a static contribution, $\ovb{\mathcal{L}}$, and a remaining dynamic contribution, $\ovb{\mathcal{T}}$. The approximate expressions for the periodic Green functions $\ovb{\mathcal{G}}^{\rm Em}$ and $\ovb{\mathcal{G}}^{\rm Eq}$  follow directly from the procedure outlined above, and they take a particularly simple form since the self-interaction term vanishes by symmetry. The derivation of the dyadic $\ovb{\mathcal{G}}^{\rm Fq}$ is slightly more involved. Here the non-radiative  dynamic contribution ${}^{\rm N}\ovb{\mathcal{T}}^{\rm Fq}$ describes the long-range interactions with $1/r$ dependence that need to be summed in Fourier space, as well as the short-range interactions with $1/r^3$ dependence that need to be summed in real space; these two types of contributions we treat separately. Following these steps we arrive at the approximate expressions,
\begin{eqnarray}
\quad \ovb{\mathcal{G}}^{\rm Em} \approx {}^{\rm R}\ovb{\mathcal{G}}^{\rm Em}_0,\label{GEMAPP}\\
\quad \ovb{\mathcal{G}}^{\rm Eq} \approx {}^{\rm R}\ovb{\mathcal{G}}^{\rm Eq}_0 +\ovb{\mathcal{L}}^{\rm Eq}+{}^{\rm N} \ovb{\mathcal{T}}^{\rm Eq}_{\rm app},\label{GEQAPP}\\
\quad \ovb{\mathcal{G}}^{\rm Fq} \approx {}^{\rm R}\ovb{\mathcal{G}}^{\rm Fq}_0 + \ovb{\mathcal{L}}^{\rm Fq} + {}^{\rm N}\ovb{\mathcal{T}}^{\rm Fq}_{\rm app},\label{GFQAPP}
\end{eqnarray}
where the non-radiative dynamic contribution to $\ovb{\mathcal{G}}^{\rm Em}$ vanishes; we have
\begin{eqnarray}
\quad {}^{\rm N}\ovb{\mathcal{T}}^{\rm Eq}_{\rm app} = \frac{\rmi\mu \kappa_0^2}{\epsilon_0 n^2} \frac{  2\bm{\hat{z}} \left\{ \bm{\hat{z}},\bm{\hat{\kappa}}_0 \right\} +\bm{\hat{\kappa}}_0 \left( \bm{\hat{z}}\bm{\hat{z}} - \bm{\hat{\kappa}}_0\bm{\hat{\kappa}}_0  \right) }{2A_c},
\end{eqnarray}
and the non-radiative dynamic contribution to $\ovb{\mathcal{G}}^{\rm Fq}$ can be written as a sum of a term evaluated exactly in real space and terms evaluated approximately in Fourier space,
\begin{equation}
\quad {}^{\rm N}\ovb{\mathcal{T}}^{\rm Fq}_{\rm app} = \ovb{\mathfrak{S}} + \ovb{\mathfrak{F}}_{\rm F} - \ovb{\mathfrak{F}}_{\rm S}.
\end{equation}
A real-space summation describing contributions to the dyadic from frequency-dependent but short-range interactions is given by
\begin{equation}
\quad \ovb{\mathfrak{S}} = \frac{\mu}{4\pi \epsilon_0 n^2}\frac{(\tilde{\omega}n)^2}{4} \sum_{\bm{R}\neq 0} \rme^{\rmi\bm{\kappa}_0\cdot \bm{R}} \frac{\mathfrak{d}}{R^3},
\end{equation}
where we identified a fourth-rank unit tensor
\begin{eqnarray}
\quad \mathfrak{d}_{ijkl} &= 6\left( \delta_{kl}\hat{R}_i\hat{R}_j + \delta_{ij}\hat{R}_k\hat{R}_l \right)- 2\delta_{ij}\delta_{kl}   \nonumber\\
& -30 \hat{R}_i\hat{R}_j\hat{R}_k\hat{R}_l  + 3 \delta_{jl}\hat{R}_i \hat{R}_k \nonumber\\
&+ 3\bigg( \delta_{ik} \hat{R}_{j}\hat{R}_l + \delta_{jk}\hat{R}_i\hat{R}_l + \delta_{il}\hat{R}_j\hat{R}_k \bigg).
\end{eqnarray}
A Poisson term at $\boldsymbol{\kappa}_0$ due to a Fourier transform of the dyadic describing short- and medium-ranged interactions takes the form,
\begin{eqnarray}
&\quad \ovb{\mathfrak{F}}_{\rm F} =  \frac{\mu \kappa_0}{2\epsilon_0 n^2 A_c} \bigg[- \tilde{\omega}^2n^2 \{\bm{\hat{s}},\bm{\hat{\kappa}}_0 \} \{ \bm{\hat{s}}, \bm{\hat{\kappa}}_0  \} \nonumber\\
&\quad+ \tilde{\omega}^2n^2 \{\bm{\hat{s}},\bm{\hat{z}}\} \{\bm{\hat{s}},\bm{\hat{z}}\}
\nonumber\\
&\quad+\left(2\tilde{\omega}^2n^2 -4\kappa_0^2\right)  \left\{ \bm{\hat{\kappa}}_0,\bm{\hat{z}}\right\}  \left\{ \bm{\hat{\kappa}}_0,\bm{\hat{z}} \right\}\nonumber\\
&\quad + \left( \kappa_0^2 - \frac{1}{2}\tilde{\omega}^2n^2\right)  (\bm{\hat{z}}\bm{\hat{z}} - \bm{\hat{\kappa}}_0\bm{\hat{\kappa}}_0) (\bm{\hat{z}}\bm{\hat{z}} - \bm{\hat{\kappa}}_0\bm{\hat{\kappa}}_0) \bigg],
\end{eqnarray} 
while an approximate contribution to the self-interaction term is
\begin{eqnarray}
\quad \ovb{\mathfrak{F}}_S &= \frac{1}{4}\left( \mathfrak{F}_S^a + \mathfrak{F}_S^c\right) \left( \boldsymbol{\hat{x}}\boldsymbol{\hat{x}} - \boldsymbol{\hat{y}}\boldsymbol{\hat{y}} \right) \left( \boldsymbol{\hat{x}}\boldsymbol{\hat{x}} - \boldsymbol{\hat{y}}\boldsymbol{\hat{y}} \right) \nonumber\\
&+\left(\mathfrak{F}_S^a - \mathfrak{F}_S^c \right) \left\{ \boldsymbol{\hat{x}},\boldsymbol{\hat{y}} \right\} \left\{ \boldsymbol{\hat{x}},\boldsymbol{\hat{y}}  \right\} \nonumber\\
&+ \left( \mathfrak{F}_S^b +4\mathfrak{F}_S^c \right)  \left\{ \boldsymbol{\hat{x}},\boldsymbol{\hat{z}} \right\}\left\{ \boldsymbol{\hat{x}},\boldsymbol{\hat{z}} \right\}   \nonumber\\
&+ \left( \mathfrak{F}_S^b +4\mathfrak{F}_S^c \right) \left\{ \boldsymbol{\hat{y}},\boldsymbol{\hat{z}} \right\}\left\{ \boldsymbol{\hat{y}},\boldsymbol{\hat{z}} \right\}    \nonumber\\
&-\mathfrak{F}_S^c (\boldsymbol{\hat{x}}\boldsymbol{\hat{x}} - \boldsymbol{\hat{z}}\boldsymbol{\hat{z}}) (\boldsymbol{\hat{x}}\boldsymbol{\hat{x}} - \boldsymbol{\hat{z}}\boldsymbol{\hat{z}}) \nonumber\\
& -\mathfrak{F}_S^c (\boldsymbol{\hat{y}}\boldsymbol{\hat{y}} - \boldsymbol{\hat{z}}\boldsymbol{\hat{z}}) (\boldsymbol{\hat{y}}\boldsymbol{\hat{y}} - \boldsymbol{\hat{z}}\boldsymbol{\hat{z}}),
\end{eqnarray} 
where 
\begin{eqnarray}
\quad \mathfrak{F}_S^a =  \frac{\mu \tilde{\omega}^2}{24\epsilon_0 \pi}  \left[w_R \left( R^2 + 2\tilde{\omega}^2n^2 \right) - R^3 \right]  \\
\quad \mathfrak{F}_S^b =  \frac{\mu \tilde{\omega}^2}{24\epsilon_0 \pi}  \left[ w_R  \left( 4\tilde{\omega}^2n^2 - R^2  \right) + R^3 \right] \\
\quad \mathfrak{F}_S^c = \frac{\mu\tilde{\omega}^2}{40\epsilon_0 \pi} \left[ \frac{R^2w_R^3}{(\tilde{\omega}n)^2} + \frac{2}{3}w_R^3 -\frac{R^5}{(\tilde{\omega}n)^2} +\frac{5}{6} R^3 \right].
\end{eqnarray}

We now compare the range of validity of the approximate expressions  (\ref{GEPAPP}-\ref{GFQAPP}) with the range of validity of the usual static approximation, and discuss the dipole periodic Green function dyadic first. 
\begin{figure}[htb]
\centering
\includegraphics[scale = 0.8]{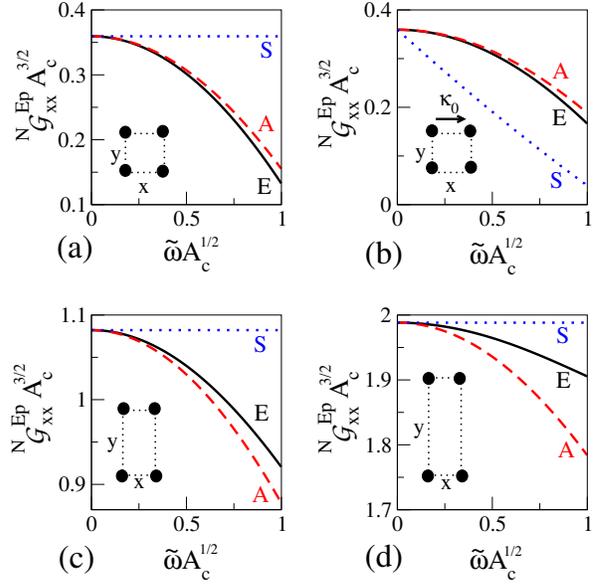}
\caption{A comparison of the exact (E), approximate (A), and static (S) dipole dyadic ${}^{\rm N}\mathcal{G}^{\rm Ep}_{xx}$ for the following scenarios: (a) a square lattice illuminated with normally incident light,  (b) a square lattice illuminated at $\theta_0=45^0$ with an in-plane wave vector $\bm{\kappa}_0\propto\bm{\hat{x}}$, and rectangular lattices with basis lattice vectors (c) $2a_x = a_y$ and (d) $3a_x = a_y$ illuminated at normal incidence.}
\label{dipcomp}
\end{figure}
Figures~\ref{dipcomp}(a) and~\ref{dipcomp}(b) show the in-plane dyadic component ${}^{\rm N}\mathcal{G}^{\rm Ep}_{xx}$ describing a square lattice evaluated using the exact expression (\ref{Gepex}), the approximate expression (\ref{GEPAPP}), and the usual static approximation. The approximate expression (\ref{GEPAPP}) is in an excellent agreement with the exact result at frequencies up to  \mbox{$\tilde{\omega}A_c^{1/2}\approx 1$,} while the validity of the static approximation is much more limited. Limitations of the static approximation are especially pronounced for light at non-normal incidence, for which the static approximation predicts an incorrect slope of the dyadic considered as a function of frequency. This discrepancy leads to clear deviations from the exact result even in an extreme long-wavelength limit, $\tilde{\omega}A_c^{1/2}\ll1$. We carry out a similar comparison for lattices of different structure. A triangular lattice shows a similar degree of agreement between the approximate and the exact result found for a square lattice. For lattices of lower symmetry, however, the approximate expression (\ref{GEPAPP}) shows more significant deviations from the exact result. This is illustrated in Figures~\ref{dipcomp}(c) and~\ref{dipcomp}(d), where we plot ${}^{\rm N}\mathcal{G}^{\rm Ep}_{xx}$ describing rectangular lattices with the basis vectors $2a_x = a_y$ and $3a_x = a_y$ respectively, calculated within different approximations. A failure of the approximate expression (\ref{GEPAPP}) to describe a strongly anisotropic lattice $3a_x = a_y$ is due to a failure of the approximation (\ref{gepI}) to respect the symmetry of the unit cell, and due a slower convergence of the summation over the reciprocal vectors in the summation (\ref{T}). Other components of the dipole periodic Green function not plotted in Figure~\ref{dipcomp} show a similar level of agreement between the exact and approximate expressions. 

\begin{figure}[htb]
\centering
\includegraphics[scale=0.8]{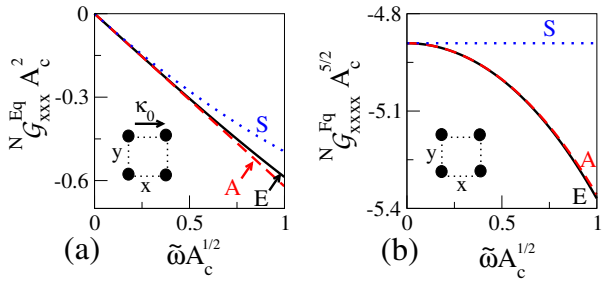}
\caption{A comparison of the exact (E), approximate (A), and static (S) dyadics describing a square lattice: (a) ${}^{\rm N}\mathcal{G}^{\rm Eq}_{xxx}$ for light incident at $\theta=30^0$ with  $\bm{\kappa}_0\propto \bm{\hat{x}}$,  (b) ${}^{\rm N}\mathcal{G}^{\rm Fq}_{xxxx}$ for normally incident light.}
\label{multcomp}
\end{figure}

For completeness we show an analogous comparison of the exact and approximate expressions for the non-radiative contributions to the periodic Green function dyadics $\ovb{\mathcal{G}}^{\rm Eq}$ and $\ovb{\mathcal{G}}^{\rm Fq}$. We compute the in-plane dyadic components ${}^{\rm N}\mathcal{G}_{xxx}^{\rm Eq}$ and ${}^{\rm N}\mathcal{G}^{\rm Fq}_{xxxx}$ describing a square lattice using the exact expressions (\ref{GEqex}, \ref{GFqex}), the approximate expressions (\ref{GEQAPP}, \ref{GFQAPP}), and the static approximation; see Figure~\ref{multcomp}.
Again the approximate expressions (\ref{GEQAPP}, \ref{GFQAPP}) are in a close agreement with the exact results at frequencies $\tilde{\omega}A_c^{1/2}<1$, with poorer agreement when the lattice has a lower symmetry (not shown); similar conclusions hold for the remaining components of the dyadics. 

Finally, we comment on the importance of the dynamic corrections in evaluating  non-radiative periodic Green function dyadics of different polarity. From Figure~\ref{dipcomp} and Figure~\ref{multcomp} we see that  the dynamic corrections give significant contributions to the periodic Green function dyadics $\ovb{\mathcal{G}}^{\rm Ep}$ and $\ovb{\mathcal{G}}^{\rm Fq}$, but are less significant in evaluation of the dyadic $\ovb{\mathcal{G}}^{\rm Eq}$, which is well-approximated by its static limit, $\ovb{\mathcal{L}}^{\rm Eq}$; the static and the leading dynamic contribution to $\ovb{\mathcal{G}}^{\rm Em}$ vanish and thus are not considered. A contribution from the dynamic correction ${}^{\rm N}\ovb{\mathcal{T}}^{\rm Eq}$ is small as compared to the corresponding corrections to the other dyadics, due to vanishing of the dynamic self-interaction term that is associated with the multipole at the origin; this leads to a smaller dynamic contribution overall. Finally we note that the static approximation gives a correct prediction for the slope of periodic Green function dyadics describing interactions with higher multipole moments, irrespective of the illumination conditions, due to a vanishing slope of the dynamic corrections at small frequencies. 

\section{Conclusions}
\label{conclusions}
We have derived a representation, suitable for numerical calculations, of periodic Green function dyadics that enter a multipole description of a 2d array of emitters. The optical response of the emitters is treated at the level that includes their electric dipole moments, magnetic dipole moments, and electric quadrupole moments; we take into account a response of each multipole moment to the electric field, to the magnetic field, and to the symmetrized gradient of electric field at the position of the multipole. All the periodic Green function dyadics that enter the description of an array at this level are given within one formalism that explicitly shows energy balance on the lattice. We identify all the radiative contributions to periodic Green function dyadics that give rise to radiation reaction fields. We evaluate the radiative contributions exactly, and give them in a form such that radiative fields associated with s- and p-polarized beams diffracted at any order are easily identified. The remaining non-radiative contributions to periodic Green function dyadics, which do not affect the energy conservation condition, are given in a form of rapidly converging series. All the non-radiative contributions to periodic Green functions are expressed as a summation over linear combinations of the same set of functions. Our results are summarized in sections (\ref{Gepexact}-\ref{Gfqexact}) together with (\ref{i1}-\ref{i2}). 

With energy balance on a lattice explicitly identified and the radiation reaction fields evaluated in a concise analytic form, simplified descriptions of the optical response of the array can be introduced to gain more insight into the electrodynamics. As an example we presented a simplified description  of periodic Green function dyadics that describe arrays of high symmetry illuminated with light at long wavelengths. The radiative contributions to the periodic Green function dyadics are treated exactly. The non-radiative contributions are approximated by their static limit and analytic dynamic corrections. The approximate description rigorously satisfies energy conservation and  extends the range of validity of the usual static approximation, without adding much complexity to the description. 

\appendix

\section{Free-space Green functions}
\setcounter{section}{1}

We use  $g_0(\bm{r})$ to denote the scalar free-space Green function, 
\begin{equation}
\quad g_0(\bm{r}) = \frac{1}{4\pi r} e^{\rmi \tilde{\omega}nr},
\end{equation}
which is a solution to the free-space Helmholtz equation,
\begin{equation}
\quad \left[ \nabla^2 + (\tilde{\omega}n)^2 \right] g_0(\bm{r}) = -\delta(\bm{r}).\label{g0}
\end{equation}
In terms of the scalar Green function, the four independent dyadic Green functions are given by the usual expressions
\begin{eqnarray}
\quad g^{\rm Ep}_{ij}(\bm{r}) =& \frac{\mu}{\epsilon_0 n^2}\slashed{\partial}_{ij} g_0(\bm{r}),\label{g1}\\
\quad g_{ij}^{\rm Em}(\bm{r}) =& \frac{\rmi\tilde{\omega}\mu}{\epsilon_0 c}\epsilon_{isj}\partial_s g_0(\bm{r}),\\
\quad g^{\rm Eq}_{ijk}(\bm{r}) = &-\frac{\mu}{\epsilon_0 n^2} \frac{1}{2} \left( \partial_k \slashed{\partial}_{ij} + \partial_j \slashed{\partial}_{ik}   \right) g_0(\bm{r}),\\
\quad g^{\rm Fq}_{ijkl}(\bm{r}) = &-\frac{\mu}{\epsilon_0 n^2} \frac{1}{4}\left( \partial_j \partial_l \slashed{\partial}_{ik} + \partial_i \partial_l \slashed{\partial}_{jk}  \right)g_0(\bm{r}) \nonumber\\
&-\frac{\mu}{\epsilon_0 n^2} \frac{1}{4}\left(\partial_j \partial_k \slashed{\partial}_{il} + \partial_i \partial_k \slashed{\partial}_{jl} \right) g_0(\bm{r}), \label{g2}
\end{eqnarray}
where we introduce a notation 
\begin{equation}
\quad \slashed{\partial}_{ij} = \partial_i \partial_j + (\tilde{\omega}n)^2 \delta_{ij},
\end{equation}
and the remaining dyadics follow from (\ref{g1}-\ref{g2}) in an obvious manner.

The expressions for the Fourier transformed free-space Green function dyadics can be found using the Green function formalism for planar structures \cite{Sipe1987}, generalized to include the quadrupole moments. At points in space such that $z\neq 0$ all the Green functions are of the form,
\begin{eqnarray}
\quad \ovb{g}(\bm{\kappa},z) &= \ovb{g}_+(\bm{\kappa})  e^{\rmi wz} \theta(z) \nonumber\\
&+ \ovb{g}_-(\bm{\kappa})  e^{-\rmi wz} \theta(-z), \label{amp}
\end{eqnarray}
where $\ovb{g}_{\pm}(\boldsymbol{\kappa})$ are the Green function amplitudes of the upward and downward propagating electromagnetic fields. The amplitudes of the dipole Green functions are given by \cite{Sipe1987},
\begin{eqnarray}
\quad \ovb{g}^{\rm Ep}_+(\bm{\kappa}) &= \frac{\rmi \tilde{\omega}^2\mu}{2\epsilon_0 w} \left( \bm{\hat{s}}\bm{\hat{s}}+\bm{\hat{p}}_+ \bm{\hat{p}} _+  \right),\label{gepFa}\\
\quad \ovb{g}^{\rm Ep}_-(\bm{\kappa}) &= \frac{\rmi \tilde{\omega}^2\mu}{2\epsilon_0 w} \left( \bm{\hat{s}}\bm{\hat{s}}+\bm{\hat{p}}_- \bm{\hat{p}} _-  \right),\label{gepFb}\\
\quad \ovb{g}^{\rm Em}_+(\bm{\kappa}) &= \frac{\rmi \tilde{\omega}^2n\mu}{2\epsilon_0 cw} \left( \bm{\hat{p}}_+\bm{\hat{s}}-\bm{\hat{s}} \bm{\hat{p}}_+  \right),\\
\quad \ovb{g}^{\rm Em}_+(\bm{\kappa}) &= \frac{\rmi \tilde{\omega}^2n\mu}{2\epsilon_0 cw} \left( \bm{\hat{p}}_-\bm{\hat{s}}-\bm{\hat{s}} \bm{\hat{p}}_-  \right).
\end{eqnarray}
The amplitudes of the quadrupole Green functions are given by
\begin{eqnarray}
\quad \ovb{g}^{\rm Eq}_+(\bm{\kappa}) &= \frac{\tilde{\omega}^3 n\mu}{2\epsilon_0 w} \left( \bm{\hat{s}}\left\{ \bm{\hat{s}}, \bm{\hat{v}}_+ \right\} +\bm{\hat{p}}_+ \left\{\bm{\hat{p}}_+, \bm{\hat{v}}_+\right\} \right),\\
\quad \ovb{g}^{\rm Eq}_-(\bm{\kappa}) &= \frac{\tilde{\omega}^3 n\mu}{2\epsilon_0 w} \left( \bm{\hat{s}}\left\{ \bm{\hat{s}}, \bm{\hat{v}}_- \right\} +\bm{\hat{p}}_- \left\{\bm{\hat{p}}_-, \bm{\hat{v}}_-\right\} \right), 
\end{eqnarray}
and 
\begin{eqnarray}
\quad \ovb{g}^{\rm Fq}_+(\bm{\kappa}) &= \frac{\rmi \tilde{\omega}^4 n^2\mu}{2\epsilon_0  w}  \left\{ \bm{\hat{s}}, \bm{\hat{v}}_+ \right\} \left\{ \bm{\hat{s}}, \bm{\hat{v}}_+ \right\} \nonumber\\
& + \frac{\rmi \tilde{\omega}^4 n^2\mu}{2\epsilon_0  w}\left\{ \bm{\hat{p}}_+, \bm{\hat{v}}_+ \right\}\left\{ \bm{\hat{p}}_+, \bm{\hat{v}}_+ \right\}, 
\end{eqnarray}
\begin{eqnarray}
\quad \ovb{g}^{\rm Fq}_-(\bm{\kappa}) &= \frac{\rmi \tilde{\omega}^4 n^2\mu}{2\epsilon_0 w}  \left\{ \bm{\hat{s}}, \bm{\hat{v}}_- \right\} \left\{ \bm{\hat{s}}, \bm{\hat{v}}_- \right\} \nonumber\\
&+ \frac{\rmi \tilde{\omega}^4 n^2\mu}{2\epsilon_0 w}\left\{\bm{\hat{p}}_-, \bm{\hat{v}}_-\right\} \left\{ \bm{\hat{p}}_-, \bm{\hat{v}}_- \right\}. \label{gfqFb}
\end{eqnarray}

\section{Energy balance on a lattice}

Here we verify the optical theorem for a 2d array of emitters.  We find that  the crucial role is played by the radiative contributions to the periodic Green function dyadics, with the non-radiative contributions not affecting the energy conservation condition.

\begin{figure}[htb]
\centering
\includegraphics[scale=0.2]{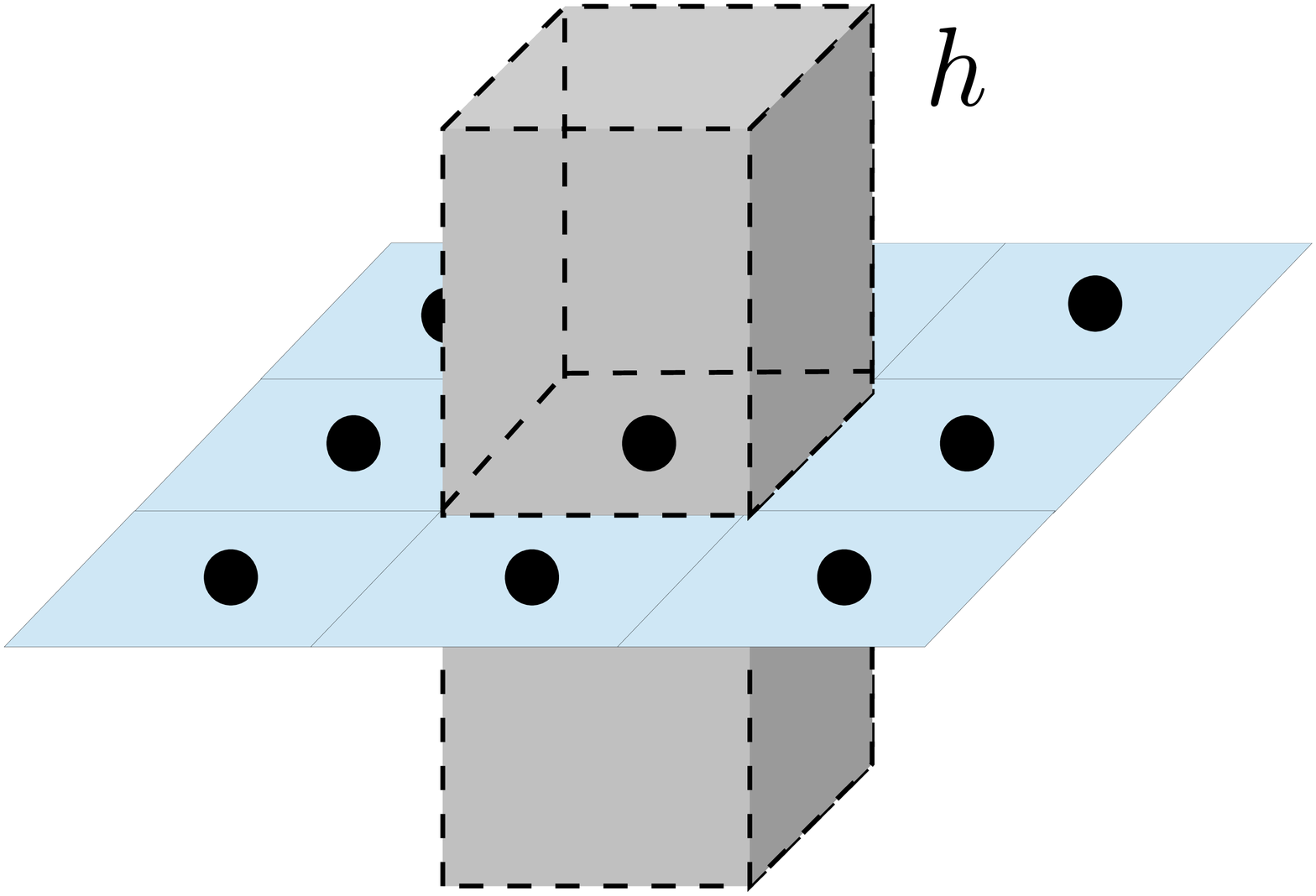}
\caption{Parallelepiped enclosing one unit cell of the array.}
\label{diagram}
\end{figure}

We consider energy balance for a volume of space enclosing one unit cell of the array. We divide the space into parallelepipeds of height $h$ such that the projection of each parallelepiped on the $z=0$ plane corresponds to one unit cell of the array; see Figure~\ref{diagram}.
The energy conservation condition for a volume of space enclosed by one parallelepiped is identified by the usual condition,
\begin{equation}
\quad W^{\rm A} = -\Phi,\label{econs}
\end{equation}
where $W^{\rm A}$ is the average power absorbed by the emitter enclosed within the volume of the parallelepiped and $\Phi$ is the time averaged flux of the Poynting vector through the surface of the parallelepiped.

We identify the restrictions on the periodic Green functions imposed by the condition (\ref{econs}) and show that they are satisfied. To do so, we evaluate both sides of (\ref{econs}) within the multipolar model. For simplicity we do the calculation for the parallelepiped that encloses the emitter at the origin of the lattice, $\bm{R}_{\bm{n}}=0$.  A straightforward calculation gives  
\begin{eqnarray}
\quad W^{\rm A} = 2\omega\mathrm{Im}H^{\rm A}, \label{abs}
\end{eqnarray}
for the total time-averaged power absorbed by the emitter, where
\begin{eqnarray}
\quad H^{\rm A} &= \left(\bm{E}^{\rm tot}\right)^* \cdot \bm{p} + \left(\bm{B}^{\rm tot}\right)^* \cdot \bm{m} \nonumber\\
& + \left(\ovb{F}^{\rm tot}\right)^* :\ovb{q},
\end{eqnarray}
and where the total fields at the position of the emitter are given by (\ref{Etot}-\ref{Ftot}). The flux of the Poynting vector involves three kinds of contributions, 
\begin{eqnarray}
\quad \Phi = \Phi^{\rm inc} + \Phi^{\rm scatt}+\Phi^{\rm si}, \label{twoterms}
\end{eqnarray}
where the flux $\Phi^{\rm inc}$ is due to the incident fields,  the flux $\Phi^{\rm scatt}$ is due to the scattered fields, and the flux $\Phi^{\rm si}$ involves cross-terms between the incident and the scattered fields. The flux due to the incident fields vanishes when an array is embedded in a lossless dielectric, $\Phi^{\rm inc} = 0$. To find the remaining contributions we find the electric field scattered by the array. We have
\begin{eqnarray}
\quad \bm{E}^{\rm scat}(\bm{\kappa},z) = \frac{(2\pi)^2}{A_c} \sum_{\bm{n}} \delta(\bm{\kappa}-\bm{\kappa}_{\bm{n}}) \bm{f}^{\rm E}(\bm{\kappa},z),\label{EA}
\end{eqnarray}   
where the vector $\bm{f}^{\rm E}(\bm{\kappa},z)$ is given by the products of the Fourier-transformed free-space Green functions  (see (\ref{amp})-(\ref{gfqFb})) with the multipole moments,
\begin{eqnarray}
\quad \bm{f}^{\rm E}(\bm{\kappa},z) &= \ovb{g}^{\rm Ep}(\bm{\kappa},z)\cdot \bm{p} + \ovb{g}^{\rm Em}(\bm{\kappa},z)\cdot \bm{m}\nonumber\\
&+\ovb{g}^{\rm Eq}(\bm{\kappa},z):\ovb{q}.\label{FE}
\end{eqnarray}
The time-averaged power scattered through the surface of one parallelepiped follows immediately from (\ref{FE}),
\begin{eqnarray}
\quad \Phi^{\rm scatt} &= \frac{2}{\mu\mu_0} \frac{1}{\omega} \sum_{w_{\bm{n}}\in\mathbb{R}} \frac{w_{\bm{n}}}{A_c} \left|\bm{f}^{\rm E}_{\bm{n}} \right|^2,\label{Iscat}
\end{eqnarray}
where we identified 
\begin{equation}
\quad \left|\bm{f}^{\rm E}_{\bm{n}} \right|^2 = \left|\bm{f}^{\rm E}(\bm{\kappa}_{\bm{n}},0_+) \right|^2 + \left|\bm{f}^{\rm E}(\bm{\kappa}_{\bm{n}},0_-) \right|^2.
\end{equation}
We simplify expression (\ref{Iscat}) using the following identities,
\begin{eqnarray}
\quad \left[\ovb{g}^{\rm Ep}_{\pm}(\bm{\kappa})\right]^{\dagger}\cdot \ovb{g}^{\rm Ep}_{\pm}(\bm{\kappa}) &= -\frac{\rmi \tilde{\omega}^2\mu}{2\epsilon_0 w} \ovb{g}^{\rm Ep}_{\pm}(\bm{\kappa}), \label{reli}  \\
\quad \left[\ovb{g}^{\rm Em}_{\pm}(\bm{\kappa})\right]^{\dagger}\cdot \ovb{g}^{\rm Em}_{\pm}(\bm{\kappa}) &=  - \frac{\rmi \tilde{\omega}^2\mu}{2\epsilon_0 w} \ovb{g}^{\rm Bm}_{\pm}(\bm{\kappa}),  \\
\quad \left[\ovb{g}^{\rm Eq}_{\pm}(\bm{\kappa})\right]^{\dagger}\cdot \ovb{g}^{\rm Eq}_{\pm}(\bm{\kappa}) &=  - \frac{\rmi \tilde{\omega}^2\mu}{2\epsilon_0 w} \ovb{g}^{\rm Fq}_{\pm}(\bm{\kappa}),  \\
\quad \left[\ovb{g}^{\rm Ep}_{\pm}(\bm{\kappa})\right]^{\dagger}\cdot \ovb{g}^{\rm Em}_{\pm}(\bm{\kappa}) &= -\frac{\rmi \tilde{\omega}^2\mu}{2\epsilon_0 w} \ovb{g}^{\rm Em}_{\pm}(\bm{\kappa}),  \\
\quad \left[\ovb{g}^{\rm Ep}_{\pm}(\bm{\kappa})\right]^{\dagger}\cdot \ovb{g}^{\rm Eq}_{\pm}(\bm{\kappa}) &= -\frac{\rmi \tilde{\omega}^2\mu}{2\epsilon_0 w} \ovb{g}^{\rm Eq}_{\pm}(\bm{\kappa}),  \\
\quad \left[\ovb{g}^{\rm Em}_{\pm}(\bm{\kappa})\right]^{\dagger}\cdot \ovb{g}^{\rm Eq}_{\pm}(\bm{\kappa}) &= -\frac{\rmi \tilde{\omega}^2\mu}{2\epsilon_0 w} \ovb{g}^{\rm Bq}_{\pm}(\bm{\kappa}), \label{relf}  
\end{eqnarray}
and arrive at
\begin{eqnarray}
\quad \Phi^{\rm scatt} = 2\omega\mathrm{Im}  \frac{1}{A_c}\sum_{w_{\bm{n}}\in\mathbb{R}} H^{\rm scatt}_{\boldsymbol{n}},\label{scatt}
\end{eqnarray}
where 
\begin{eqnarray}
\quad H^{\rm scatt}_{\boldsymbol{n}} &= \bm{p}^*\cdot \ovb{g}_{\bm{n}}^{\rm Ep}\cdot \bm{p} +   \bm{m}^*\cdot \ovb{g}_{\bm{n}}^{\rm Bm}\cdot \bm{m} \nonumber\\
&+\ovb{q}^*: \ovb{g}_{\bm{n}}^{\rm Fq}:\ovb{q} + 2\bm{p}^*\cdot \ovb{g}_{\bm{n}}^{\rm Em}\cdot \bm{m} \nonumber\\
& + 2\bm{p}^*\cdot \ovb{g}_{\bm{n}}^{\rm Eq}: \ovb{q} + 2\bm{m}^*\cdot \ovb{g}_{\bm{n}}^{\rm Bq}: \ovb{q},
\end{eqnarray}
and we defined $\ovb{g}_{\bm{n}} = [\ovb{g}_+(\bm{\kappa}_{\bm{n}})+\ovb{g}_-(\bm{\kappa}_{\bm{n}})]/2$. Finally we find the total power removed from an incident beam, $-\Phi^{\rm si}$. Using the expression (\ref{EA}) for the scattered field and the expression (\ref{Einc}) for the incident field, and noting that for a transverse incident field the following identities hold,
\begin{eqnarray}
\quad \left(\bm{E}_{\pm}^{\rm inc}\right)^*\cdot \ovb{g}^{\rm Ep}_{\pm}(\bm{\kappa}_0) &= \frac{\rmi \tilde{\omega}^2\mu}{2\epsilon_0 w} \left(\bm{E}_{\pm}^{\rm inc}\right)^* \label{ida}\\
\quad \left(\bm{E}_{\pm}^{\rm inc}\right)^* \cdot \ovb{g}^{\rm Em}_{\pm}(\bm{\kappa}_0) &= \frac{\rmi \tilde{\omega}^2\mu}{2\epsilon_0 w}\left(\bm{B}_{\pm}^{\rm inc}\right)^* \\
\quad \left(\bm{E}_{\pm}^{\rm inc}\right)^* \cdot \ovb{g}^{\rm Eq}_{\pm}(\bm{\kappa}_0) &= \frac{\rmi \tilde{\omega}^2\mu}{2\epsilon_0 w}\left(\ovb{F}_{\pm}^{\rm inc}\right)^*, \label{idb}
\end{eqnarray}
we find that
\begin{eqnarray}
\quad -\Phi^{\rm si} &= 2\omega \mathrm{Im} H^{\rm si} ,\label{crossf}
\end{eqnarray}
where 
\begin{eqnarray}
\quad H^{\rm si} &= \left(\bm{E}^{\rm inc}\right)^*\cdot\bm{p} + \left(\bm{B}^{\rm inc}\right)^*\cdot \bm{m} \nonumber\\
& + \left(\ovb{F}^{\rm inc}\right)^*:\ovb{q},
\end{eqnarray}
and where $\bm{E}^{\rm inc} = \bm{E}_+^{\rm inc}+\bm{E}_-^{\rm inc}$ is a total incident electric field at the position of the emitter, and so for the remaining fields. 

Now we return to (\ref{econs}) and first we find the difference between the power removed from an incident beam, (\ref{crossf}), and power absorbed by an emitter, (\ref{abs}). After simplifying the expression with the use of (\ref{Etot}-\ref{Ftot}) we find, 
\begin{eqnarray}
\quad -\Phi^{\rm si}-W^{\rm A} 
=2\omega\mathrm{Im} H' ,\label{Ga}
\end{eqnarray}
where
\begin{eqnarray}
\quad H' &= \bm{p}\cdot{}^ {\rm R}\ovb{\mathcal{G}}^{\rm Ep} \cdot\bm{p}^* + \bm{m}\cdot{}^{\rm R}\ovb{\mathcal{G}}^{\rm Bm}\cdot \bm{m}^* \nonumber\\
&+ \ovb{q}:{}^{\rm R}\ovb{\mathcal{G}}^{\rm Fq}:\ovb{q}^* + 2\bm{p}\cdot{}^{\rm R}\ovb{\mathcal{G}}^{\rm Em}\cdot \bm{m}^* \nonumber\\
&- 2\bm{p}\cdot {}^{\rm R}\ovb{\mathcal{G}}^{\rm Eq}:\ovb{q}^* -2\bm{m}\cdot{}^{R}\ovb{\mathcal{G}}^{\rm Bq}:\ovb{q}^*.  \label{Hprime}
\end{eqnarray}
In deriving (\ref{Hprime}) we used the symmetry conditions satisfied by the periodic Green function dyadics that follow from reciprocity relations \cite{Sersic2011}, and we used the relations ${}^{\rm R}\ovb{\mathcal{G}}^{\rm Ep} = \rmi\mathrm{Im}\ovb{\mathcal{G}}^{\rm Ep}$, ${}^{\rm R}\ovb{\mathcal{G}}^{\rm Em} = \rmi\mathrm{Im}\ovb{\mathcal{G}}^{\rm Em}$, ${}^{\rm R}\ovb{\mathcal{G}}^{\rm Eq} = \mathrm{Re}\ovb{\mathcal{G}}^{\rm Eq}$, ${}^{\rm R}\ovb{\mathcal{G}}^{\rm Fq} = \rmi\mathrm{Im}\ovb{\mathcal{G}}^{\rm Fq}$, etc., for the remaining periodic Green functions (see section \ref{exact exp}). For the optical theorem (\ref{econs}) to hold the difference between extinction and absorbed power, (\ref{Ga}), needs to be equal to the scattered power, (\ref{scatt}). Comparing the expressions (\ref{Ga}) and (\ref{scatt}) we find that this condition is satisfied when the radiative periodic Green functions ${}^{\rm R}\ovb{\mathcal{G}}$ satisfy 
\begin{equation}
\quad {}^{\rm R}\ovb{\mathcal{G}} = \frac{1}{A_c} \sum_{w_{\bm{n}}\in \mathbb{R}} \ovb{g}_{\bm{n}},  \label{fcd}
\end{equation}
irrespective of the form that the non-radiative periodic Green functions take as long they satisfy the symmetry conditions that follow from reciprocity. A straightforward calculation verifies that the radiative contributions to the periodic Green functions found in section \ref{exact exp} indeed satisfy the condition (\ref{fcd}).

\bibliography{References}

\onecolumn

\title{Supplementary information}

\appendix
\setcounter{section}{19}

\section*{}
We illustrate the intermediate steps that lead to periodic Green function dyadics in a form presented in the paper. We focus here on the quadrupole dyadic $\bm{\mathcal{G}}^{\mathrm{Fq}}$. The derivation of the remaining periodic Green function dyadics is simpler but analogous. The notation is the same as used in the paper. 

\section*{Lattice coordinate system}
We work in a skewed coordinate system associated with the basis vectors of the lattice,  $\bm{\hat{a}}_i$, and identify coordinates along those basis vectors as $u_i$, 
\begin{equation}
\quad \bm{r} =  \sum_i u_i\bm{\hat{a}_i}+z\bm{\hat{z}}.
\label{r}
\end{equation}
The transformation from the coordinates associated with the lattice to Cartesian coordinates is 
\begin{eqnarray}
\quad x &= u_1(\bm{\hat{a}}_1\cdot \bm{\hat{x}} ) + u_2 (\bm{\hat{a}}_2\cdot \bm{\hat{x}}), \nonumber\\
\quad y &= u_1 (\bm{\hat{a}}_1\cdot \bm{\hat{y}} ) + u_2 (\bm{\hat{a}}_2\cdot \bm{\hat{y}}),
\end{eqnarray}
and the inverse transformation is given by
\begin{eqnarray}
\quad u_1 &= \frac{ x(\bm{\hat{a}}_2\cdot \bm{\hat{y}}) - y(\bm{\hat{a}}_2\cdot \bm{\hat{x}}) }{ (\bm{\hat{a}}_1\times\bm{\hat{a}}_2)\cdot \bm{\hat{z}} },\nonumber\\
\quad u_2 &= \frac{ y(\bm{\hat{a}}_1\cdot \bm{\hat{x}}) - x(\bm{\hat{a}}_1\cdot \bm{\hat{y}}) }{ (\bm{\hat{a}}_1\times\bm{\hat{a}}_2)\cdot \bm{\hat{z}} }.
\end{eqnarray}
Derivatives taken with respect to Cartesian coordinates are related to those taken with respect to coordinates associated with the lattice by, 
\begin{eqnarray}
\quad \partial_x &= \frac{1}{(\bm{\hat{a}}_1 \times\bm{\hat{a}}_2)\cdot \bm{\hat{z}}}\left[ (\bm{\hat{a}}_2\cdot\bm{\hat{y}})\partial_{u_1} - (\bm{\hat{a}}_1 \cdot \bm{\hat{y}})\partial_{u_2} \right],\label{pd1}\\
\quad \partial_y &= \frac{1}{(\bm{\hat{a}}_1 \times\bm{\hat{a}}_2)\cdot \bm{\hat{z}}}\left[- (\bm{\hat{a}}_2\cdot\bm{\hat{x}})\partial_{u_1} + (\bm{\hat{a}}_1\cdot \bm{\hat{x}})\partial_{u_2}  \right],\label{pd2}
\end{eqnarray}
where analogous relations for higher order derivatives can be obtained by a repeated application of (\ref{pd1},\ref{pd2}). We also identify the scalar free-space Green function written in coordinate system associated with the lattice
\begin{equation}
\quad g_0(u_1,u_2,z) = \frac{1}{4\pi}\frac{\mathrm{e}^{\mathrm{i}\tilde{\omega}n \sqrt{ u_1^2+u_2^2+2u_1u_2\mathrm{cos}\phi+z^2 } } }{ \sqrt{ u_1^2+u_2^2+2u_1u_2\mathrm{cos}\phi+z^2 }},\label{A_G0}
\end{equation}
where $\phi$ is the angle between the two basis lattice vectors. 
 
\section*{Real space definition}
The quadrupole periodic Green function is a sum of two terms,
\begin{equation}
\quad \bm{\mathcal{G}}^{\mathrm{Fq}} =  \bm{G}^{\mathrm{Fq}} + \bm{\mathcal{R}}^{\mathrm{Fq}},
\end{equation}
where $\bm{\mathcal{R}}^{\mathrm{Fq}}$ is known (see (26) in the paper) and the first contribution involves a summation over lattice sites 
\begin{equation}
\quad \bm{G}^{\mathrm{Fq}} = \lim_{z\rightarrow 0} \sum_{\bm{n}\neq (0,0)} \mathrm{e}^{\mathrm{i}\bm{\kappa}_0\cdot \bm{R}_{\bm{n}}} \bm{g}^{\mathrm{Fq}}(-\bm{R}_{\bm{n}}+z\bm{\hat{z}}),\label{Gdefin}
\end{equation}
that we need to accelerate. As a first step we write the free-space Green function $\bm{g}^{\mathrm{Fq}}(\bm{r})$ in coordinate system associated with the lattice. We illustrate the method for the $g^{\mathrm{Fq}}_{xxxx}(\bm{r})$ component; the derivation for the remaining components is analogous. Starting from the definition of the quadrupole free-space Green function (see (A.6) in the paper), we use (\ref{pd1}) to rewrite the derivatives with respect to $x$ in terms of derivatives with respect to the lattice coordinates $u_1,u_2$ . We arrive at
\begin{eqnarray}
\quad g^{\mathrm{Fq}}_{xxxx}(u_1,u_2,z) = (\tilde{\omega}n)^2 g_{xx}^{\mathrm{Ep}}(u_1,u_2,z) -\frac{\mu}{\epsilon_0 n^2}\frac{\mathcal{D} g_0(u_1,u_2,z)}{\left[ (\bm{\hat{a}}\times\bm{\hat{a}}_2)\cdot \bm{\hat{z}} \right]^4 } ,\label{glatt}
\end{eqnarray}
where the differential operator acting on the free-space Green function is 
\begin{eqnarray}
\quad\mathcal{D} &=  (\bm{\hat{a}}_2\cdot\bm{\hat{y}})^4 \slashed{\partial}_{u_1}^2 \slashed{\partial}_{u_1}^2 + (\bm{\hat{a}}_1\cdot\bm{\hat{y}})^4 \slashed{\partial}_{u_2}^2 \slashed{\partial}_{u_2}^2 \nonumber\\
&\quad+ 2(\bm{\hat{a}}_1\cdot \bm{\hat{y}})^2 (\bm{\hat{a}}_2\cdot\bm{\hat{y}})^2 \left[ \slashed{\partial}_{u_1}^2 \slashed{\partial}_{u_2}^2 + 2 \slashed{\partial}_{u_1u_2} \slashed{\partial}_{u_1u_2} \right]\nonumber\\
&\quad-4(\bm{\hat{a}}_1\cdot\bm{\hat{y}})(\bm{\hat{a}}_2\cdot\bm{\hat{y}})^3 \slashed{\partial}_{u_1}^2 \slashed{\partial}_{u_1u_2} - 4(\bm{\hat{a}}_2\cdot\bm{\hat{y}})(\bm{\hat{a}}_1\cdot\bm{\hat{y}})^3 \slashed{\partial}^2_{u_2} \slashed{\partial}_{u_1u_2}, 
\end{eqnarray}
and where we have defined
\begin{eqnarray}
&\quad\slashed{\partial}_{u_1}^2 = \partial_{u_1}^2 + (\tilde{\omega}n)^2, \nonumber\\
&\quad\slashed{\partial}_{u_2}^2 = \partial_{u_2}^2 + (\tilde{\omega}n)^2, \nonumber\\
&\quad\slashed{\partial}_{u_1u_2} = \partial_{u_1}\partial_{u_2} + \bm{\hat{a}}_1\cdot\bm{\hat{a}}_2 (\tilde{\omega}n)^2.
\end{eqnarray}
Using expression (\ref{glatt}) in the definition (\ref{Gdefin}) we arrive at
\begin{equation}
\quad G^{\mathrm{ Fq}}_{xxxx} = \left(\tilde{\omega}n\right)^2G^{\mathrm{ Ep}}_{xx} + \frac{\mu}{\epsilon_0 n^2}  \sum_{(lmnp)} \frac{C^{(lmnp)}_{xxxx} \tilde{\mathcal{S}}_{(lmnp)}}{\left[\left(\bm{\hat{a}}_1\times\bm{\hat{a}}_2\right)\cdot \bm{\hat{z}} \right]^4},\label{GxxxxS}
\end{equation}
with the coefficients $C^{(lmnp)}_{xxxx}$ identified in Table 4 in the paper. The $\tilde{\mathcal{S}}_{(lmnp)}$ are complex partial sums defined in terms of partial functions
\begin{equation}
\quad \tilde{\mathcal{S}}_{(lmnp)} = \lim_{z\rightarrow 0}\sum_{(n,n')\neq (0,0)}  \tilde{\mathcal{S}}_{(lmnp)}(na_1,n'a_2,z),
\end{equation}
where $a_i = |\bm{a}_i|$. The partial functions for $l=m=n=p$ are   
\begin{eqnarray}
\quad\tilde{\mathcal{S}}_{(iiii)}(u_1,u_2,z) &=  \mathrm{e}^{-\mathrm{i}\bm{\kappa}_0\cdot \left(u_1\bm{\hat{a}}_1+u_2\bm{\hat{a}}_2  \right)}  \slashed{\partial}_{u_i}^2 \slashed{\partial}_{u_i}^2 g_0(u_1,u_2,z), \label{part1}
\end{eqnarray}
the partial functions for $l=m=n\neq p$ are 
\begin{eqnarray}
\quad\tilde{\mathcal{S}}_{(iii\bar{i})}(u_1,u_2,z) &= \mathrm{e}^{-\mathrm{i}\bm{\kappa}_0\cdot \left(u_1\bm{\hat{a}}_1+u_2\bm{\hat{a}}_2  \right)} \slashed{\partial}_{u_i}^2 \slashed{\partial}_{u_iu_{\bar{i}}} g_0(u_1,u_2,z), \label{part2}
\end{eqnarray}
and the partial functions for $l=m$ and $n=p$ are
\begin{eqnarray}
\quad\tilde{\mathcal{S}}_{(1122)}(u_1,u_2,z) &= \mathrm{e}^{-\mathrm{i}\bm{\kappa}_0\cdot \left(u_1\bm{\hat{a}}_1+u_2\bm{\hat{a}}_2  \right)} \frac{1}{3} \left(  \slashed{\partial}_{u_1}^2 \slashed{\partial}_{u_2}^2 + 2 \slashed{\partial}_{u_1u_2}^2  \right) g_0(u_1,u_2,z). \label{part3} 
\end{eqnarray}
The complex partial sums, $\tilde{\mathcal{S}}_{(lmnp)}$, are related to the partial sums identified in the paper, $\mathcal{S}_{(lmnp)}$, by
\begin{eqnarray}
\quad\mathrm{Re}\tilde{\mathcal{S}}_{(lmnp)} = \mathcal{S}_{(lmnp)}.
\end{eqnarray}
The expressions for the remaining components of the dyadic take the form similar to (\ref{GxxxxS}) and we do not give them here explicitly. The main difficulty in accelerating the summations that enter the quadrupole periodic Green function definition lies in evaluating the partial sums (\ref{part1}-\ref{part3}). We illustrate the acceleration method for these sums in the next section.

\section*{Evaluation of partial sums}
\label{partsum}
\subsection*{Evaluation of $\tilde{\mathcal{S}}_{(iiii)}$}
We illustrate the acceleration method for $\tilde{\mathcal{S}}_{(1111)}$. The acceleration of the partial sum $\tilde{\mathcal{S}}_{(2222)}$ follows the same steps, but with all the indices that indicate direction along the lattice vectors interchanged. 

We start by decomposing the partial sum into two contributions, 
\begin{equation}
\quad \tilde{\mathcal{S}}_{(1111)} = \tilde{\mathcal{S}}^c_{(1111)} + \tilde{\mathcal{S}}^p_{(1111)},\label{decomp}
\end{equation}
where the first term involves a summation over a 1d chain along the direction of $\bm{\hat{a}}_1$, 
\begin{eqnarray}
\quad \tilde{\mathcal{S}}^c_{(1111)} = \lim_{z \rightarrow 0} \sum_{n\neq 0} \tilde{\mathcal{S}}_{(1111)} (na_1,0,z),
\end{eqnarray}
and the remaining contribution is a summation over a 2d array with that chain removed,
\begin{eqnarray}
\quad \tilde{\mathcal{S}}^p_{(1111)} &= \lim_{z \rightarrow 0} \sum_{n=-\infty}^{\infty}\sum_{n'\neq 0} \tilde{\mathcal{S}}_{(1111)} (na_1,n'a_2,z).\label{Sp}
\end{eqnarray}
We consider first a contribution from the chain. Taking the derivatives of the free-space Green function with respect to $u_1$ and identifying the real and imaginary contributions we arrive at
\begin{equation}
\quad \tilde{\mathcal{S}}^c_{(1111)} = \mathrm{Re}\tilde{\mathcal{S}}^c_{(1111)} +\mathrm{i} \mathrm{Im}\tilde{\mathcal{S}}^c_{(1111)},
\end{equation}
with the real part 
\begin{eqnarray}
\quad \mathrm{Re} \tilde{\mathcal{S}}^c_{(1111)} &= \frac{1}{4\pi a_1^5}\sum_{n=1}^{\infty} \left[-8(\tilde{\omega}na_1)^2\frac{\mathrm{cos}\left(ns_{+}^{(1)}\right)}{n^3} + 24\frac{\mathrm{cos}\left(ns_{+}^{(1)}\right)}{n^5}  + 24\left(\tilde{\omega}na_1\right)\frac{\mathrm{sin}\left(ns_{+}^{(1)}\right)}{n^4}  \right] \nonumber\\
&+ \left(s_{+}^{(1)}\leftrightarrow s_{-}^{(1)}\right),\label{real}
\end{eqnarray}
and the imaginary part
\begin{eqnarray}
\quad \mathrm{Im} \tilde{\mathcal{S}}^c_{(1111)} &= \frac{1}{4\pi a_1^5}\sum_{n=1}^{\infty} \left[ 24\frac{\mathrm{sin}\left(ns_{+}^{(1)}\right)}{n^5} -24\left(\tilde{\omega}na_1\right)\frac{\mathrm{cos}\left(ns_{+}^{(1)}\right)}{n^4} -8\left(\tilde{\omega}na_1\right)^2\frac{\mathrm{sin}\left(ns_{+}^{(1)}\right)}{n^3} \right] \nonumber\\
&+ \left(s_{+}^{(1)} \leftrightarrow s_{-}^{(1)}\right).\label{imag}
\end{eqnarray}
The sums that enter expression (\ref{real}) do not admit a simple analytic form and we evaluate them numerically. The summand in (\ref{real}) drops-off as $1/n^3$ and can be evaluated directly, but we accelerate the summation using the dominant part extraction method. We use the identity
\begin{equation}
\quad \frac{1}{n^3} = \frac{3n+2}{n^3(n+1)(n+2)} + \frac{1}{n(n+1)(n+2)}.\label{domin}
\end{equation}
When (\ref{domin}) is used in (\ref{real}), the first term on the right-hand side of (\ref{domin}) gives a contribution that drops-off as $1/n^4$, while the second term gives the contribution  
\begin{eqnarray}
\quad \sum_{n=1}^{\infty} \frac{\mathrm{cos}\left(s_{\pm}^{(1)} n\right)}{n(n+1)(n+2)} &= \frac{1}{2}\sum_{n=1}^{\infty}\frac{\mathrm{cos}\left(s_{\pm}^{(1)} n\right)}{n}-\sum_{n=1}^{\infty}\frac{\mathrm{cos}\left(s_{\pm}^{(1)} n\right)} {n+1}+\frac{1}{2}\sum_{n=1}^{\infty}\frac{\mathrm{cos}\left(s_{\pm}^{(1)} n\right)}{n+2}, \label{chainD}
\end{eqnarray}
that can be evaluated in a simple analytic form. Using the identities (see 1.441 in [50]),
\begin{eqnarray}
\quad \sum_{n=1}^{\infty} \frac{\mathrm{cos}\left(s_{\pm}^{(1)} n\right)}{n} &= -\mathrm{ln}\left(2\left|\mathrm{sin}\frac{s_{\pm}^{(1)}}{2}\right| \right), \label{cos}\\
\quad \sum_{n=1}^{\infty}\frac{\mathrm{sin}\left(s_{\pm}^{(1)} n\right)}{n} &= \frac{\pi}{2}\mathrm{sgn}(s_{\pm}^{(1)} ) - \frac{ s_{\pm}^{(1)} }{2}, 
\end{eqnarray}
and shifting the summation indices we arrive at 
\begin{equation}
\quad \sum_{n=1}^{\infty} \frac{\mathrm{cos}(s_{\pm}^{(1)} n)}{n(n+1)(n+2)} =-\tilde{\mathcal{A}}^{(1)}_{\pm}. \label{A}
\end{equation}
Using (\ref{A}) in (\ref{real}) we arrive at contribution from a chain of the form,
\begin{eqnarray}
\quad \mathrm{Re}\tilde{\mathcal{S}}^c_{(1111)} = \sum_{j=\pm} \left( \tilde{\omega}^2n^2 \frac{ \tilde{\mathcal{A}}^{(1)}_j - 2B_j^{(1)} }{\pi a_1^3}  + \tilde{\omega}n \frac{6 C_j^{(1)}}{\pi a_1^4} +  \frac{ 6D_j^{(1)}}{\pi a_1^5}  \right),\label{Re_tot}
\end{eqnarray}
which converges rapidly. The imaginary part of the contribution from the chain can be evaluated exactly using  (see 1.443 in [50]),
\begin{eqnarray}
&\quad \sum_{n=1}^{\infty}\frac{\mathrm{sin}\left(s_{\pm}^{(1)} n\right)}{n^3} = \frac{1}{12}\left(s_{\pm}^{(1)}\right)^3-\frac{\pi}{4}s_{\pm}^{(1)} \left|s_{\pm}^{(1)}\right|+\frac{\pi^2}{6}s_{\pm}^{(1)}, \nonumber\\
&\quad \sum_{n=1}^{\infty}\frac{\mathrm{cos}\left(n s_{\pm}^{(1)}\right)}{n^4} = \frac{\pi^4}{90}-\frac{\pi^2 }{12}\left(s_{\pm}^{(1)}\right)^2 +\frac{\pi}{12}\left|s_{\pm}^{(1)}\right|^3-\frac{\left(s_{\pm}^{(1)}\right)^4}{48}, \nonumber\\
&\quad \sum_{n=1}^{\infty}\frac{\mathrm{sin}\left(s_{\pm}^{(1)} n\right)}{n^5} = \frac{\pi^4}{90}s_{\pm}^{(1)} - \frac{\pi^2}{36}\left(s_{\pm}^{(1)}\right)^3 + \frac{\pi}{48}\left(s_{\pm}^{(1)}\right)^3\left|s_{\pm}^{(1)}\right| - \frac{\left(s_{\pm}^{(1)}\right)^5}{240}.
\end{eqnarray}
The result is
\begin{eqnarray}
\quad \mathrm{Im} \tilde{\mathcal{S}}^c_{(1111)} = \sum_{j = \pm}\left[ f\left(s^{(1)}_{j} \right)+ g\left(s^{(1)}_{j} \right) +  h\left(s^{(1)}_{j} \right) +  k\left(s^{(1)}_{j} \right) \right],\label{Im_tot}
\end{eqnarray}
where 
\begin{eqnarray}
\quad f\left(s^{(1)}_{\pm} \right) =& -24\frac{\tilde{\omega}n}{4\pi a_1^4} \frac{\pi^4}{90} +\frac{24}{4\pi a_1^5} \frac{\pi^4}{90}s^{(1)}_{\pm}, \\
\quad g\left(s^{(1)}_{\pm} \right) =& -24\frac{\tilde{\omega}n}{4\pi a_1^4} \frac{\pi}{12}\left|s^{(1)}_{\pm}\right|^3 +8\frac{(\tilde{\omega}n)^2}{4\pi a_1^3} \frac{\pi}{4}s^{(1)}_{\pm}\left|s^{(1)}_{\pm}\right|+\frac{24}{4\pi a_1^5} \frac{\pi}{48}\left(s^{(1)}_{\pm}\right)^3\left|s^{(1)}_{\pm}\right|, \\
\quad h\left(s^{(1)}_{\pm} \right) =& 24\frac{\tilde{\omega}n}{4\pi a_1^4} \frac{\pi^2}{12}\left(s^{(1)}_{\pm}\right)^2 - 8\frac{(\tilde{\omega}n)^2}{4\pi a_1^3}\frac{\pi^2}{6}s^{(1)}_{\pm} -\frac{24}{4\pi a_1^5}\frac{\pi^2}{36}\left(s^{(1)}_{\pm}\right)^3, \\
\quad k\left(s^{(1)}_{\pm} \right) =& 24\frac{\tilde{\omega}n}{4\pi a_1^4} \frac{1}{48}\left (s^{(1)}_{\pm}\right)^4 - 8\frac{(\tilde{\omega}n)^2}{4\pi a_1^3} \frac{1}{12}\left(s^{(1)}_{\pm}\right)^3 -\frac{24}{4\pi a_1^5} \frac{\left(s^{(1)}_{\pm}\right)^5}{240}.
\end{eqnarray} 

We next consider the contribution from a 2d summation with the chain removed; see (\ref{Sp}). The summation converges slowly in both indices. We accelerate the summation using the Poisson transformation with respect to an unrestricted index,
\begin{eqnarray}
\quad \sum_{n}  \tilde{\mathcal{S}}_{(1111)}(na_1,n'a_2, z) = \frac{1}{a_1}\sum_{n}  \tilde{\mathcal{S}}_{(1111)}^{\rm F}\left(K^{(1)}_{n} , n'a_2 ,z\right).\label{poisson_sub}
\end{eqnarray}
The Fourier transform of the partial sum function is
\begin{eqnarray}
\quad \tilde{\mathcal{S}}_{(1111)}^{\rm F}\left(K^{(1)}_{n} , n'a_2 ,z\right) = \left( w^{(1)}_n \right)^4 \mathrm{e}^{-\mathrm{i}n' \bm{\kappa}_0\cdot \bm{a}_2} g_0^F\left( -Q^{(1)}_n, n'a_2,z  \right),
\end{eqnarray}
where the Fourier transform of the free-space scalar Green function with respect to its first coordinate in the skewed coordinate system is
\begin{eqnarray}
\quad g_0^F\left( -Q^{(1)}_n, n'a_2,z  \right) = \frac{\mathrm{i}}{4}\mathrm{e}^{\mathrm{i}Q^{(1)}_n n' a_2 \mathrm{cos}\phi} H_0^{(1)}\left( w^{(1)}_n \sqrt{\left(  d^{(2)}_{n'} \right)^2+z^2}  \right), \label{g0F}
\end{eqnarray}
and we used the identities 6.616 in [50] to arrive at (\ref{g0F}). Using (\ref{poisson_sub}-\ref{g0F}) in (\ref{Sp}) we arrive at
\begin{eqnarray}
\quad \tilde{\mathcal{S}}^p_{(1111)} &= \lim_{z\rightarrow 0}\frac{\mathrm{i}}{4a_1}\sum_{n'\neq 0}\sum_{n} \left( w^{(1)}_n \right)^4 \mathrm{e}^{-\mathrm{i}\psi^{(1)}_{nn'}} H_0^{(1)} \left( w^{(1)}_n \sqrt{ \left(d^{(2)}_{n'}\right)^2+z^2}  \right).\label{sumi}
\end{eqnarray}
We identify terms in the summation (\ref{sumi}) that correspond to cylindrical waves that are  exponentially decaying away from the 1d array identified by $\bm{\hat{a}}_1$. The evanescent waves are identified by the indices $n\notin (n^{(i)}_-,n^{(i)}_+)$, for which $w^{(i)}_n\in \mathbb{I}$. Separating out terms with those indices  and rewriting the Hankel function in terms of the modified Bessel function of the second kind of a real argument (see 9.6.4 in [51]), we have
\begin{eqnarray}
\quad \tilde{\mathcal{S}}^p_{(1111)} =& \sum_{n\notin (n^{(i)}_-,n^{(i)}_+)} \frac{  \left( w^{(1)}_n \right)^4}{2\pi a_1} \sum_{n'\neq 0} \mathrm{e}^{-\mathrm{i}\psi^{(1)}_{nn'}} K_0\left( \frac{w^{(1)}_n}{\mathrm{i}} d^{(2)}_{n'} \right)\nonumber\\
&\quad + \lim_{z\rightarrow 0} \sum_{n=n^{(i)}_-}^{n^{(i)}_+} \frac{\mathrm{i}  \left( w^{(1)}_n \right)^4}{4a_1} \sum_{n'\neq 0} \mathrm{e}^{-\mathrm{i}\psi^{(1)}_{nn'}} H_0^{(1)}\left( w^{(1)}_n \sqrt{ \left(d^{(2)}_{n'}\right)^2+z^2}  \right).\label{K}
\end{eqnarray}
The first line in (\ref{K}) converges rapidly in both indices $n$ and $n'$. The first summation in the second line is typically over few terms only, but the infinite summation in the second line converges slowly. To accelerate the infinite summation we write
\begin{eqnarray}
&\quad \sum_{n'\neq 0} \mathrm{e}^{-\mathrm{i}\psi^{(1)}_{nn'}} H_0^{(1)}\left( w^{(1)}_n \sqrt{ \left(d^{(2)}_{n'}\right)^2+z^2}  \right) = \sum_{n'} \mathrm{e}^{-\mathrm{i}\psi^{(1)}_{nn'}} H_0^{(1)}\left( w^{(1)}_n \sqrt{ \left(d^{(2)}_{n'}\right)^2+z^2}  \right) -  H_0^{(1)}\left( w^{(1)}_n |z|  \right).\label{ssP}
\end{eqnarray}
We use a Poisson transformation for the unrestricted summation,
\begin{eqnarray}
\quad \sum_{n'} \mathrm{e}^{-\mathrm{i}\psi^{(1)}_{nn'}} H_0^{(1)}\left( w^{(1)}_n \sqrt{ \left(d^{(2)}_{n'}\right)^2+z^2}  \right) = \frac{2}{a_2|\mathrm{sin}\phi|}\sum_{n'} \frac{\mathrm{e}^{\mathrm{i}|z|w_{nn'}^{(12)}} }{w^{(12)}_{nn'}},\label{interim}
\end{eqnarray}
where we used identities 6.677 in [50] to arrive at (\ref{interim}). Using (\ref{interim}) in (\ref{ssP}) we find
\begin{eqnarray}
\quad \sum_{n'\neq 0} \mathrm{e}^{-\mathrm{i}\psi^{(1)}_{nn'}} H_0^{(1)}\left( w^{(1)}_n \sqrt{ \left(d^{(2)}_{n'}\right)^2+z^2}  \right) = \frac{2}{a_2|\mathrm{sin}\phi|}\sum_{n'} \frac{\mathrm{e}^{\mathrm{i}|z|w_{nn'}^{(12)}} }{w^{(12)}_{nn'}} -  H_0^{(1)}\left( w^{(1)}_n |z|  \right).\label{sing}
\end{eqnarray}
We note that even though the summation (\ref{sing}) is well defined in the $z\rightarrow 0$ limit, both terms on the right-hand side of (\ref{sing}) when taken separately are singular in that limit.  We isolate the singularity in the first term on the right-hand side of (\ref{sing}) using the dominant part extraction method. At large $n'$ the leading contribution to the summand is of the form
\begin{eqnarray}
 \quad \frac{\mathrm{e}^{\mathrm{i}|z|w^{(12)}_{nn'}} }{w^{(12)}_{nn'}} \approx \frac{a_2|\mathrm{sin}\phi|}{2\pi \mathrm{i} |n'|} \mathrm{exp}\left( \quad-\frac{2\pi|z||n'|}{a_2|\mathrm{\mathrm{sin}}\phi|}\right).
\end{eqnarray}
Extracting that dominant contribution and evaluating the summation over the dominant contribution in an analytic form using (\ref{cos}), we have
\begin{eqnarray}
\quad \sum_{n'} \frac{\mathrm{e}^{\mathrm{i}|z|w^{(12)}_{nn'}} }{w^{(12)}_{nn'}} &= \frac{\mathrm{e}^{\mathrm{i}|z|w^{(12)}_{n0}}}{w^{(12)}_{n0}}+\sum_{n' \neq 0}\left[ \frac{\mathrm{e}^{\mathrm{i}|z|w^{(12)}_{nn'}}}{w^{(12)}_{nn'}} - \frac{a_2|\mathrm{sin}\phi|}{2\pi \mathrm{i} |n'|} \mathrm{e}^{-\frac{2\pi|z||n'|}{a_2|\mathrm{\mathrm{sin}}(\phi)|}}  \right]- \frac{a_2}{\pi \mathrm{i}}\mathrm{ln}\left(1-\mathrm{e}^{-\frac{2\pi |z|}{a_2|\mathrm{sin}\phi|}} \right)|\mathrm{sin}\phi|.\label{ssdx}
\end{eqnarray}
Using now (\ref{ssdx}) in (\ref{sing}) we find
\begin{eqnarray}
\quad \sum_{n'\neq 0} \mathrm{e}^{-\mathrm{i}\psi^{(1)}_{nn'}} H_0^{(1)}\left( w^{(1)}_n \sqrt{ \left(d^{(2)}_{n'}\right)^2+z^2}  \right) &=\frac{2}{b|\mathrm{sin}\phi|}\sum_{n' \neq 0}\left[ \frac{\mathrm{e}^{\mathrm{i}|z|w^{(12)}_{nn'}}}{w^{(12)}_{nn'}} - \frac{a_2|\mathrm{sin}\phi|}{2\pi \mathrm{i} |n'|} \mathrm{e}^{-\frac{2\pi|z||n'|}{a_2|\mathrm{\mathrm{sin}}(\phi)|}}  \right]\nonumber\\
&\quad +\frac{2}{a_2|\mathrm{sin}\phi|}\frac{\mathrm{e}^{\mathrm{i}|z|w^{(12)}_{n0}}}{w^{(12)}_{n0}}-\frac{2}{\pi \mathrm{i}}\mathrm{ln}\left(1-\mathrm{e}^{-\frac{2\pi |z|}{a_2|\mathrm{sin}\phi|}} \right)-H_0^{(1)}\left( w^{(1)}_n |z|  \right).\label{sigr}
\end{eqnarray}
The last two terms in (\ref{sigr}) are singular at $z=0$ when consider separately. We eliminate the singularity in the $z\rightarrow 0$ limit using the small argument expansion of the Hankel function,
\begin{equation}
\quad H_0^{(1)}\left(w^{(1)}_n|z|\right) = 1+\frac{2\mathrm{i}}{\pi}\mathrm{Log}\left( \frac{w^{(1)}_n|z|}{2}  \right)+\frac{2\mathrm{i}\gamma}{\pi}+O(z^2), \label{hnkx}
\end{equation}
where $\mathrm{Log}(z)$ is the principal value of the complex logarithm and  $\gamma$ is the Euler-Mascheroni constant, $\gamma\approx 0.577$. Using the expansion (\ref{hnkx}) we find
\begin{eqnarray}
\quad \lim_{z\rightarrow 0}\left[-\frac{2}{\pi \mathrm{i}}\mathrm{ln}\left(1-\mathrm{e}^{-\frac{2\pi |z|}{a_2|\mathrm{sin}\phi|}} \right)-H_0^{(1)}\left(w^{(1)}_n|z|\right)\right] = -\frac{2\mathrm{i}}{\pi}\mathrm{Log}\left(\frac{w^{(1)}_n a_2|\mathrm{sin}\phi|}{4\pi}\right)-\frac{2\mathrm{i}\gamma}{\pi}-1.\label{sngr}
\end{eqnarray}
Finally using (\ref{sngr}) in (\ref{sigr}) we arrive at,
\begin{eqnarray}
\quad \lim_{z\rightarrow 0}\sum_{n'\neq 0} \mathrm{e}^{-\mathrm{i}\psi^{(1)}_{nn'}} H_0^{(1)}\left( w^{(1)}_n \sqrt{ \left(d^{(2)}_{n'}\right)^2+z^2}  \right) &= \frac{2}{a_2|\mathrm{sin}\phi|}\frac{1}{w_{n0}}-1 - \frac{2\mathrm{i}}{\pi}\mathrm{Log}\left(\frac{w^{(1)}_n a_2|\mathrm{sin}\phi|}{4\pi}\right)\nonumber\\
&\quad  -\frac{2\mathrm{i}\gamma}{\pi} + \frac{2}{\mathrm{i} a_2|\mathrm{sin}\phi|}\sum_{n'\neq 0} \left[ \frac{\mathrm{i}}{w^{(12)}_{nn'}} - \frac{a_2|\mathrm{sin}\phi|}{2\pi |n'|}  \right].\label{interim2}
\end{eqnarray}
The summand in the last term on the right-hand side of (\ref{interim2}) drops-off like $1/n^3$. We accelerate the summation using the dominant part extraction method, 
\begin{eqnarray}
\quad \sum_{n'\neq 0} \left[ \frac{\mathrm{i}}{w^{(12)}_{nn'}} - \frac{a_2|\mathrm{sin}\phi|}{2\pi |n'|}  \right] = \left( \tau^{(1)}\right)^3 \zeta(3) p^{(12)}_{nn'} + \sum_{n'\neq 0} \left[ \frac{\mathrm{i}}{w^{(12)}_{nn'}} - \frac{a_2|\mathrm{sin}\phi|}{2\pi |n'|}  - \frac{1}{2}\left( \frac{\tau^{(1)}}{|n'|} \right)^3 p^{(12)}_{nn'} \right], \label{interim3}
\end{eqnarray}
where $\zeta(k) = \sum_{n=1}^{\infty}1/n^k$ is the Riemann-zeta function, $\zeta(3)\approx 1.202$. Using (\ref{interim3}) in (\ref{interim2}) and then using (\ref{K})  we arrive at
\begin{equation}
\quad \tilde{\mathcal{S}}^p_{(1111)} = \mathrm{Re}\tilde{\mathcal{S}}^p_{(1111)} +\mathrm{i} \mathrm{Im}\tilde{\mathcal{S}}^p_{(1111)},
\end{equation}
with the real part
\begin{eqnarray}
\quad \mathrm{Re}\tilde{\mathcal{S}}^p_{(1111)} =  \sum_{n\notin (n^{(i)}_-,n^{(i)}_+)} \frac{  \left( w^{(1)}_n \right)^4}{2A_c} M^{(1)}_{n} + \sum_{n=n^{(i)}_-}^{n^{(i)}_+} \frac{\left( w^{(1)}_n \right)^4}{2A_c} \left( P^{(1)}_n + \tilde{P}^{(1)}_{n}  \right) ,\label{Re_tot2}
\end{eqnarray}
and the imaginary part
\begin{eqnarray}
\quad \mathrm{Im}\tilde{\mathcal{S}}^p_{(1111)} =  \sum_{n=n^{(i)}_-}^{n^{(i)}_+} \frac{\left( w^{(1)}_n \right)^4}{2A_c}  \mathrm{Im} &\bigg[ 2 \tau^{(1)} \mathrm{Log}\left( \frac{\tau^{(1)}}{2}  w^{(1)}_n  \right)-\mathrm{i}\pi \tau^{(1)} + \sum_{n'}\frac{\mathrm{i}}{w^{(12)}_{nn'}}   \bigg].\label{Im_tot2}
\end{eqnarray}
Adding (\ref{Re_tot}) and (\ref{Re_tot2}) gives the real part of the complex partial sum $\tilde{\mathcal{S}}_{(1111)}$ in the form identified in the paper,
\begin{equation}
\quad \mathrm{Re} \tilde{\mathcal{S}}_{(1111)} = \mathrm{Re} \tilde{\mathcal{S}}^p_{(1111)} + \mathrm{Re} \tilde{\mathcal{S}}^c_{(1111)} = \mathcal{S}_{(1111)}.
\end{equation}
The imaginary contribution to $\tilde{\mathcal{S}}_{(1111)}$ is given by the sum of terms 
(\ref{Im_tot}) and (\ref{Im_tot2}), which we now simplify. We consider first the contribution from the first term in (\ref{Im_tot2}). Since $w^{(1)}_{n}$ is either purely real or imaginary we have
\begin{equation}
\quad \mathrm{Im} \mathrm{Log} \left( \frac{\tau^{(1)}}{2}  w^{(1)}_n  \right) = \frac{\pi}{2} \Theta \left[ - \left( w^{(1)}_n \right)^2  \right],
\end{equation}
 where we note that terms with index $n\in \left(n^{(i)}_-,n^{(i)}_+ \right)$ give no contribution. We thus find 
\begin{eqnarray}
 &\quad \tau^{(1)} \sum_{n=n^{(i)}_-}^{n^{(i)}_+} \frac{\left( w^{(1)}_n \right)^4}{A_c}  \mathrm{Im}   \mathrm{Log}\left( \frac{\tau^{(1)}}{2}  w^{(1)}_n  \right) = \frac{\pi \tau^{(1)}}{2 A_c} \sum_{n=n^{(i)}_-, n^{(i)}_+} \left( w^{(1)}_n \right)^4 \Theta \left[ - \left( w^{(1)}_n \right)^2  \right],\label{con1}
\end{eqnarray} 
Now we evaluate the contribution from the second term on the right-hand side of (\ref{Im_tot2}). Directly from the definition of $w^{(1)}_{n}$ we find
\begin{eqnarray}
&\quad -\pi \tau^{(1)}\sum_{n=n^{(i)}_-}^{n^{(i)}_+}\frac{\left(w_n^{(1)} \right)^4}{2 A_c} = \tilde{f} + \tilde{g} + \tilde{h} +\tilde{k},\label{con2}
\end{eqnarray}
where
\begin{eqnarray}
\quad \tilde{f} = &-\frac{1}{4a_1^5}\left[ (a_1\tilde{\omega}n)^2 - (\bm{\kappa}_0\cdot\bm{a}_1)^2 \right]^2 -\frac{1}{4a_1^5}\frac{(2\pi)^4}{30}\left(-n^{(i)}_++ n^{(i)}_- \right), \label{f}\\
\quad \tilde{g} = &-\frac{(2\pi)}{2a_1^5} \left(\bm{\kappa}_0\cdot\bm{a}_1\right) \left[ (\tilde{\omega}na_1)^2 - \left (\bm{\kappa}_0\cdot\bm{a}_1\right )^2 \right] \left (n^{(i)}_++n^{(i)}_-\right) \nonumber\\
&-\frac{(2\pi)^2}{4 a_1^5}\left[ 3(\bm{\kappa}_0\cdot\bm{a}_1)^2 - (\tilde{\omega}na_1)^2 \right] \left( \left(n^{(i)}_+\right)^2+\left(n^{(i)}_-\right)^2-\frac{4}{3} n^{(i)}_+n^{(i)}_- \right)\nonumber\\
&+\frac{(2\pi)^3}{2a_1^5}(\bm{\kappa}_0\cdot\bm{a}_1)  \left(n^{(i)}_+ +n^{(i)}_-\right)\left( \left(n^{(i)}_+\right)^2+\left(n^{(i)}_-\right)^2-n^{(i)}_+n^{(i)}_- \right)\nonumber\\
&-\frac{(2\pi)^4}{8a_1^5}\left[ \left(n^{(i)}_+\right)^4 + \left(n^{(i)}_-\right)^4  \right], \label{g}\\
\quad \tilde{h} =  &-\frac{(2\pi)^2}{12 a_1^5}\left[ 3(\bm{\kappa}_0\cdot\bm{a}_1)^2 - (\tilde{\omega}na_1)^2 \right] \left( n^{(i)}_+ -n^{(i)}_- \right)\nonumber\\
&+\frac{(2\pi)^3}{4a_1^5}\left(\bm{\kappa}_0\cdot\bm{a}_1\right)  \left(n^{(i)}_+ +n^{(i)}_-\right)\left(n^{(i)}_+ -n^{(i)}_-\right)\nonumber\\
&-\frac{(2\pi)^4 }{12a_1^5}\left[ \left(n^{(i)}_+\right)^3 - \left(n^{(i)}_-\right)^3  \right],\label{h}\\
\quad \tilde{k} = &-\frac{1}{4a_1^5}\left[ (a_1\tilde{\omega}n)^2 - (\bm{\kappa}_0\cdot\bm{a}_1)^2 \right]^2\left[ n^{(i)}_+ -n^{(i)}_- \right] \nonumber\\
&-\frac{\pi}{a_1^5} (\bm{\kappa}_0\cdot\bm{a}_1) \left[ (\tilde{\omega}na_1)^2 - (\bm{\kappa}_0\cdot\bm{a}_1)^2 \right] \left(n^{(i)}_+ +n^{(i)}_-\right)\left(n^{(i)}_+ -n^{(i)}_-\right) \nonumber\\
&+\frac{(2\pi)^3}{4a_1^5}(\bm{\kappa}_0\cdot\bm{a}_1)  \left(n^{(i)}_++n^{(i)}_-\right)\left(n^{(i)}_+-n^{(i)}_-\right)\left( \left(n^{(i)}_+\right)^2+\left(n^{(i)}_-\right)^2\right)\nonumber\\
&-\frac{1}{4a_1^5}\frac{(2\pi)^4}{5}\left[ \left(n^{(i)}_+\right)^5 - \left(n^{(i)}_-\right)^5 \right] \nonumber\\
&-\frac{(2\pi)^2}{6 a_1^5}\left[ 3(\bm{\kappa}_0\cdot\bm{a}_1)^2 - (\tilde{\omega}na_1)^2 \right]  \left(n^{(i)}_+ -n^{(i)}_-\right)\left(\left(n^{(i)}_+\right)^2+\left(n^{(i)}_-\right)^2+n^{(i)}_+n^{(i)}_-\right ).
\end{eqnarray}
The contribution from the third term in (\ref{Im_tot}) we leave in its current form. We add the contributions (\ref{con1},\ref{con2}) with (\ref{Im_tot}) and simplify the expressions. This gives,
\begin{eqnarray}
\quad \tau^{(1)} \sum_{n=n^{(i)}_-}^{n^{(i)}_+} \frac{\left( w^{(1)}_n \right)^4}{A_c}  \mathrm{Im}   \mathrm{Log}\left( \frac{\tau^{(1)}}{2}  w^{(1)}_n  \right)   + \sum_{j=\pm} g\left(s^{(1)} _j\right) +\tilde{g} =  \frac{1}{4a_1^5}\left[ (a_1\tilde{\omega}n)^2 - (\bm{\kappa}_0\cdot\bm{a}_1)^2 \right]^2, \label{smpl1}
\end{eqnarray}
and
\begin{eqnarray}
&\quad \sum_{j=\pm} h\left(s^{(1)} _j\right)+\tilde{h} = 0, \\
&\quad \sum_{j=\pm} k\left(s^{(1)} _j\right) +\tilde{k}=-\frac{8}{15}\frac{(\tilde{\omega}n)^5}{4\pi a_1^4}, \\
&\quad \sum_{j=\pm} f\left(s^{(1)} _j\right) +\tilde{f}  = - \frac{1}{4a_1^5}\left[ (a_1\tilde{\omega}n)^2 - (\bm{\kappa}_0\cdot\bm{a}_1)^2  \right]^2. \label{smpl2}
\end{eqnarray}
Using (\ref{smpl1}-\ref{smpl2}) we find
\begin{equation}
\quad \mathrm{Im}\tilde{\mathcal{S}}_{(1111)} = -\frac{8}{15}\frac{(\tilde{\omega}n)^5}{4\pi a_1^4} +  \sum_{n n'} \frac{\left( w^{(1)}_n \right)^4}{2A_c}  \mathrm{Im} \frac{\mathrm{i}}{w^{(12)}_{nn'}},\label{imf1}
\end{equation}
where in the second term on the right-hand side of (\ref{imf1}) we dropped the restriction on the index $n$; the terms identified by indices $n\notin (n^{(i)}_-, n^{(i)}_+) $ of course give a vanishing contribution.

\subsection*{Evaluation of $\tilde{\mathcal{S}}_{(iii\bar{i})}$}
Here we show the calculation for $\tilde{\mathcal{S}}_{(1112)}$. The calculation of $\tilde{\mathcal{S}}_{(2221)}$ follows the same steps but with the indices identifying the lattice basis vectors interchanged. As before we use the decomposition 
\begin{equation}
\quad \tilde{\mathcal{S}}_{(1112)} = \tilde{\mathcal{S}}^c_{(1112)} + \tilde{\mathcal{S}}^p_{(1112)},
\end{equation}
where 
\begin{eqnarray}
\quad \tilde{\mathcal{S}}^c_{(1112)} &= \lim_{z \rightarrow 0} \sum_{n\neq 0} \quad \tilde{\mathcal{S}}_{(1112)} (na_1,0,z),\\
\quad \tilde{\mathcal{S}}^p_{(1112)} &= \lim_{z \rightarrow 0} \sum_{n=-\infty}^{\infty}\sum_{n'\neq 0} \tilde{\mathcal{S}}_{(1112)} (na_1,n'a_2,z).\label{Sp1112}
\end{eqnarray}
We consider first a contribution from the chain. Directly from the definitions of the Green functions we find 
\begin{eqnarray}
\quad \tilde{\mathcal{S}}^c_{(1112)} =  \quad \tilde{\mathcal{S}}^c_{(1111)}\mathrm{cos}\phi,\label{ch}
\end{eqnarray}
after taking the derivatives of the free-space Green function with respect to $u_1$ and $u_2$. Next we consider the contribution (\ref{Sp1112}) from a 2d summation with the chain removed. As before we carry out the Poisson transformation for the unrestricted summation, and arrive at
\begin{eqnarray}
\quad \tilde{\mathcal{S}}^p_{(1112)} =  \tilde{\mathcal{S}}^p_{(1111)}\mathrm{cos}\phi -\lim_{z\rightarrow 0}&\sum_{n=-\infty}^{\infty} \frac{Q^{(1)}_n \left( w^{(1)}_n \right)^2 }{4a_1} \sum_{n'\neq 0} \mathrm{e}^{-\mathrm{i}\psi^{(1)}_{nn'}} \mathfrak{f}_{nn'}(z), \label{allevs}
\end{eqnarray}
where we have put
\begin{equation}
\quad \mathfrak{f}_{nn'}(z) = \partial_{u_2} H_0^{(1)}\left(  w^{(1)}_n \sqrt{ u_2^2\mathrm{sin}^2\phi +z^2 }   \right)\big|_{u_2=n'a_2}.
\end{equation}
In (\ref{allevs}) we isolate terms that describe evanescent cylidrical waves, rewrite Hankel function in terms of modified Bessel functions of the second kind, and use 
\begin{eqnarray}
\quad \partial_{u_2} K_0\left(\frac{w^{(1)}_n}{\mathrm{i}}\sqrt{ u_2^2\mathrm{sin}^2\phi+z^2 }   \right) =-K_1\left( \frac{w^{(1)}_n}{\mathrm{i}}\sqrt{ u_2^2\mathrm{sin}^2\phi+z^2 } \right)\frac{w^{(1)}_n}{\mathrm{i}}\frac{u_2\mathrm{sin}^2\phi}{\sqrt{u_2^2\mathrm{sin}^2\phi+ z^2} }\nonumber
\end{eqnarray}
to arrive at
\begin{eqnarray}
\quad \tilde{\mathcal{S}}^p_{(1112)} &=  \tilde{\mathcal{S}}^p_{(1111)}\mathrm{cos}\phi-\sum_{n\notin (n_-^{(1)},n_+^{(1)})} \frac{Q^{(1)}_{n} \left(w^{(1)}_{n}\right)^2}{2A_c} N^{(1)}_n  -\lim_{z\rightarrow 0}\sum_{n=n_-^{(1)}}^{n_+^{(1)}} \frac{Q^{(1)}_n \left( w^{(1)}_n \right)^2 }{4a_1} \sum_{n'\neq 0} \mathrm{e}^{-\mathrm{i}\psi^{(1)}_{nn'}} \mathfrak{f}_{nn'}(z).\label{afp}
\end{eqnarray}
We accelerate the infinite summation in the third term on the right-hand side of (\ref{afp}) using the Poisson summation. We have
\begin{eqnarray}
\quad \sum_{n'\neq 0} \mathrm{e}^{-\mathrm{i}\psi^{(1)}_{nn'}} \mathfrak{f}_{nn'}(z) 
= \frac{2\mathrm{i}}{a_2 |\mathrm{sin}\phi|}  \frac{\tilde{\kappa}^{(12)}_n}{w^{(12)}_{n0}} \mathrm{e}^{\mathrm{i}|z|w^{(12)}_{n0}} + \frac{2\mathrm{i}}{a_2} |\mathrm{sin}\phi| \sum_{n'=1}^{\infty} \left( \frac{\tilde{\kappa}^{(12)}_{nn';2}}{w^{(12)}_{nn'}} \mathrm{e}^{\mathrm{i}|z|w^{(12)}_{nn'}} + \left(n' \rightarrow -n'\right) \right).\label{asp}
\end{eqnarray}
To accelerate the summation further  we use the dominant part extraction method. We rewrite
\begin{eqnarray}
\quad \mathrm{i}\sum_{n'=1}^{\infty} \left( \frac{\tilde{\kappa}^{(12)}_{nn';2}}{w^{(12)}_{nn'}} \mathrm{e}^{\mathrm{i}|z|w^{(12)}_{nn'}} + \left(n' \rightarrow -n'\right) \right) &=\mathrm{i}\sum_{n'=1}^{\infty} \frac{\tilde{\kappa}^{(12)}_{nn';2}}{w^{(12)}_{nn'}} \left( \mathrm{e}^{\mathrm{i}|z|w^{(12)}_{nn'}} - \mathrm{e}^{\mathrm{i}|z|w^{(12)}_{n,-n'}}  \right) \nonumber\\
&+ \mathrm{i}\sum_{n'=1}^{\infty} \mathrm{e}^{\mathrm{i}|z|w^{(12)}_{n,-n'}} \left( \frac{\tilde{\kappa}^{(12)}_{nn';2}}{w^{(12)}_{nn'}} + \frac{\tilde{\kappa}^{(12)}_{n,-n';2}}{w^{(12)}_{n,-n'}} \right), \label{dec_dom}
\end{eqnarray}
and we identify the $z\rightarrow 0$ limit of the expression (\ref{dec_dom}). Here care needs to be taken as the first term on the right-hand side of (\ref{dec_dom}) does not vanish in that limit; at large $n$ the dominant contribution to the summand in that term is
\begin{eqnarray}
\quad \frac{\mathrm{i}\kappa^{(12)}_{nn';2}}{w^{(12)}_{nn'}} \left[ \mathrm{e}^{\mathrm{i}|z|w^{(12)}_{nn'}} - \mathrm{e}^{\mathrm{i}|z|w^{(12)}_{n,-n'}} \right] \approx -\frac{2}{|\mathrm{sin}\phi|} \mathrm{e}^{-\frac{2\pi \left|n'\right||z|}{a_2|\mathrm{sin}\phi|}} \mathrm{sinh}\left( |z|\frac{\tilde{\kappa}^{(12)}_n}{|\mathrm{sin}\phi|}  \right).\label{dom_2}
\end{eqnarray}
Extracting the dominant contribution (\ref{dom_2}) from the first sum and summing it in an analytic form we arrive at
\begin{eqnarray}
&\quad \sum_{n'=1}^{\infty} \mathrm{i}\frac{\tilde{\kappa}^{(12)}_{nn';2}}{w^{(12)}_{nn'}} \left( \mathrm{e}^{\mathrm{i}|z|w^{(12)}_{nn'}} - \mathrm{e}^{\mathrm{i}|z|w^{(12)}_{n,-n'}}  \right) = -\frac{2}{|\mathrm{sin}\phi|} \frac{\mathrm{sinh}\left( |z| \frac{\tilde{\kappa}^{(12)}_n}{|\mathrm{sin}\phi|} \right) }{1 - \mathrm{e}^{-\frac{2\pi|z|}{a_2|\mathrm{sin}\phi|} }} \nonumber\\
&\qquad\qquad\qquad\qquad\qquad\qquad+\sum_{n'=1}^{\infty} \bigg[ \mathrm{i}\frac{\tilde{\kappa}^{(12)}_{nn';2}}{w^{(12)}_{nn'}} \left( \mathrm{e}^{\mathrm{i}|z|w^{(12)}_{nn'}} - \mathrm{e}^{\mathrm{i}|z|w^{(12)}_{n,-n'}}  \right) \nonumber\\
&\qquad\qquad\qquad\qquad\qquad\qquad\qquad+\frac{2}{|\mathrm{sin}\phi|} \mathrm{e}^{-\frac{2\pi \left|n'\right||z|}{a_2|\mathrm{sin}\phi|}} \mathrm{sinh}\left( |z|\frac{\tilde{\kappa}^{(12)}_n}{|\mathrm{sin}\phi|}  \right) \bigg]. \label{summ}
\end{eqnarray}
We take the $z\rightarrow 0$ limit of (\ref{summ}). The summation in the second term on the right-hand side of (\ref{summ}) vanishes in that limit, and the only contribution comes from the first term on the right-hand side of (\ref{summ}). We get
\begin{eqnarray}
\quad \lim_{z\rightarrow 0}\sum_{n'=1}^{\infty} \mathrm{i}\frac{\tilde{\kappa}^{(12)}_{nn';2}}{w^{(12)}_{nn'}} \left( \mathrm{e}^{\mathrm{i}|z|w^{(12)}_{nn'}} - \mathrm{e}^{\mathrm{i}|z|w^{(12)}_{n,-n'}}  \right) = -\frac{a_2 \tilde{\kappa}^{(12)}_n}{\pi|\mathrm{sin}\phi|}.\label{z1}
\end{eqnarray}
Using the result (\ref{z1}) in (\ref{dec_dom}) we find
\begin{eqnarray}
\quad \lim_{z\rightarrow 0}\sum_{n'=1}^{\infty} \mathrm{i} \left( \frac{\tilde{\kappa}^{(12)}_{nn';2}}{w^{(12)}_{nn'}} \mathrm{e}^{\mathrm{i}|z|w^{(12)}_{nn'}} + \left(n' \rightarrow -n'\right) \right) =-\frac{a_2 \tilde{\kappa}^{(12)}_n}{\pi|\mathrm{sin}\phi|} + \mathrm{i}\sum_{n'=1}^{\infty}  \left( \frac{\tilde{\kappa}^{(12)}_{nn';2}}{w^{(12)}_{nn'}} + \frac{\tilde{\kappa}^{(12)}_{n,-n';2}}{w^{(12)}_{n,-n'}} \right). \label{fin_int}
\end{eqnarray}
The summand on the right-hand side of (\ref{fin_int}) drops-off as $1/n^3$ and we accelerate that sum extracting the dominant contribution. We arrive at
\begin{eqnarray}
\quad \lim_{z\rightarrow 0}\sum_{n'=1}^{\infty} \mathrm{i}\left( \frac{\tilde{\kappa}^{(12)}_{nn';2}}{w^{(12)}_{nn'}} \mathrm{e}^{\mathrm{i}|z|w^{(12)}_{nn'}} + \left(n' \rightarrow -n'\right) \right) = &-\frac{a_2 \tilde{\kappa}^{(12)}_n}{\pi|\mathrm{sin}\phi|} -2\zeta(3)\left( \tau^{(1)} \right)^3 t^{(12)}_n \nonumber\\
& + \sum_{n'=1}^{\infty}  \left( \frac{\mathrm{i}\tilde{\kappa}^{(12)}_{nn';2}}{w^{(12)}_{nn'}} + \frac{\mathrm{i}\tilde{\kappa}^{(12)}_{n,-n';2}}{w^{(12)}_{n,-n'}} +2\left( \frac{\tau^{(1)}}{\left|n'\right|}  \right)^3t_n^{(12)} \right). \label{snigaaab}
\end{eqnarray}
Using (\ref{snigaaab}) in (\ref{asp}) we find
\begin{eqnarray}
\quad \lim_{z\rightarrow 0}\sum_{n'\neq 0} \mathrm{e}^{-\mathrm{i}\psi^{(1)}_{nn'}} \mathfrak{f}_{nn'}(z) 
&= \frac{2\mathrm{i}}{a_2 |\mathrm{sin}\phi|}  \frac{\tilde{\kappa}^{(12)}_n}{w^{(12)}_{n0}} + \frac{2}{a_2} |\mathrm{sin}\phi| \left( -\frac{a_2 \tilde{\kappa}^{(12)}_n}{\pi|\mathrm{sin}\phi|} -2\zeta(3)\left( \tau^{(1)} \right)^3 t^{(12)}_n \right) \nonumber\\
&+ \frac{2}{a_2} |\mathrm{sin}\phi| \sum_{n'=1}^{\infty}  \left( \frac{\mathrm{i}\tilde{\kappa}^{(12)}_{nn';2}}{w^{(12)}_{nn'}} + \frac{\mathrm{i}\tilde{\kappa}^{(12)}_{n,-n';2}}{w^{(12)}_{n,-n'}} +2\left( \frac{\tau^{(1)}}{\left|n'\right|}  \right)^3t_n^{(12)} \right).\label{dgsrt}
\end{eqnarray}
Finally using (\ref{dgsrt}) in (\ref{afp}) we get
\begin{eqnarray}
\quad \tilde{\mathcal{S}}^p_{(1112)} &=  \tilde{\mathcal{S}}^p_{(1111)}\mathrm{cos}\phi -\sum_{n\notin (n_-^{(1)},n_+^{(1)})} \frac{Q^{(1)}_{n} \left(w^{(1)}_{n}\right)^2}{2A_c} N^{(1)}_n \nonumber\\
& -\sum_{n=n_-^{(1)}}^{n_+^{(1)}} \frac{Q^{(1)}_n \left( w^{(1)}_n \right)^2 }{2A_c} \left[ T^{(1)}_n +\tilde{T}^{(1)}_n  + \mathrm{i}\mathrm{Im} \left(\sum_{n'}\frac{\mathrm{i} \kappa^{(12)}_{nn';2}}{w^{(12)}_{nn'}} \right) \mathrm{sin}^2\phi \right]. \label{p_fin}
\end{eqnarray}
Adding the contributions (\ref{ch}) and (\ref{p_fin}) and taking the real part we find the real contribution to $\tilde{\mathcal{S}}_{(1112)}$ in the form identified in the paper,
\begin{equation}
\quad \mathrm{Re} \tilde{\mathcal{S}}_{(1112)} = \mathrm{Re} \tilde{\mathcal{S}}^p_{(1112)} + \mathrm{Re} \tilde{\mathcal{S}}^c_{(1112)} =  \mathcal{S}_{(1112)}.
\end{equation}
For the imaginary part we find
\begin{eqnarray}
&\quad \mathrm{Im}\tilde{\mathcal{S}}_{(1112)} =  \mathrm{Im}\tilde{\mathcal{S}}_{(1111)}\mathrm{cos}\phi - \mathrm{sin}^2\phi \sum_{nn'} \frac{Q^{(1)}_n \left( w^{(1)}_n \right)^2 \kappa^{(12)}_{nn';2} }{2A_c}  \mathrm{Im}\left( \frac{\mathrm{i} }{w^{(12)}_{nn'}}\right).\label{imf2}
\end{eqnarray}

\subsection*{Evaluation of $\tilde{\mathcal{S}}_{(1122)}$}
This partial sum is symmetric with respect to the two lattice directions, and so we do the decomposition in a symmetric way. We write
\begin{equation}
\quad \tilde{\mathcal{S}}_{(1122)} = \frac{1}{2}\left( \tilde{\mathcal{S}}^{(1)}_{(1122)} + \tilde{\mathcal{S}}^{(2)}_{(1122)}  \right),
\end{equation}
where $\tilde{\mathcal{S}}^{(i)}_{(1122)}$ indicates a representation of the partial sum found by dividing the 2d sum into the summation over a chain along  $\bm{\hat{a}}_i$ and a summation over a 2d plane with that chain removed; the representations $\tilde{\mathcal{S}}^{(1)}_{(1122)}$ and $\tilde{\mathcal{S}}^{(2)}_{(1122)}$ are of course equivalent, and two representations are introduced to write the final result in a form that is explicitly symmetric with respect to the two lattice directions. Below we compute $\tilde{\mathcal{S}}^{(1)}_{(1122)}$, the derivation of $\tilde{\mathcal{S}}^{(2)}_{(1122)}$ follows the same steps but with the lattice directions interchanged.  As before we write
\begin{equation}
\quad \tilde{\mathcal{S}}^{(1)}_{(1122)} = \tilde{\mathcal{S}}^{(1)c}_{(1122)} + \tilde{\mathcal{S}}^{(1)p}_{(1122)},
\end{equation}
where 
\begin{eqnarray}
\quad \tilde{\mathcal{S}}^{(1)c}_{(1122)} &= \lim_{z \rightarrow 0} \sum_{n\neq 0} \tilde{\mathcal{S}}_{(1122)} (na_1,0,z),\label{Sc1122} \\
\quad \tilde{\mathcal{S}}^{(1)p}_{(1122)} &= \lim_{z \rightarrow 0} \sum_{n=-\infty}^{\infty}\sum_{n'\neq 0} \tilde{\mathcal{S}}_{(1122)} (na_1,n'a_2,z).
\end{eqnarray}
First we consider the contribution (\ref{Sc1122}) from the chain. Directly from the partial sum definition we find
\begin{eqnarray}
\quad \tilde{\mathcal{S}}^{(1)c}_{(1122)} = 2\tilde{\mathcal{S}}^c_{(1112)} \mathrm{cos}\phi -\frac{1}{2} \tilde{\mathcal{S}}^c_{(1111)} \left( \mathrm{cos}^2\phi +1 \right) + \frac{2}{3} \tilde{\mathcal{S}}^c_{(11)}\left( \tilde{\omega}n\right)^2  \mathrm{sin}^2\phi.\label{Saabbchain}
\end{eqnarray}
Now we consider the summation over a 2d plane with the chain removed. We do a Poisson transformation with respect to the unrestricted index and find
\begin{eqnarray}
\quad \tilde{\mathcal{S}}^{(1)p}_{(1122)} &= 2\tilde{\mathcal{S}}^p_{(1112)} \mathrm{cos}\phi -\frac{1}{2} \tilde{\mathcal{S}}^p_{(1111)} \left( \mathrm{cos}^2\phi +1 \right) + \frac{2}{3} \tilde{\mathcal{S}}^p_{(11)}\left( \tilde{\omega}n\right)^2  \mathrm{sin}^2\phi \nonumber\\
& +\lim_{z\rightarrow 0}\sum_n \frac{\mathrm{i}\mathfrak{c}^{(1)}_n}{12a_1} \sum_{n'\neq 0} \mathrm{e}^{-\mathrm{i}\psi^{(1)}_{nn'}} \mathfrak{g}_{nn'}(z), \label{aabb_afp}
\end{eqnarray}
where we have put
\begin{equation}
\quad \mathfrak{g}_{nn'}(z) = \left[ \partial_{u_2}^2 + \frac{1}{2}\mathrm{sin}^2\phi\left( w^{(1)}_n \right)^2 \right] 
 H_0^{(1)}\left(  w^{(1)}_n \sqrt{ u_2^2\mathrm{sin}^2\phi +z^2 }   \right)\big|_{u_2=n'a_2}.
\end{equation}
Isolating the evanescent cylindrical waves in the summation on the right-hand side of (\ref{aabb_afp}) we get
\begin{eqnarray}
\quad \tilde{\mathcal{S}}^{(1)p}_{(1122)} &= 2\tilde{\mathcal{S}}^p_{(1112)} \mathrm{cos}\phi -\frac{1}{2} \tilde{\mathcal{S}}^p_{(1111)} \left( \mathrm{cos}^2\phi +1 \right)  + \frac{2}{3} \tilde{\mathcal{S}}^p_{(11)}\left( \tilde{\omega}n\right)^2\mathrm{sin}^2\phi \nonumber\\
& -\sum_{n\notin (n_-^{(1)},n_+^{(1)})} \frac{\mathfrak{c}^{(1)}_n}{6A_c} L^{(1)}_n +\lim_{z\rightarrow 0}\sum_{n=n_-^{(1)}}^{n_+^{(1)}} \frac{\mathrm{i}\mathfrak{c}^{(1)}_n}{12a_1} \sum_{n'\neq 0} \mathrm{e}^{-\mathrm{i}\psi^{(1)}_{nn'}} \mathfrak{g}_{nn'}(z).\label{ffr}
\end{eqnarray}
To accelerate the summation on the right-hand side of (\ref{ffr}) we carry out a second Poisson transformation,
\begin{eqnarray}
\quad \sum_{n'\neq 0} \mathrm{e}^{-\mathrm{i}\psi^{(1)}_{nn'}} \mathfrak{g}_{nn'}(z) 
 &= \frac{2|\mathrm{sin}\phi|}{a_2} \sum_{n'} \left[ \frac{1}{2} \left( w^{(1)}_n \right)^2 - \left( \kappa^{(12)}_{lm;2} \right)^2 |\mathrm{sin}\phi|^2 \right] \frac{\mathrm{e}^{\mathrm{i}|z|w^{(12)}_{nn'}}}{w^{(12)}_{nn'}} \nonumber\\
&-\frac{|\mathrm{sin}\phi|^2}{2} \left( w^{(1)}_n \right)^2 \left[ H_0^{(1)} \left( w^{(1)}_n|z| \right) - \frac{2}{|z|w^{(1)}_n} H_1^{(1)}\left( w^{(1)}_n|z| \right) \right].\label{aabb_asp}
\end{eqnarray}
We want to  find the $|z|\rightarrow 0$ limit  of (\ref{aabb_asp}). To do that we find the small parameter expansion of both lines in (\ref{aabb_asp}). While the expression (\ref{aabb_asp}) is well defined in the $|z|\rightarrow 0$ limit, when taken separately the two terms on the right hand side of (\ref{aabb_asp}) involve a singularity at $z= 0$. In the second line of (\ref{aabb_asp}) we use the small parameter expansion of the Hankel functions,
\begin{eqnarray}
\quad -\frac{|\mathrm{sin}\phi|^2}{2} \left( w^{(1)}_n \right)^2 \left[ H_0^{(1)} \left( w^{(1)}_n|z| \right) - \frac{2}{|z|w^{(1)}_n} H_1^{(1)}\left( w^{(1)}_n|z| \right) \right] = \frac{|\mathrm{sin}\phi|^2}{2\pi \mathrm{i}} \left[ \left(w^{(1)}_n\right)^2+ \frac{4}{|z|^2 } \right] + O(|z|),
\end{eqnarray}
and then we find a small $z$ expansion of the first line in (\ref{aabb_asp}). We extract the dominant part in the sum on the right-hand side,
\begin{eqnarray}
&\quad \sum_{n'} \left[ \frac{1}{2} \left( w^{(1)}_n \right)^2 - \left( \kappa^{(12)}_{nn';2} \right)^2|\mathrm{sin}\phi|^2 \right] \frac{\mathrm{e}^{\mathrm{i}|z|w^{(12)}_{nn'}}}{w^{(12)}_{nn'}} =  \nonumber\\
&\quad =\left[ \frac{1}{2} \left( w^{(1)}_n \right)^2 - \left( \kappa^{(12)}_{n0;2} \right)^2|\mathrm{sin}\phi|^2 \right] \frac{\mathrm{e}^{\mathrm{i}|z|w^{(12)}_{n0}}}{w^{(12)}_{n0}} +\sum_{n'\neq 0} \left( X_{nn'} + Y_{nn'}  \right) \nonumber\\
&\quad +\sum_{n'\neq 0}\left( \left[ \frac{1}{2} \left( w^{(1)}_n \right)^2 - \left( \kappa^{(12)}_{nn';2} \right)^2 |\mathrm{sin}\phi|^2 \right] \frac{\mathrm{e}^{\mathrm{i}|z|w^{(12)}_{nn'}}}{w^{(12)}_{nn'}} -X_{nn'} - Y_{nn'} \right),
\end{eqnarray}
where we put
\begin{eqnarray}
\quad X_{nn'} &= \frac{1}{\mathrm{i}}\kappa^{(12)}_{n0;2}|\mathrm{sin}\phi|\mathrm{sgn}(n')\mathrm{exp}\left(  -\frac{2\pi|z||n'|}{a_2|\mathrm{sin}\phi|} + |z|\kappa^{(12)}_{n0;2}|\mathrm{sin}\phi|\mathrm{sgn}(n') \right), \\
\quad Y_{nn'} &= -\frac{2\pi|n'|}{\mathrm{i}a_2|\mathrm{sin}\phi|}\mathrm{exp}\left(  -\frac{2\pi|z||n'|}{a_2|\mathrm{sin}\phi|} + |z|\kappa^{(12)}_{n0;2}|\mathrm{sin}\phi|\mathrm{sgn}(n') +|z|\left(w^{(1)}_n\right)^2 \frac{a_2 |\mathrm{sin}\phi|}{4\pi |n'|} \right).
\end{eqnarray}
We can evaluate the sum over the extracted terms $X_{nn'}$ and $Y_{nn'}$ in an analytic form. Neglecting terms of the order $|z|$ and higher we find
\begin{eqnarray}
\quad \sum_{n'\neq 0}  X_{nn'}  =& \frac{1}{2\mathrm{i}}\tau^{(1)}|\mathrm{sin}\phi|^2\left(\kappa^{(12)}_{n0;2}\right)^2 +O(|z|), \\
\quad \sum_{n'\neq 0}  Y_{nn'}  =& -\frac{1}{\mathrm{i}}\frac{a_2|\mathrm{sin}\phi|}{\pi |z|^2} + \frac{1}{\mathrm{i}}\frac{1}{6 \tau^{(1)}} -\frac{1}{\mathrm{i}}\tau^{(1)}\left[ \left(w^{(1)}_n\right)^2 +|\mathrm{sin}\phi|^2\left(\kappa^{(12)}_{n0;2}\right)^2 \right] +O(|z|).\label{fn}
\end{eqnarray}
Using (\ref{aabb_asp}-\ref{fn}) we eventually find
\begin{eqnarray}
\quad \lim_{z\rightarrow 0}\sum_{n'\neq 0} \mathrm{e}^{-\mathrm{i}\psi^{(1)}_{nn'}} \mathfrak{g}_{nn'}(z) 
 &=\frac{2|\mathrm{sin}\phi|}{\mathrm{i}a_2}\frac{1}{6\tau^{(1)}} \nonumber\\
 &+ \frac{2|\mathrm{sin}\phi|}{\mathrm{i}a_2} \left[ \left(w^{(12)}_{n0}\right)^2  - \frac{1}{2}\left(w^{(1)}_n\right)^2 \right] \left( \frac{\mathrm{i}}{w^{(12)}_{n0}} - \tau^{(1)}  \right) \nonumber\\
 &+ \frac{2|\mathrm{sin}\phi|}{\mathrm{i}a_2} \sum_{n'\neq 0}\left( \left[ \left(w^{(12)}_{nn'}\right)^2  - \frac{1}{2}\left(w^{(1)}_n\right)^2 \right]  \frac{\mathrm{i}}{w^{(12)}_{nn'}} + \frac{|n'|}{\tau^{(1)}} \right),\label{ast}
\end{eqnarray}
where we used
\begin{equation}
\quad \frac{1}{2}\left(w^{(1)}_n\right)^2 - |\mathrm{sin}\phi|^2 \left( \kappa^{(12)}_{nn';2} \right)^2 = \left(w^{(12)}_{nn'}\right)^2 - \frac{1}{2}\left(w^{(1)}_n\right)^2.
\end{equation}
Finally we accelerate the summation on the right-hand side of (\ref{ast}). The summand in the last term on the right-hand side of (\ref{ast}) drops-off as $1/n^3$. We accelerate the summation using the dominant part extraction method, and find
\begin{eqnarray}
&\quad \sum_{n'\neq 0}\left( \left[ \left(w^{(12)}_{nn'}\right)^2  - \frac{1}{2}\left(w^{(1)}_n\right)^2 \right]  \frac{\mathrm i}{w^{(12)}_{nn'}} + \frac{|n'|}{\tau^{(1)}} \right) = \nonumber\\
&\quad = \sum_{n'\neq 0}\left( \left[ \left(w^{(12)}_{nn'}\right)^2  - \frac{1}{2}\left(w^{(1)}_n\right)^2 \right]  \frac{\mathrm i}{w^{(12)}_{nn'}} + \frac{|n'|}{\tau^{(1)}} +\frac{1}{2} \left( \frac{\tau^{(1)}}{|n'|} \right)^3 o^{(12)}_{n} \right) -\left( \tau^{(1)} \right)^3  \zeta(3) o^{(12)}_{n}.\label{aabbd}
\end{eqnarray}
Using (\ref{aabbd}) in (\ref{ast}) we eventually find
\begin{eqnarray}
\quad \lim_{z\rightarrow 0}\sum_{n'\neq 0} \mathrm{e}^{-\mathrm{i}\psi^{(1)}_{nn'}} \mathfrak{g}_{nn'}(z) &= \frac{1}{\mathrm i}\frac{2}{a_2|\mathrm{sin}\phi|}\left( O^{(1)}_n + \tilde{O}^{(1)}_n  \right) + \frac{2 |\mathrm{sin}\phi|}{a_2}\sum_{n'\neq 0} \left[ \left(w^{(12)}_{nn'}\right)^2  - \frac{1}{2}\left(w^{(1)}_n\right)^2 \right]  \mathrm{Im}\frac{\mathrm i}{w^{(12)}_{nn'}}. \label{inf}
\end{eqnarray}
Using (\ref{inf}) in (\ref{ffr}) and adding (\ref{Saabbchain}), we find the real part of the complex partial sum in the form identified in the paper,
\begin{equation}
\quad \mathrm{Re}\tilde{\mathcal{S}}^{(1)}_{(1122)}  =  \mathrm{Re}\tilde{\mathcal{S}}^{(1)p}_{(1122)}+ \mathrm{Re}\tilde{\mathcal{S}}^{(1)c}_{(1122)} = \mathcal{S}^{(1)}_{(1122)}.
\end{equation}
Taking the imaginary part we find
\begin{eqnarray}
\quad \mathrm{Im}\tilde{\mathcal{S}}^{(1)}_{(1122)} &= 2  \mathrm{Im}\tilde{\mathcal{S}}_{(1112)}\mathrm{cos}\phi -\frac{1}{2} \mathrm{Im}\tilde{\mathcal{S}}_{(1111)} \left( \mathrm{cos}^2\phi +1 \right) + \frac{2}{3} \mathrm{Im}\tilde{\mathcal{S}}_{(11)}\left( \tilde{\omega}n\right)^2 \mathrm{sin}^2\phi \nonumber\\
&+\sum_{nn'} \frac{\mathfrak{c}^{(i)}_n}{6A_c} \mathrm{sin}^2\phi \left[ \left(w^{(12)}_{nn'}\right)^2  - \frac{1}{2}\left(w^{(1)}_n\right)^2 \right]  \mathrm{Im}\frac{\mathrm i}{w^{(12)}_{nn'}}.\label{imf3}
\end{eqnarray}

\section*{Evaluation of the periodic Green function dyadic}
We evaluate the $\mathcal{G}^{\mathrm{Fq}}_{xxxx}$ component of the periodic Green function dyadic using the representations of the partial sums. The remaining components can be found in an analogous manner. 

We find the non-radiative contribution first. Noting that the non-radiative contribution to the periodic Green function dyadic $\bm{\mathcal{G}}^{\mathrm{Fq}}$ is identified by its real contribution (see Appendix B in the paper) we find 
\begin{eqnarray}
\quad {}^{\mathrm{N}}\mathcal{G}^{\mathrm{Fq}}_{xxxx} = \mathrm{Re}G^{\mathrm{Fq}}_{xxxx}= \left(\tilde{\omega}n\right)^2{}^{\mathrm{N}}\mathcal{G}^{\mathrm{Ep}}_{xx} + \frac{\mu}{\epsilon_0 n^2}  \sum_{(lmnp)} \frac{C^{(lmnp)}_{xxxx} \mathcal{S}_{(lmnp)}}{\left[\left(\bm{\hat{a}}_1\times\bm{\hat{a}}_2\right)\cdot \bm{\hat{z}} \right]^4},\label{finalxxxxre}
\end{eqnarray}
where to arrive at the second line we used (\ref{GxxxxS}) and the relations
\begin{eqnarray}
&\quad \mathrm{Re}\bm{G}^{\mathrm{Ep}} = {}^{\mathrm{N}}\bm{\mathcal{G}}^{\mathrm{Ep}}, \\
&\quad \mathrm{Re} \tilde{\mathcal{S}}_{(lmnp)} = \mathcal{S}_{(lmnp)}.
\end{eqnarray}
Expression (\ref{finalxxxxre}) is the representation given in the paper.

We find the radiative contribution to the quadrupole dyadic which is identified by its imaginary part, ${}^{\mathrm{R}}\bm{\mathcal{G}}^{\mathrm{Fq}} = i\mathrm{Im}\bm{\mathcal{G}}^{\mathrm{Fq}} $. We consider here the $\mathcal{G}^{\mathrm{Fq}}_{xxxx}$ component. The radiative contribution is a sum of two terms,
\begin{equation}
\quad {}^{\mathrm{R}}\mathcal{G}^{\mathrm{Fq}}_{xxxx} = {}^{\mathrm{R}}G_{xxxx}^{\mathrm{Fq}} + \mathcal{R}_{xxxx}^{\mathrm{Fq}}.\label{twoterms_sup}
\end{equation}
The first contribution to (\ref{twoterms_sup}) we find using (\ref{GxxxxS}),
\begin{equation}
\quad {}^{\mathrm{R}} G^{\mathrm{ Fq}}_{xxxx} = \left(\tilde{\omega}n\right)^2{}^{\mathrm{R}} G^{\mathrm{ Ep}}_{xx} + \frac{i\mu}{\epsilon_0 n^2}  \sum_{(lmnp)} \frac{C^{(lmnp)}_{xxxx} \mathrm{Im}\tilde{\mathcal{S}}_{(lmnp)}}{\left[\left(\bm{\hat{a}}_1\times\bm{\hat{a}}_2\right)\cdot \bm{\hat{z}} \right]^4},
\end{equation}
where we used ${}^{\mathrm{R}}\bm{\mathcal{G}}^{\mathrm{Ep}} = i\mathrm{Im}\bm{\mathcal{G}}^{\mathrm{Ep}} $. Now using in (\ref{fin_int}) the expression
\begin{eqnarray}
\quad {}^{\mathrm{R}}G^{\mathrm{Ep}}_{xx} = -\frac{i\mu}{\epsilon_0 n^2}\frac{1}{4\pi}\frac{2}{3}(\tilde{\omega}n)^3 + \frac{i\mu}{\epsilon_0 n^2}\sum_{nn'}\frac{\tilde{\omega}^2n^2-\left(\bm{\kappa}^{(12)}_{nn'}\cdot\bm{\hat{x}}\right) ^2}{2A_c} \mathrm{Im} \frac{i}{w^{(12)}_{nn'}},
\end{eqnarray}
which was not derived here but can be found following the steps outlined here for the quadrupole Green function, and then using the expressions (\ref{imf1}, \ref{imf2}, \ref{imf3}) for the imaginary contributions to the partial sums, we find
\begin{eqnarray}
\quad {}^{\mathrm{R}} G^{\mathrm{Fq}}_{xxxx} = &- \frac{2}{15}\frac{i\mu}{\epsilon_0 n^2}\frac{(\tilde{\omega}n)^5}{4\pi}+\frac{i\mu}{\epsilon_0 n^2}\sum_{nn'} \frac{ \left(\bm{\kappa}^{(12)}_{nn'}\cdot \bm{\hat{x}}\right)^2 \left[\tilde{\omega}^2n^2-\left(\bm{\kappa}^{(12)}_{nn'}\cdot\bm{\hat{x}}\right) ^2\right] }{2A_c}\mathrm{Im}\frac{i}{w^{(12)}_{nn'}}. 
\end{eqnarray}
Adding now the radiation reaction term from the quadrupole at the origin we arrive at the total contribution to the radiative periodic Green function,
\begin{eqnarray}
\quad {}^{\mathrm{R}} \mathcal{G}^{\mathrm{Fq}}_{xxxx} = \frac{i\mu}{\epsilon_0 n^2}\sum_{nn'} \frac{ \left(\bm{\kappa}^{(12)}_{nn'}\cdot \bm{\hat{x}}\right)^2 \left[\tilde{\omega}^2n^2-\left(\bm{\kappa}^{(12)}_{nn'}\cdot\bm{\hat{x}}\right) ^2\right] }{2A_c}\mathrm{Im}\frac{i}{w^{(12)}_{nn'}}. \label{fin_im}
\end{eqnarray}
Expression (\ref{fin_im}) can be easily seen to be the Cartesian component of the representation given in the paper (see (95) together with (73) in the paper) in the basis associated with the wave vectors $(\bm{\hat{s}}_{\bm{n}}, \bm{\hat{\kappa}}_{\bm{n}}, \bm{\hat{z}})$.

\end{document}